\documentclass[11pt]{article}
\pdfoutput=1
\usepackage{jcapmod}

\usepackage{tikz}
\usetikzlibrary{calc}
\usepackage{booktabs}
\usepackage{bookmark}
\usepackage{dsfont}
\usepackage{geometry}
\usepackage[english]{babel}
\usepackage{amsmath,amssymb,amsbsy,amstext, amsthm, simplewick}
\usepackage{hyperref}
\usepackage{graphicx}
\usepackage{amsfonts}
\usepackage[small]{caption}
\usepackage{upgreek}
\usepackage{hyperref}
\usepackage{slashed}
\usepackage[makeroom]{cancel}
\usepackage{tabularx}
\usepackage[normalem]{ulem}

\usepackage[export]{adjustbox}
\usepackage{bm}


\definecolor{Red}{RGB}{214, 39, 40}
\definecolor{Blue}{RGB} {31, 119, 180}
\definecolor{Orange}{RGB}{255, 153, 51}
\definecolor{Purple}{RGB}{178, 102, 255}
\definecolor{Green}{RGB}{44, 160, 44}
\definecolor{regal}{RGB}{90,0,120}

\definecolor{darkblue}{rgb}{0.15,0.35,0.55}
\definecolor{reddish}{rgb}{0.65, 0.2, 0.2}
\usepackage[linktocpage=true]{hyperref}
\hypersetup{
colorlinks=true,
citecolor=darkblue,
linkcolor=reddish,
urlcolor=darkblue,
pdfauthor={},
pdftitle={},
pdfsubject={}
}

\newcolumntype{L}[1]{>{\raggedright\arraybackslash}p{#1}}
\newcolumntype{C}[1]{>{\centering\arraybackslash}p{#1}}
\newcolumntype{R}[1]{>{\raggedleft\arraybackslash}p{#1}}
\newcolumntype{M}[1]{>{\centering\arraybackslash}m{#1}}
\newcolumntype{N}{@{}m{0pt}@{}}

\makeatletter
\g@addto@macro\bfseries{\boldmath}
\makeatother

\def\d{{\rm d}}

\usepackage{fontawesome5}

\makeatletter
\newcommand{\github}[1]{%
  \href{#1}{\faGithub}%
}
\makeatother

\NewDocumentCommand{\colornucleus}{omme{_^}}{%
 \begingroup\colorlet{currcolor}{.}%
  \IfValueTF{#1}
  {\textcolor[#1]{#2}}
  {\textcolor{#2}}
  {%
 #3% the nucleus
  \IfValueT{#4}{_{\textcolor{currcolor}{#4}}}% subscript
 \IfValueT{#5}{^{\textcolor{currcolor}{#5}}}% superscript
 }%  \endgroup
}

\newcommand{\la}{\langle}
\newcommand{\ra}{\rangle}

\def \cD {{\cal D}}
\def \cE {{\cal E}}
\def \cP {{\cal P}}

\newcommand \LA {{\langle}}
\newcommand \RA {{\rangle}}

\usepackage{colortbl}
\definecolor{lightgreen}{cmyk}{0.2, 0, 0.2, 0.2}
\definecolor{lightgray}{cmyk}{0.1,0.2,0,0.1}
\definecolor{lightgray2}{cmyk}{0.1,0.1,0,0.1}
\definecolor{repBlue}{RGB}{31, 119, 180}
\definecolor{repRed}{RGB}{	214, 39, 40}
\definecolor{repGreen}{RGB}{44, 160, 44}
\definecolor{repOrange}{RGB}{255, 127, 14}
\definecolor{repViolet}{RGB}{102,0,204}

\makeatletter
\newlength{\apb@width}
\newcommand{\autoparbox}[2][c]{\settowidth{\apb@width}{#2}\parbox[#1]{\apb@width}{#2}}

\makeatother

\usepackage[framemethod=default]{mdframed}
\newmdenv[skipabove=7pt,
skipbelow=7pt,
rightline=true,
leftline=true,
topline=true,
bottomline=true,
backgroundcolor=gray!10,
linecolor=black,
innerleftmargin=5pt,
innerrightmargin=5pt,
innertopmargin=5pt,
innerbottommargin=5pt,
leftmargin=0cm,
rightmargin=0cm,
linewidth=1pt]{eBox}

\usetikzlibrary{shapes.misc}

\numberwithin{equation}{section}

\usepackage{slashed}

\def\beq{\begin{equation}}
	\def\eeq{\end{equation}}
\def\bea{\begin{eqnarray}}
	\def\eea{\end{eqnarray}}
\def\be{\begin{equation}}
	\def\ee{\end{equation}}
	
\def\x{x}
\def\z{z}
\def\k{k}
\def\Z{Z}
\def\O{O}

\def\hs{\hskip 1pt}

\allowdisplaybreaks

\usepackage{tocloft}

\begin{document}

	\newgeometry{top=2cm, bottom=2cm, left=2cm, right=2cm}

	\begin{titlepage}
		\setcounter{page}{1} \baselineskip=15.5pt 
		\thispagestyle{empty}

		\begin{center}
			{\fontsize{19.}{18} \bf Constraints on  Long-Range Forces in De Sitter Space}
		\end{center}

		\vskip 20pt
		\begin{center}
			\noindent
			{\fontsize{14}{18}\selectfont 
				Daniel Baumann\hs$^{1,2}$, Kurt Hinterbichler\hs$^{3}$, Callum R.~T.~Jones\hs$^{1}$,\\[8pt]  
				Austin Joyce\hs$^{4,5}$, Hayden Lee\hs$^{6}$, 
				Jiajie Mei\hs$^{1}$ and Nathan Meurrens\hs$^{1,2}$}
		\end{center}

		\begin{center}
			\vskip8pt
			\textit{$^1$ Institute of Physics, University of Amsterdam, Amsterdam, 1098 XH, The Netherlands}
			
			\vskip8pt
			\textit{$^2$  Leung Center for Cosmology and Particle Astrophysics,
				Taipei 10617, Taiwan}
			
			\vskip8pt
			\textit{$^3$  CERCA, Department of Physics,
				Case Western Reserve University, Cleveland, OH 44106, USA}
			
			\vskip 8pt
			\textit{$^4$ 
				Department of Astronomy and Astrophysics,
				University of Chicago, Chicago, IL 60637, USA}
			
			\vskip 8pt
			\textit{$^5$ Kavli Institute for Cosmological Physics, 
				University of Chicago, Chicago, IL 60637, USA}
			
			\vskip 8pt
			\textit{$^6$ Center for Particle Cosmology, Department of Physics and Astronomy,\\[3pt] University of Pennsylvania, Philadelphia, PA 19104, USA}
			
		\end{center}

%=========================================
\vspace{0.4cm}
\begin{center}{\bf Abstract}
\end{center}
\noindent	
The representation theory of de Sitter space admits partially massless (PM) particles, 
but whether such particles can participate in consistent interacting theories remains unclear.
 We investigate the consistency of theories containing PM fields, particularly when these fields are coupled to gravity. Our strategy exploits the fact that PM fields correspond to partially conserved currents on the spacetime boundary, which generate symmetries. These symmetries place stringent constraints on correlation functions of charged operators, allowing us to test the consistency of a proposed bulk spectrum. When the assumed operator content violates these constraints, the corresponding bulk theory is ruled out. Applying this framework, we show that, in four-dimensional de Sitter space, PM fields of spin 2 or 3 (at depth~0) cannot couple consistently to gravity: such couplings necessitate additional massive fields, which are inevitably non-unitary. In higher dimensions, however, the constraints can be satisfied without violating unitarity if further PM fields are included. The resulting structure leads to additional charge conservation laws, which suggests that consistency may ultimately require an infinite tower of higher-spin PM fields, akin to the situation for ordinary higher-spin symmetries.  The methods developed here provide powerful constraints on possible long-range interactions in de Sitter space and delineate the landscape of consistent quantum field theories in cosmological spacetimes.

\end{titlepage}
\restoregeometry

% Re-number pages
\clearpage
\setcounter{page}{2}
\makeatletter
\def\ps@plain{
  \renewcommand\@oddfoot{\hfill\thepage\hfill}
  \renewcommand\@evenfoot{\hfill\thepage\hfill}
}
\pagestyle{plain}
\makeatother

% TOC
\linespread{1.2}
\setcounter{tocdepth}{2}
\tableofcontents
\linespread{1.1}

	\newpage
\section{Introduction}\label{sec: Introduction}

The principles of quantum mechanics and relativity put remarkably stringent constraints on the possible laws of Nature. In flat space, this is manifested by the fact that consistent long-range forces---mediated by massless particles with integer spins---are almost unique~\cite{Weinberg:1965nx}.
There is a finite menu of possibilities: the interactions mediated by massless spin-1 particles must take the form of Yang--Mills theory, while those mediated by massless spin-2 particles lead to Einstein gravity with a universal coupling to matter~\cite{Benincasa:2007xk,Wald:1986bj}.  There are no long-range forces mediated by massless particles with spins greater than two~\cite{Weinberg:1965nx,Benincasa:2007xk,Wald:1986bj, Weinberg:1980kq,Porrati:2008rm,McGady:2013sga}.

\vskip 4pt
Constraints on long-range forces in flat space are most invariantly understood by studying the infrared structure of scattering amplitudes~\cite{Weinberg:1965nx,Benincasa:2007xk}. However, in curved spacetimes, the long-distance dynamics is inevitably sensitive to the large-scale curvature of the space. In general, not only is Lorentz invariance violated, but there is not even an S-matrix to imagine constraining.
It is therefore natural to wonder if long-range interactions are still restricted or if additional possibilities are available.
Indeed, it is known that the curvature of spacetime allows for new kinds of particles that have no counterparts in flat space. For example, in (anti)-de Sitter space, higher-spin fields can exhibit a higher-derivative kind of gauge invariance at special discrete mass values. These  ``partially massless" (PM) fields interpolate between massless fields (with full gauge symmetry) and massive fields (with none)~\cite{Deser:1983mm,Higuchi:1986py,Brink:2000ag,Deser:2001pe,Deser:2001us,Zinoviev:2001dt}. They are part of the unitary representations of the de Sitter isometry group, but have no analog in flat space.  If present during inflation, these PM particles would imprint distinctive signatures on inflationary correlators~\cite{Baumann:2017jvh,Goon:2018fyu,Baumann:2019oyu}, offering a new window into the de Sitter era of the early universe.

\vskip 4pt
Given this interesting family of representations that exist only in curved spacetime, 
the goal is to determine which, if any, can consistently interact. Such interacting theories would be similar to those of massless particles in flat spacetime, since PM particles have fixed masses of order one in units of the curvature scale and therefore would mediate long-range forces inside the cosmological horizon.
From this perspective of local fields, the challenge is to find interactions between the representations of interest that respect the gauge invariances of the PM particles. While cubic interactions exist for any choice of particles~\cite{Joung:2012rv}, 
they typically cannot be consistently extended to quartic order~\cite{Konstein:1989ij,deRham:2013wv, Joung:2014aba,Sleight:2021iix}, sharply constraining the space of viable theories. The space of theories has been mapped out in some cases, but working in terms of fields is algebraically quite involved, and is
subject to field reparametrization ambiguities. 

\vskip 4pt
In this paper, we will instead constrain PM theories in ${\rm (A)dS}_{d+1}$ by requiring the consistency of their dual boundary correlators. In the standard holographic dictionary, a PM field of spin $s$ and depth $t$ is dual to a partially conserved, symmetric tensor operator~\cite{Dolan:2001ih}:
\beq
\partial_{\mu_1 } \cdots \partial_{\mu_{s-t}} X^{\mu_1 \cdots \mu_s} = 0\,,
\eeq 
where $t \in \{0,1,\cdots,s-1\}$ and the case $t=s-1$ corresponds to an ordinary singly conserved current.
These partially conserved currents can be used to construct symmetry generators (charges), $Q_{(s,t)}^{{\mu_1...\mu_t}}$, whose presence greatly constrains correlation functions of charged operators. 
As explained by Maldacena and Zhiboedov~\cite{Maldacena:2011jn}, these charges lead to identities when they are inserted in
correlation functions of other operators. For an $n$-point function
of operators $\O_i$ (which may include the current itself), we get a ``charge conservation identity" of the form
\begin{equation}
		 \sum_{i=1}^n \langle \O_1 \ldots [Q,\O_i] \ldots \O_n \rangle = 0\,.
		 \label{equ:CCI}
	\end{equation}
	The action of the charge on any operator is a linear combination of local operators (assumed to be primaries and descendants), and can therefore be written schematically as $[Q,\O_i] = \sum_j a_{ij}\, \partial^{n_{ij}}\O_j$, where $a_{ij}$ are constant coefficients. Inserting this ansatz for the action of the charge into (\ref{equ:CCI}) then leads  to non-trivial relations between $n$-point functions. In some cases, these relations cannot be satisfied, ruling out the existence of the charge $Q$.
   
\vskip 4pt
We will apply this logic to three-point functions, which are largely determined by symmetry.  
The charge conservation identity~\eqref{equ:CCI} then becomes
\beq
	\label{equ:sumrule}
	\sum_j \Big[a_{1j} \partial^{n_{1j}} \langle \O_j\O_2\O_3 \rangle + a_{2j} \partial^{n_{2j}} \langle \O_1\O_j\O_3 \rangle  + a_{3j} \partial^{n_{3j}}\langle \O_1\O_2\O_j \rangle \Big] = 0\,.
	\eeq
Writing the three-point functions as a sum over conformally invariant structures,  $\langle \O_i \O_j \O_k \rangle = \sum_n c^{(n)}_{ijk} F_{ijk}^{(n)}$, the constraint~\eqref{equ:sumrule} becomes a bilinear in the coefficients $a_{ij}$ and $c_{ijk}^{(n)}$.  Given an assumed spectrum of operators, we then look for non-trivial solutions to the charge conservation identity. If no such solution can be found, a theory with this spectrum is rigorously ruled out.
This strategy can be thought of as a synthesis of the Weinberg soft constraints~\cite{Weinberg:1965nx} and the four-particle test~\cite{Benincasa:2007xk,Schuster:2008nh,McGady:2013sga} from flat space, adapted to curved backgrounds. The constraint~\eqref{equ:sumrule} can be obtained by integrating a four-point function involving a current insertion to produce the corresponding charge operator. Integrating over all of space extracts a piece of the kinematic information, precisely like the soft limit of a four-particle amplitude. In detail,
this involves two ingredients, similar in spirit to the four-particle test. We start with a set of allowed three-point functions and then impose consistency conditions that effectively require it to be possible for higher-point functions to be constructed from these building blocks.  Much as in flat space, the power of this approach stems from the fact that three-point functions are largely fixed by symmetry. In the present case, given a choice of operators, conformal symmetry limits the possible three-point functions to a finite list of possibilities. It is this finite space of possibilities that makes the constraints challenging to satisfy, and thus rules out many theories.

	\vskip 4pt
	A trivial way to satisfy the constraints for any spectrum is to set all three-point functions to zero, which is equivalent to turning off all cubic interactions in the bulk. The situation is different, however, if we assume that our model contains Einstein gravity, since the equivalence principle then mandates a cubic coupling of the graviton to every field with uniform strength---it is not possible to turn off gravity.  In the boundary CFT, this implies the existence of a conserved spin-2 operator $T_{\mu \nu}$, (the stress tensor), with non-vanishing three-point functions $\langle T_{\mu \nu}\O \O\rangle$, for all local operators~$\O$. 
For the theory of a PM field coupled to gravity to be consistent, the constraint~\eqref{equ:sumrule} therefore must have a non-trivial solution.
If there is no solution for an assumed spectrum of operators, we can conclude that no consistent model of such fields exists. 
	
	\vskip 4pt
For $d=3$, we find a strict no-go result for PM fields of spin 2 or 3 (and depth 0) coupled to gravity. In particular, we find that the charge conservation condition requires the addition of extra massive fields, but that these fields are non-unitary. For $d>3$, we derive consistent ``candidate theories" involving these particles without any violation of unitarity. We call these candidate theories, because we have only checked 
the charge conservation constraints for a subset of operators in the spectrum. 
In fact, we conjecture that these theories must involve an infinite tower of higher-spin PM fields, like in the case of ordinary higher-spin currents~\cite{Maldacena:2011jn}.
	
\paragraph{Outline}
	The outline of the paper is as follows: In Section~\ref{sec:background}, we review the properties of partially massless fields and explain how to construct ans\"atze for their conformally invariant three-point functions.
	 In Section~\ref{sec:Dynamics}, we then explain in more detail our strategy for using the charge conservation identity (\ref{equ:sumrule})  to constrain PM theories. 
	In Section~\ref{sec:PM-electrodynamics},  we apply this methodology to the case of partially massless electrodynamics, deriving no-go results for the coupling of PM fields to scalars.
	In Section~\ref{sec:PM-gravity}, we show that the couplings of PM fields to gravity are very constrained. For $d=3$, we prove rigorous no-go results, while, for $d>3$, we find candidate theories that may be consistent.  In Section~\ref{sec:SlightlyBroken}, we discuss a generalization of our treatment to theories where the PM gauge symmetry is Higgsed, which is relevant for understanding the case of conformal gravity.
 Finally, our conclusions are stated in Section~\ref{sec:conclusions}.
	
	\vskip 4pt
	Two appendices contain technical details: In Appendix~\ref{sec: Current Algebra of Box-Squared CFT}, 
	we present the current algebra for the $\square^2$ generalization of the large-$N$ free bosonic $U(N)$ vector model, which is used in Section~\ref{subsec:spin3extra}. 
		 In Appendix~\ref{app:confgrav}, we derive the cubic action for conformal gravity and compute the associated three-point functions.  
		 
		 \vskip4pt
		Finally, a supplemental {\sc Mathematica} notebook 
		 with explicit details on the results presented in Section~\ref{sec:PM-gravity} can be found at a GitHub repository for this paper~\href{https://github.com/CRTJones/PM-Consistency}{\faGithub}.

\begin{table}[t!]
\renewcommand{\arraystretch}{1.6}
\centering
\begin{tabular}{| C{0.05\textwidth}|L{0.12\textwidth}|  C{0.1\textwidth}|  C{0.12\textwidth}| C{0.1\textwidth}| L{0.15\textwidth}|}
\hline
 \cellcolor{gray!30}   $(s,t)$ &   \cellcolor{gray!30}   {Name}  &   \cellcolor{gray!30}   {Bulk} &  \cellcolor{gray!30}   {Boundary}  & \cellcolor{gray!30}   {Dual} &  \cellcolor{gray!30}   {Conservation} \\
\hline
 (0,0)& Scalar & $\Phi$  & $\phi$ & $O$ & --  \\\hline
  (1,0)& Photon & $A_M$  & $A_\mu$ & $J^\mu$ & $\partial_\mu J^\mu = 0$ \\\hline
    (2,1)& Graviton & $g_{MN}$  & $g_{\mu\nu}$ & $T^{\mu\nu}$ & $\partial_\mu T^{\mu\nu} = 0$ \\\hline
        (2,0)& PM spin-2 & $\Phi_{MN}$  & $\phi_{\mu\nu}$ & $X^{\mu\nu}$ & $\partial_\mu \partial_\nu X^{\mu\nu} = 0$ \\\hline
\end{tabular}
\caption{Examples of (partially) massless fields and associated (partially) conserved currents.}
\label{tab:notation}
\end{table}

	\paragraph{Notation}  Throughout the paper, we use natural units, $\hbar = c \equiv 1$, and the mostly plus convention for the metric. We will work in $D \ge 4$ spacetime dimensions and $d=D-1$ spatial dimensions.
	The coordinates of the bulk spacetime are $X^M$, with $M=0,1,\cdots, d$, while those of the boundary are $x^\mu$, with $\mu=1,\,\cdots, d$.  Position vectors are denoted by $\x$ and their magnitudes are $|\x|$. Embedding-space coordinates are $P^A$, with $A=0,1,\cdots, d+1$.
	Examples of bulk fields, as well as their boundary values and associated dual operators, are shown in Table~\ref{tab:notation}. 
	We will often represent the operators in an index-free notation:
	\be
O^{(s)}(\x,\z) \equiv O_{\mu_1 \ldots \mu_s}(\x) \,z^{\mu_1} \cdots z^{\mu_s}\,,
\ee
where  $z$ is an auxiliary null vector.

\newpage
\section{Review of Partial Masslessness} 
\label{sec:background}
	
We begin with a brief review of the properties of partially massless fields in (anti-)de Sitter space and show how they lead to partially conserved currents in a dual boundary description. Correlators in the dual theory are constrained by conformal symmetry and we describe how to construct three-point functions involving the partially conserved currents.  We explain that PM fields are necessarily below the unitarity bound in AdS, but are unitary in dS.  
	
\subsection{Partially Massless Fields}
\label{ssec:PM-Fields}
	
 We consider de Sitter space in the flat slicing, so that the line element is
\beq
\d s^2 = \frac{L^2}{\eta^2}\left(-\d \eta^2+\d \x^2\right) ,
\label{equ:dS-metric}
\eeq
where $-\infty < \eta <0$ is conformal time and $L$ is the curvature radius.
We first describe the dynamics of bulk fields and then discuss their behavior near the boundary at $\eta \to 0$.

\paragraph{Bulk fields}
Particles with spin $s$ are described by symmetric tensor fields $\Phi_{M_1 \cdots M_s}$, 
which are both transverse ($\nabla^M\Phi_{M \,N_2\cdots N_s} = 0$) and traceless ($\Phi^M_{\ \ M \,N_3\cdots N_{s}} = 0$), and obey the following equation of motion\footnote{It is possible to write an action that produces this equation of motion~\cite{Singh:1974qz,Hallowell:2005np}, but this requires auxiliary fields which will play no role in the following, so it is simpler to directly consider the equation of motion.} 
\beq
\Big(\square-\big[d-(s-1)(s+d-4)\big] L^{-2} -m^2\Big)\Phi_{M_1\cdots M_s} = 0\,,
\eeq
where $D\equiv d+1$ is the spacetime dimension  and $\square$ is the d'Alembertian operator.
For generic values of the mass parameter $m^2$, this describes the propagation of  $N_s \equiv (D-3+2 s) (D-4+s)!/[s!(D-3)!]$ degrees of freedom---the same number as a massive spin-$s$ field in flat space.

\vskip4pt
In flat space, there is a single distinguished value of the mass, $m^2=0$, where the theory develops a gauge invariance and consequently propagates fewer degrees of freedom. 
However, a new phenomenon occurs in (A)dS space: In that case, there are additional special mass values, where the field has a smaller gauge invariance and propagates a number of degrees of freedom intermediate between the massive and massless cases. Because of this, these exceptional points are called {\it partially massless} fields.

\vskip4pt
We can describe both the massless and partially massless points by introducing a parameter, $t \in \{0,1,\,\cdots,s-1\}$, called the {\it depth} of partial masslessness. When the mass takes the discrete set of values
\beq
m^2 L^2 = (s-t-1)(s+t+d-3)\,,
\label{eq:PMmassvalues}
\eeq
the theory develops a gauge invariance, with a $t$-index symmetric tensor gauge parameter, of the  form
\beq
\delta_\xi \Phi_{M_1\cdots M_s} = \nabla_{(M_{t+1}}\cdots\nabla_{M_s}\xi_{M_1\cdots M_t)}+ \text{trace terms}\,,
\label{equ:gauge}
\eeq
where the detailed form of the  ``trace terms'' will not be needed in the following.
The gauge redundancy~\eqref{equ:gauge} removes the degrees of freedom with helicity  $0,\,\cdots,\pm t$.\footnote{A depth-$t$ field therefore propagates  
$N_s-N_t$ degrees of freedom (in general dimension).} 
Note that a spin-$s$ field has $s$ partially massless or massless points, with the depth $t = s-1$ corresponding to a massless field. Since all of the distinguished mass values (aside from the massless point) are proportional to the curvature scale of the manifold, they all collapse to $m^2 = 0$ in the flat-space limit, so that the phenomenon of partial masslessness has no direct analog in flat space. Partially massless fields exist in both de Sitter space and anti-de Sitter space (as well as their analytic continuations) and furnish irreducible representations of their group of symmetries. However, as we review in Section~\ref{ssec:dS-Rep}, these representations are only unitary in de Sitter space, where they fit into the exceptional series of unitary irreducible representations in generic dimension.\footnote{In $D=4$, they are instead elements of the discrete series.}
In addition, in de Sitter space,
 the $t=0$ point serves as a lower bound on the mass of a generic massive spinning representation (called the Higuchi bound~\cite{Higuchi:1986py}).
This makes the study of partially massless fields particularly interesting to cosmology, since they are so intrinsically connected to de Sitter space.

\paragraph{Boundary operators}  We will place constraints on the interactions of partially massless fields using the properties of their boundary correlation functions. In order to do this, we need to study how the spinning fields $\Phi_{M_1\cdots M_s}$ behave near the boundary of the spacetime (\ref{equ:dS-metric}). 

\vskip 4pt
Asymptotically (as $\eta\to 0$), we are interested in the components of the field that lie parallel to the boundary. On shell (or in an appropriate gauge), the transverse components vanish $\Phi_{0M_2\cdots M_s} = 0$ and the components of the field along the boundary $\Phi_{\mu_1\cdots \mu_s}$ can be taken to be transverse and traceless.
This transverse-traceless spatial part of the field then has two characteristic fall-offs
\beq
\Phi_{\mu_1\cdots \mu_s}(\eta, \x) \xrightarrow{\ \eta\to 0\ } ~\phi_{\mu_1\cdots\mu_s}(\x) \,\eta^{\Delta_--s}+ j_{\mu_1\cdots\mu_s}(\x)\,\eta^{\Delta_+-s}\,,
\label{eq:phifalloff}
\eeq
where the powers $\Delta_\pm$ are determined by the mass of the field as~\cite{Deser:2003gw}\footnote{Scalar fields ($s=0$) obey a different scaling, which coincides with~\eqref{eq:deltascaling} evaluated at $s=2$.}
\beq
\Delta_\pm = \frac{d}{2}\pm \sqrt{\left(\frac{d}{2}+s-2\right)^2- m^2L^2}\,.
\label{eq:deltascaling}
\eeq
The spacetime (A)dS isometries act as conformal transformations on the $\eta = 0$ boundary, under which $\phi_{\mu_1\cdots\mu_s}$ transforms like a spin-$s$ conformal primary of weight $\Delta_-$.
We view $\phi_{\mu_1\cdots\mu_s}$ as a source in a ``dual" description~\cite{Witten:1998qj,Maldacena:2002vr}.\footnote{In the AdS context, these sources are literally sources in a CFT partition function, while, in dS, they are boundary values of the bulk field, appearing in the cosmological wavefunction.} By studying the properties of the operators sourced by $\phi_{\mu_1\cdots\mu_s}$, we will be able to place constraints on the bulk interactions of $\Phi_{\mu_1\cdots \mu_s}$.

\vskip4pt
To efficiently deal with the index structure of $\phi_{\mu_1\cdots\mu_s}$, it is convenient to introduce an index-free notation~\cite{Dobrev:1975ru,Costa:2011mg}: 
\be
\phi^{(s)}(\x,\z) \equiv   \phi_{\mu_1 \ldots \mu_s}(\x) \,z^{\mu_1} \cdots z^{\mu_s}\,,
\ee
where  $\z$ is an auxiliary null vector  (which satisfies $\z^2 = 0$). 
This homogeneous polynomial in $\z$ of order $s$ naturally captures the symmetric and traceless part of tensors. One can extract the original tensor by acting with the  \textit{Thomas--Todorov operator}~\cite{thomas,Dobrev:1975ru,tractors} 
\begin{equation}\label{eq: Todorov operator}
		D^{\mu}_z \equiv \left(\frac{d}{2}-1 + \z \cdot \frac{\partial}{\partial \z}\right)\frac{\partial}{\partial z_{\mu}}-\frac{1}{2}z^{\mu} \frac{\partial^2 }{\partial \z \cdot \partial \z}\,.
\end{equation}
The slightly strange form of this operator guarantees that the tensor obtained from stripping off the auxiliary $\z$ vectors is traceless.

\vskip4pt
At the partially massless points~\eqref{eq:PMmassvalues}, the bulk field has the gauge invariance~\eqref{equ:gauge}. Choosing the gauge parameter to be $\xi^{(t)}(\eta,\x,\z) = \xi^{(t)}(\x,\z) \hs \eta^{\Delta_--s}$ (where  $\Delta_- = 1-t$ at the PM points), we see that, under a gauge transformation, the coefficient of the leading fall-off shifts as\footnote{Here, and in the following, $\partial$ (without a subscript) denotes the partial derivative with respect to the position~$x^\mu$. We will use $\partial_z$ and $D_z$ (with the subscript) when we want to highlight that the derivative is with respect to $z^\mu$.}
\beq
\delta_\xi \phi^{(s)} = (\z\cdot \partial)^{s-t} \xi^{(t)}\,,
\label{eq:boundarygauge}
\eeq
while $j^{(s)}$, the coefficient of the subleading fall-off,  is gauge invariant. The gauge transformation~\eqref{eq:boundarygauge} implies that $\phi^{(s)}$ must source an operator $J_{(s,t)}$ that satisfies the partial conservation condition~\cite{Dolan:2001ih,Deser:2003gw}
\beq
(\partial \cdot D_z)^{s-t} J_{(s,t)} = 0\,.
\label{eq:partiallymasslessconservation}
\eeq
Similar to the source $\phi^{(s)}$, the current $J^{(s)}$ transforms like a conformal primary field of spin $s$. Its conformal dimension is 
\beq
\Delta = d-1+t\,,
\label{eq:shorteningcondition}
\eeq
which coincides with $\Delta_+ = d-\Delta_-$. This weight~\eqref{eq:shorteningcondition} is the only value compatible with both the shortening condition~\eqref{eq:partiallymasslessconservation} and conformal invariance~\cite{Dolan:2001ih}. Note that this value lies below the unitarity bound for a Lorentzian conformal field theory~\cite{Minwalla:1997ka,Rychkov:2016iqz,Simmons-Duffin:2016gjk}, which is another statement of the fact that partially massless fields are non-unitary in AdS space. In contrast, as we review in Section~\ref{ssec:dS-Rep}, PM fields correspond to unitary Euclidean CFT representations.

\paragraph{Wavefunction}  Our goal is to constrain theories of PM fields in (A)dS using the consistency of the dual boundary CFT. This is most straightforward in AdS, where the partition function of bulk fields, thought of as a function of the boundary values of these fields, can be interpreted as a CFT partition function of dual operators~\cite{Gubser:1998bc,Witten:1998qj}.\footnote{The association between bulk gauge fields and boundary conserved currents can be slightly subtle. While the global symmetries of the bulk and boundary descriptions must match, gauge invariances in the bulk do not necessarily imply the existence of boundary global symmetries. The simplest counterexample are confining gauge theories, but we will encounter another example in Section~\ref{sec:confgrav}.} The constraints on CFTs with partially conserved currents can  therefore be straightforwardly related to AdS physics. However, since partially massless fields are of greatest interest in cosmology, we will also want to interpret the CFT constraints that we derive in terms of de Sitter physics. 

\vskip 4pt
In de Sitter space, the natural object of interest is the field theory wavefunction
\beq
\Psi[\phi] \ =  \hspace{-8pt}\overset{{~~\Phi\,=\,\phi}}{\int} \hspace{-6pt}\mathcal{D} \Phi\, e^{iS[\Phi]}\,,
\label{eq:wfpathint}
\eeq
which
encodes the quantum state of the system. 
The wavefunction is a functional of the boundary field profile $\phi(\x)$, which arises in the final state boundary condition by fixing the coefficient of the ``non-normalizable" mode as in (\ref{eq:phifalloff}), so that $\Phi(\eta,x) \sim \phi(x) \hs \eta^{\Delta_-}$, as $\eta \rightarrow 0$.
We choose the
 initial state boundary condition of the path integral to select the \textit{Bunch--Davies vacuum} \cite{Bunch:1978yq} (i.e.~the unique de Sitter-invariant state that reduces to the Minkowski vacuum at short distances). In momentum space, this state is defined by imposing
\begin{equation}
    \label{eq:BDinitial}
\Phi(\eta,k) \sim \frac{1}{\sqrt{2|k|}}e^{+i|k|\eta}\,, \hspace{6mm} \text{as} \hspace{6mm} \eta \rightarrow -\infty\,.
\end{equation}
The wavefunctional $\Psi[\phi]$ has a similar path integral construction to the partition function in Euclidean AdS, and indeed the two objects are related by analytic continuation~\cite{Maldacena:2002vr,Harlow:2011ke,Anninos:2014lwa}. 

\vskip 4pt
In perturbation theory, it is natural to parameterize the wavefunction as
\begin{equation}\label{eq: wavefunction coefficients}
		\Psi[\phi(\x)] \approx \text{exp} \left( - \sum_{n = 2}^\infty \frac{1}{n!}\int \d^d x_1 \dots \d^d x_n\, \la \O(\x_1) \dots \O(\x_n)\ra \phi(\x_1) \cdots \phi(\x_n) \right) ,
\end{equation}
where the kernels $\la \O(\x_1) \dots \O(\x_n)\ra$ are called {\it wavefunction coefficients}, and contain information about the correlations in the state $\Psi$. At \textit{separated points}, in position space, the wavefunction coefficients are invariant under local field redefinitions of the sources $\phi(x)$; similar to S-matrix elements in Minkowski space. The constraints we derive are therefore insensitive to the choice of bulk field parametrization, or even to the existence of a weakly coupled Lagrangian description. 

\vskip 4pt
For light fields in de Sitter space, wavefunction coefficients have the same kinematic properties as correlation functions in a conformal field theory, with the operators $\O$ having conformal dimensions $\Delta_+ = d-\Delta_-$, where $\Delta_-$ is the slower falloff~\eqref{eq:phifalloff} of the bulk field $\Phi$.\footnote{Essentially, we are considering a weak form of a dS/CFT correspondence~\cite{Witten:2001kn,Strominger:2001pn} relating bulk interactions to wavefunction coefficients. We do not rely on any properties of these wavefunction coefficients beyond the fact that they have the same kinematic properties as correlation functions in a Euclidean CFT.} In this language, the kinematic constraints on the operators $J^{(s)}$ in~\eqref{eq:partiallymasslessconservation}, dual to partially massless bulk fields $\phi^{(s)}$, can be thought of as a consequence of demanding that the wavefunctional be gauge invariant. From this perspective, if we find there is no set of correlation functions that satisfy the constraints coming from~\eqref{eq:partiallymasslessconservation}, then there is no possible gauge-invariant wavefunctional, and correspondingly no gauge-invariant set of bulk interactions for the dual fields.\footnote{It is also possible to take a more indirect approach by deriving constraints on bulk interactions in AdS space and then trying to analytically continue them. Our constraints can also be interpreted this way.}

\subsection{De Sitter Representations}
\label{ssec:dS-Rep}

As explained by Wigner~\cite{Wigner:1939cj}, possible elementary particles are in one-to-one correspondence with unitary irreducible representations (UIRs) of the group of symmetries of the spacetime. In $(d+1)$-dimensional Minkowski space, this leads to the familiar classification of particles into UIRs of $\text{ISO}(d,1)$: massive and massless particles labeled by representations of the corresponding \textit{little groups} SO$(d)$ and ISO$(d-1)$, respectively.  In this section, we will review how partially massless fields fit into the representation theory of the de Sitter group.

\begin{figure}[t!]
		\centering
		\includegraphics[scale=1.0]{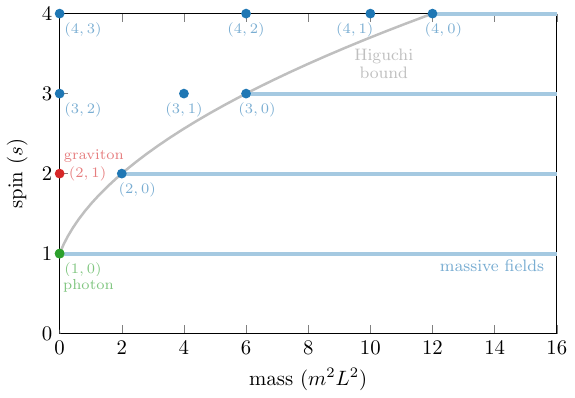}
		\caption{Unitary bosonic representations of dS$_{4}$. The solid dots denote the mass and spin values of the exceptional type-II series, corresponding to (partially) massless fields. The blue lines correspond to massive fields in the complementary and principal series.} 
		\label{fig:irreps}
	\end{figure}

\vskip 4pt
Elementary particles in $(d+1)$-dimensional de Sitter space can be organized into UIRs of the group SO$(d+1,1)$ of de Sitter isometries. These representations were classified long ago in the mathematics literature~\cite{thomasds,thomasds2,Dixmier,Hirai2,Takahashi,Dobrev:1977qv}. (More recent reviews can be found in~\cite{Basile:2016aen,Sun:2021thf,Penedones:2023uqc}.) In the following, we describe the representations of the de Sitter group that are realized by \textit{scalar fields} and \textit{traceless-symmetric tensor fields} in spacetime. These representations are labeled by the spin~$s$ and the conformal dimension $\Delta$. This latter parameter is related to the mass of the bulk field through the relation
\begin{equation}
m^2 L^2 = \begin{cases}
\Delta(d-\Delta)\,, & s=0\, ,\\ 
(\Delta+s-2)(d-\Delta +s-2)\,, & s\geq 1\,.
\end{cases}
\end{equation} 
This is the same relation as~\eqref{eq:deltascaling}, and so we see that the late-time fall-offs of bulk fields are related to the scaling dimensions of the de Sitter representations that they carry. We now list the possible bosonic UIRs of the de Sitter algebra (in generic dimension $D \ge 4$), along with the de Sitter field theory whose single-particle Hilbert space carries the representation.\footnote{In even spacetime dimensions, there is an additional family of representations, called the {\it discrete series}. For $D=4$, these representations happen to coincide with the {\it exceptional series II}. In higher dimensions, discrete series representations are necessarily carried by mixed-symmetry field theories~\cite{Basile:2016aen}, which we do not consider here.}
\begin{itemize}
\item {\bf Principal series:} These representations have conformal dimensions $\Delta = \frac{d}{2}+i\mu$, with $\mu\in{\mathbb R}_{\ge 0}$. For $s=0$, this corresponds to a bulk field with mass $m^2 L^2 \geq \frac{d^2}{4}$, while, for $s\geq 1$, the corresponding bulk fields are heavy spinning fields with 
\beq
m^2 L^2 \geq \left(\frac{d}{2}+s-2\right)^2 \,.
\eeq
In the flat-space limit ($L\to\infty$), these are the representations that correspond to massive particle representations of the Poincar\'e group.

\item  {\bf Complementary series:} These are representations with real scaling dimensions. The allowed range differs depending on the spin of the field.
For $s=0$, the conformal dimension lies in the range $0 < \Delta < d$, corresponding
to light fields with $0< m^2 L^2 <\frac{d^2}{4}$. For $s\geq 1$, the conformal dimension must be in the range $1 < \Delta < d-1$. 
These representations correspond to bulk fields with masses in the range 
\beq
(s-1)(d+s-3) < m^2 L^2 < \left(\frac{d}{2}+s-2\right)^2 \,.
\eeq 
A qualitative difference from the scalar case is that spinning representations with $s \geq 2$ have a lower bound---the {\it Higuchi bound}--- on their mass in order to remain unitary~\cite{Higuchi:1986py}.

\item {\bf Exceptional series I:}  These representations exist for $s=0$ with conformal dimension $\Delta  = d+k$, where $k\in{\mathbb Z}_{\geq 0}$.   A reasonable expectation is that the shift-symmetric scalar fields considered in~\cite{Bros:2010wa,Epstein:2014jaa,Bonifacio:2018zex} carry these representations if they are quantized treating the shift symmetries like gauge redundancies. The mass values of these fields are $m^2 L^2 = -k(d+k)$.

\item {\bf Exceptional series II:} These representations exist for $s\geq 1$ with conformal dimension $\Delta  = d-1+t$, where $t\in\{0,1,\cdots, s-1\}$ is the depth. The corresponding bulk fields have  spin $s$ and mass 
\beq
m^2 L^2 = (s-t-1)(s+t+d-3)\,. 
\eeq
The representations with $t=s-1$ corresponds to ordinary massless gauge fields. For other values of the depth, the mass values are precisely those of the partially massless fields~\eqref{eq:PMmassvalues}, which therefore carry these exceptional series representations.
\end{itemize}
The representations that will be of interest to us---the principal, complementary and exceptional type-II series---are depicted in Figure~\ref{fig:irreps}, together with the Higuchi bound. 
	
\subsection{Three-Point Functions}\label{sec: Three-Point Functions}

Fundamentally, consistent interactions between (partially) massless particles are constrained because there are only a finite number of cubic interactions compatible with conformal symmetry. The requirements of current conservation then restrict how these cubic interactions can be combined into four-point functions. To derive these consistency constraints, we therefore need to be able to explicitly construct conformal three-point functions.

\paragraph{Embedding space} 
A convenient approach to efficiently enumerate conformal three-point functions is to use embedding space~\cite{Dirac:1936fq,Costa:2011mg}. This formalism is based on the fact that the conformal algebra on $d$-dimensional Euclidean space, ${\mathbb R}^d$, is isomorphic to the algebra of Lorentz transformations on $(d+ 2)$-dimensional Minkowski space, ${\mathbb R}^{1,d+1}$. It is therefore  possible to find an embedding of ${\mathbb R}^d$ into ${\mathbb R}^{1,d+1}$, so that the Lorentz transformations in the higher-dimensional space become conformal transformations on the lower-dimensional slice. Explicitly, we uplift the $d$-dimensional position vectors $x^\mu$, with $\mu \in\{1,2,\,\cdots, d\}$, to $(d+2)$-dimensional coordinates $P^A$, with $A\in \{0,1,\,\cdots,d+1\}$, living in the embedding space $\mathbb{R}^{1, d+1}$.  Since $P^A$ has more components than $x^\mu$, we impose the following
two constraints
\begin{equation}
		P_A P^A = 0\,,\qquad P^A \sim \lambda P^A\,,
		\label{eq:lightconeconstraints}
\end{equation}
	for any $\lambda \in \mathbb{R}_{>0}$. This identifies each $x^\mu$ with a null ray in $\mathbb{R}^{1, d+1}$. These constraints define the \textit{projective null cone} in embedding space and can be solved to give the \textit{Euclidean section}:
\begin{equation}
		\big(P^+,P^-,x^\mu\big) = \big(1,|\x|^2,x^\mu\big)\,,
\end{equation}
where we have used light-cone coordinates $P^{\pm} \equiv P^0 \pm P^{d+1}$. The benefit of this construction is that SO$(1, d+1)$ is realized linearly on $P^A$, and becomes the usual nonlinear action on $x^\mu$ when we solve the constraints~\eqref{eq:lightconeconstraints}.

\vskip4pt
We can also represent operators in embedding space. In general, the higher-dimensional operators will have too many components, so we need a way to eliminate them. A spin-$s$ operator $\O_{\mu_1\cdots\mu_s}(\x)$ is uplifted to an embedding-space operator $\O_{A_1\cdots A_s}(P)$, subject to the constraints
\begin{align}
\label{eq:ambientscaling}
\O_{A_1\cdots A_s}(\lambda P) &= \lambda^{-\Delta} \O_{A_1\cdots A_s}(P)\,,\\
P^{A_1}\O_{A_1\cdots A_s}(P) &=0\,,
\label{eq:ambienttransverse}
\end{align}
where $\Delta$ is the conformal dimension of the operator, and the transversality condition~\eqref{eq:ambienttransverse} removes the additional unphysical components of the tensor. 
To deal with index structure, it is again convenient to work in an index-free formalism by introducing an auxiliary embedding-space vector $\Z^A$ satisfying $\Z^2 = \Z\cdot P = 0$. The vector $\Z^A$ is the embedding-space lift of the auxiliary vector $z^\mu$, with components $\Z^A = \left(\Z^+,\Z^-,z^\mu\right)$. The transversality condition between $\Z^A$ and $P^A$ can be solved by setting $\Z^A = \left(0,2\hs \z \cdot \x ,z^\mu\right)$. As discussed above, the real-space auxiliary vectors $z^\mu$ can then be stripped off by acting with the operator defined in~\eqref{eq: Todorov operator}.

\paragraph{Three-point structures}
We therefore associate an embedding-space operator $\O^{(s)}(P,\Z)$ to an operator in the physical space. In the index-free notation, the consistency conditions~\eqref{eq:ambientscaling} and~\eqref{eq:ambienttransverse} read $\O^{(s)}(\lambda P,\Z) = \lambda^{-\Delta} \O^{(s)}(P,\Z)$ and $\O^{(s)}( P,\Z+\alpha P) = \O^{(s)}(P,\Z)$. In addition to this, a spin-$s$ operator is homogeneous of order $s$ in the auxiliary vector $\Z$: $\O^{(s)}( P,\alpha \Z) = \alpha^s \O^{(s)}(P, \Z)$.
In order to construct the three-point functions of these operators, we therefore need to find functions of $P$ and $\Z$ that satisfy these same requirements.
Assuming parity invariance, the transversality constraint~\eqref{eq:ambienttransverse} is solved by writing the correlators in terms of the following building blocks~\cite{Costa:2011mg} 
\beq
	\begin{aligned}
		P_{ij} &\equiv P_i \cdot P_j\,, \\
		H_{ij} &\equiv -2\big[(\Z_i \cdot \Z_j)(P_i\cdot P_j) - (\Z_i \cdot P_j)(\Z_j\cdot P_i)\big]\,, \\
		V_{i,jk} &\equiv \frac{(\Z_i\cdot P_j)(P_i\cdot P_k) - (\Z_i \cdot P_k)(P_i \cdot P_j)}{P_j \cdot P_k}\,, 
	\end{aligned}
	\label{equ:conformal}
\eeq
where $i\neq j\neq k$. Notice that $H_{ij} = H_{ji}$, while $V_{i,jk} = -V_{i,kj}$. We will only need explicit expressions for three-point functions, so we will write $V_{i,jk} \equiv V_i$ as a shorthand, where the indices $j,k$ are implicit, and we assume cyclic ordering, so that for example $V_2  = V_{2,31}$.  Ultimately, we will be interested in correlation functions directly in the physical space, where the conformal structures in \eqref{equ:conformal} become~\cite{Costa:2011mg}  
\beq
	\begin{aligned}
		P_{ij} &\rightarrow  -\frac{1}{2} \x_{ij}^2\,, \\
		H_{ij} &\rightarrow  (\z_i\cdot \z_j) \x_{ij}^2 - 2(\z_i\cdot \x_{ij})(\z_j\cdot \x_{ij})\,,\\
		V_{i,jk} &\rightarrow \frac{(\z_i\cdot \x_{ik}) \x_{ij}^2 - (\z_i\cdot \x_{ij})\x_{ik}^2}{\x_{jk}^2}\,,
	\end{aligned}
\eeq
with $\x_{ij} \equiv \x_i-\x_j$.

\vskip4pt
It is now straightforward to build the three-point function of a given set of operators. We simply construct all possible functions of the form
\beq
\langle \O_1^{(s_1)}(P_1,\Z_1)\hs \O_2^{(s_2)}(P_2,\Z_2)\hs \O_3^{(s_3)}(P_3,\Z_3)\rangle = f(P_{ij},H_{ij},V_k)\,,
\eeq
which have homogeneity $\Delta_i$ under rescaling $P_i$ and homogeneity $s_i$ under rescaling $\Z_i$. For a given choice of operators, there is a finite list of possible three-point functions compatible with these requirements~\cite{Costa:2011mg}.

\vskip4pt
A small complication is that
in sufficiently low dimensions, there are dimension-dependent identities that reduce the number of independent tensor structures. For example, at three points, the embedding-space correlator is a function of the vectors $\{P_1,P_2,P_3,\Z_1,\Z_2,\Z_3\}$. However, for a three-dimensional physical spacetime, the embedding space is five-dimensional and therefore it is impossible for these six vectors to be linearly independent. Concretely, we have a so-called \textit{Gram constraint}:
\begin{equation}\label{eq: Gram constraint}
		-2 H_{12} H_{23} H_{31} = \left(V_1 H_{23} + V_2 H_{31} + V_3 H_{12} + 2 V_1 V_2 V_3\right)^2.
\end{equation}
No such constraints on the tensor structures exist in any dimension higher than three.

\paragraph{Conservation} To construct correlation functions of conserved operators, we have to impose conservation as an additional condition. As a practical matter, it is easiest to impose this conservation condition directly in the physical space after projection. For a spin-$s$, depth-$t$ operator, we require
\beq
(\partial\cdot D_z)^{s-t} \O_{(s,t)} = 0\,,
\eeq
which implies the same constraint on the possible three-point structures when $\Delta = d-1+t$. This additional constraint serves to further reduce the possible three-point structures.

\paragraph{Example}
As a concrete example of this entire construction, we will build the possible three-point functions of the stress tensor $T_{(2,1)}$ with two identical partially conserved spin-2 operators~$X_{(2,0)}$. Conformal invariance and Bose symmetry imply that this three-point function, in $d$ dimensions, is a general linear combination of eight allowed structures
	\begin{equation}
		\la T_{(2,1)}X_{(2,0)}X_{(2,0)}\ra = \frac{1}{P_{12}^{\frac{d+2}{2}}P_{23}^{\frac{d}{2}}P_{31}^{\frac{d+2}{2}}}\sum_{n = 1}^{8} c_n G_n\,, ~~{\rm where}~~G_n \equiv \begin{pmatrix}
			V_1^2 V_2^2V_3^2\\
			V_1V_2V_3^2H_{12} + (2\leftrightarrow3)\\
			V_1^2V_2V_3H_{23}\\
			V_3^2H_{12}^2 + (2\leftrightarrow3)\\
			V_1^2H_{23}^2\\
			V_1V_3H_{12}H_{23} + (2\leftrightarrow3)\\
			V_2 V_3 H_{12} H_{31}\\
			H_{12}H_{23}H_{31}
		\end{pmatrix} .
	\end{equation}
	Imposing the conservation laws for both $T_{(2,1)}$ and $X_{(2,0)}$ implies relations among the free coefficients $c_n$. For example, in $d>3$, we get
		\begin{align}
			c_6 &= \frac{4}{(d-2) (d+4)}\, c_1+\frac{(d-4) }{(d-2) (d+4)}\, c_2 +\frac{1}{2-d}\, c_3 +\frac{d^2 }{(d-2) (d+4)}\,c_4+\frac{d }{d-2}\, c_5\,,\\[4pt]
			c_7 &= \frac{4 }{(d-2) (d+4)}\, c_1 -\frac{8}{(d-2) (d+4)}\, c_2+\frac{2 d (d+2) }{(d-2) (d+4)}\,c_4\,,\\[4pt]
			c_8 &= -\frac{4 }{(d-2)^2 (d+4)}\,c_1+\frac{4  d }{(d-2)^2 (d+2) (d+4)}\,c_2 +\frac{2  d }{(d-2)^2
   (d+2)}\,c_3 \\
   &\ \ \ +\frac{2 d \left(d^2+2 d-12\right)}{(d-2)^2 (d+2) (d+4)}\,c_4 -\frac{4  d }{(d-2)^2 (d+2)}\,c_5\,. \nonumber
		\end{align}
	There are therefore five independent structures that are consistent with conformal invariance, Bose symmetry and conservation laws, matching the counting of independent, gauge-invariant interaction vertices in (A)dS$_{d+1}$~\cite{Joung:2012rv,Goon:2018fyu}. In $d=3$, there is one fewer independent structure as a consequence of the Gram constraint~\eqref{eq: Gram constraint}.
		
\subsection{Examples of PM Theories}
\label{sec: Examples of PM Theories}

Partially massless fields are somewhat unfamiliar, and their properties are a bit esoteric. Consequently, one might imagine that they just don't make sense, or cannot be made to consistently interact with other particles. In order to dispel this concern, we provide two examples of nonlinear theories involving partially massless fields, one from the boundary point of view and one from the bulk. These theories will provide useful test cases for the constraints that we derive in the rest of the paper.

\subsubsection{\texorpdfstring{$\boldsymbol{\Box^2}$}\ \hskip 2pt CFT} 
\label{ssec:Box2}
A prototypical example of the AdS/CFT correspondence is the duality between O$(N)$ vector models and Vasiliev's higher-spin theory in AdS~\cite{Klebanov:2002ja,Giombi:2016ejx}, which has an analytic continuation to de Sitter space~\cite{Anninos:2011ui,Anninos:2017eib}. In both cases, the bulk theory involves an infinite tower of massless higher-spin particles, and correspondingly the boundary theory has an infinite number of higher-spin conserved currents. By suitably generalizing the construction, it is possible to obtain theories involving an infinite number of partially massless fields (or, equivalently, partially conserved currents)~\cite{Bekaert:2013zya,Basile:2014wua,Grigoriev:2014kpa,Alkalaev:2014nsa,Joung:2015jza,Basile:2018acb,Basile:2024dcs,Brust:2016gjy,Brust:2016zns}.
	
\vskip4pt
We will be interested in generalizations of the large-$N$ free bosonic U$(N)$ 
vector model involving higher derivatives~\cite{Brust:2016gjy, Brust:2016zns, Karananas:2015ioa,Osborn:2016bev,Guerrieri:2016whh,Nakayama:2016dby,Peli:2016gio,Gwak:2016sma,Gliozzi:2016ysv,Gliozzi:2017hni,Stergiou:2022qqj}. The action of these theories is given by
\begin{equation}
		\label{eq:box2action}
		S =  \int \text{d}^d x \; \phi^\dagger_a \square^k \phi_a\,,
\end{equation}
where $\phi_a$ are a set of $N$ complex scalars transforming in a vector representation of U$(N)$ and we have left the sum over $a$ implicit. This is a free theory, with a higher-derivative equation of motion
\begin{equation}
~\square^k \phi_a =0\,.
\label{eq:HDEOM}
\end{equation}
The action~\eqref{eq:box2action} defines a CFT which, similar to the unitary $k=1$ model~\cite{Klebanov:2002ja}, is conjectured to be dual to a  Vasiliev-like higher-spin theory in (A)dS, involving both massless and partially massless fields.

\vskip4pt
In the following, we will be primarily interested in the $k=2$ version of this theory, which was studied in detail in~\cite{Brust:2016gjy,Brust:2016zns}. In this case, $\phi_a$ is an operator of dimension $\Delta_\phi = \tfrac{1}{2}(d-4)$, and the spectrum of single-trace primary operators can be organized into two infinite towers~\cite{Basile:2014wua,Brust:2016gjy}:
\begin{itemize}

\item {\it Massless higher-spin tower}\,:  The first family of primaries is of the schematic form
\beq
X_{(s,s-1)} \sim \phi^\dagger_a (\z\cdot \partial)^s \square\phi_a +\cdots\,,
\label{eq:singleconsbox2primaries}
\eeq
where $s \in \{0,1,2,\cdots\}$ and
the ellipses denote terms with derivatives that are distributed in all possible ways between the two fields. The relative coefficients are uniquely determined by requiring that the operators are symmetric, traceless and primary. For $s=0$, the operator is a scalar with $\Delta = d-2$, while for $s\geq 1$, the operator is a conserved current
\beq
s \geq 1: \quad (\partial\cdot D_z)X_{(s,s-1)} = 0\,, 
\eeq
which is compatible with the operator dimension $\Delta = d+s-2$. These singly conserved currents are similar to the higher-spin currents of the ordinary free scalar CFT~\cite{Klebanov:2002ja,Giombi:2016ejx}, but involve an additional factor of $\square$ as a consequence of the higher-derivative equation of motion~\eqref{eq:HDEOM}. Note that among these operators is the operator $X_{(2,1)}$, which is the stress tensor of the CFT.

\item {\it PM higher-spin tower}\,: The CFT also has a second infinite family of primaries, which takes the form
\beq
X_{(s,s-3)} \sim \phi^\dagger_a (\z\cdot \partial)^s\phi_a +\cdots\,,
\eeq
where $s\in \{0,1,2,\cdots\}$. Again, the subleading terms are determined by requiring that the operator is a traceless and symmetric conformal primary. These operators are quite similar in form to~\eqref{eq:singleconsbox2primaries}. They also correspond to conserved currents for $s\geq 3$, but the fewer derivatives means that we have to take more divergences in order to use the equations of motion. The currents therefore satisfy the triple-conservation shortening condition
\beq
s \geq 3: \quad (\partial\cdot D_z)^3 X_{(s,s-3)}= 0\,, 
\eeq
which is consistent with the operator dimension $\Delta = d+s-4$. Note that for $s=0,1,2$, these operators  correspond to massive spin $0$, $1$ and $2$ fields in the bulk, while the higher spins correspond to partially massless fields of depth $t=s-3$.		
\end{itemize}
We see that this CFT contains two infinite towers of conserved currents and partially conserved currents. These currents have non-trivial correlation functions, which can be computed explicitly via Wick contractions of the free fields $\phi_a$.
The bulk dual of these correlators is associated to non-trivial bulk interactions between the massless and partially massless fields.

\vskip4pt
In Section~\ref{sec:PM-gravity}, we will see that the spectrum of this theory can be reproduced directly from the current algebra generated by these conserved currents. It is worth noting that the features that we have described here are those applicable to generic dimensions. In $d=3,4,6$, various special features arise, which complicate the bulk interpretation~\cite{Brust:2016gjy,Brust:2016zns}. These and other features of the $\Box^2$ theory are described more fully in Appendix~\ref{sec: Current Algebra of Box-Squared CFT}.
	
\subsubsection{Conformal Gravity}
\label{ssec:confgrav}
Another example of an interacting theory involving a partially massless field is provided by conformal gravity~\cite{Weyl:1918ib,Weyl:1919fi,Weyl:1918pdp,Bach:1921zdq}. This is a higher-derivative theory of a symmetric tensor $h_{AB}$, which realizes a (non-unitary) irreducible representation of the conformal group SO$(4,2)$.
The action of conformal gravity is
\beq
S =-{\lambda^2\over 8}  \int \d^4 x\sqrt{-g} \ C^{MNLK}C_{MNLK} \, ,\label{conformalglage}
\eeq
where $C_{MNLK}$ is the Weyl tensor and $\lambda^2$ is a dimensionless coupling. In $D=4$, this action is invariant under both Weyl transformations and diffeomorphisms.\footnote{There is no unique generalization of conformal gravity to arbitrary dimension because in higher dimensions there are multiple Weyl invariants that one can build from the Riemann tensor. See~\cite{Boulanger:2018rxo,Joung:2019wwf}, for further discussion.} 

\vskip4pt
The theory has a maximally symmetric solution for any value of the background curvature radius, $L$. When expanded around this background, the theory propagates both a massless spin-2 degree of freedom and a partially massless spin-2 field that interact non-trivially~\cite{Maldacena:2011mk,Deser:2012qg,Deser:2012euu,Joung:2014aba}. We can see this by first noting that the action~\eqref{conformalglage} is equivalent (up to the Gauss--Bonnet total derivative) to
\beq
S =-{\lambda^2\over 4}  \int \d^4 x\sqrt{-g}\left(R_{MN}R^{MN}-{1\over 3}R^2 \right).\label{notfauxee}
\eeq 
In order to isolate the propagating degrees of freedom, we introduce an auxiliary field $f_{MN}$, so that the action becomes
\beq
 S =  \lambda^2 \int \d^4 x\sqrt{-g}\left( {\Lambda \over 6} (R-2\Lambda)+(G_{MN}-\Lambda g_{MN}) f^{MN}+f_{MN}f^{MN}-f^2\right) ,\label{flageee}
\eeq
where $G_{MN}=R_{MN}-{1\over 2}Rg_{MN}$ is the Einstein tensor, $f \equiv f^M_{\ M}$, and $\Lambda\neq0$ is a constant, which will eventually be set equal to the cosmological constant.  The  equation of motion for $f_{MN}$ can be solved to give
\beq 
f_{MN}= -{1\over 2}\left(R_{MN}-{1\over 6}Rg_{MN}\right)+{\Lambda\over 6}g_{MN}\,.
\eeq
Plugging this back into~\eqref{flageee} recovers~\eqref{notfauxee}, so that the two actions are equivalent.
Note that $f_{MN}$ vanishes on the background solution (\ref{equ:dS-metric}), with $\Lambda=3L^{-2}$. 
We can remove the term linear in $f_{MN}$ by performing the field redefinition
\beq
g_{MN} \mapsto g_{MN}-\frac{6}{\Lambda}f_{MN}\,.
\eeq
The action then takes the form
\beq
\begin{aligned}
S =  \lambda^2\int\d^4 x\sqrt{-g} \bigg[{\Lambda \over 6} (R-2\Lambda)-{3\over \Lambda}\bigg( -\frac{1}{2}&(\nabla f)^2 +{2\Lambda\over 3}\left( f^{MN}f_{MN}-{1\over 4} f^2\right)  \\[1pt]
& +\left(G_{MN}+\Lambda g_{MN}\right)F^{MN} \bigg)+{\cal O}(f^3)\bigg]\,,
\end{aligned}
\label{fafterrfredee}
\eeq
where $F^{MN}$ is a tensor quadratic in $f_{MN}$, whose form we will not need, and we have introduced the following shorthand notation for the kinetic term 
\beq
-\frac{1}{2}(\nabla f)^2 \equiv -{1\over 2}\nabla_K f_{MN} \nabla^K f^{MN}+\nabla_K f_{MN} \nabla^N f^{MK}-\nabla_M f\nabla_N f^{MN}+\frac{1}{2} \nabla_M f\nabla^M f\,.
\eeq
In order to isolate the propagating degrees of freedom in~\eqref{fafterrfredee}, we expand around a cosmological background  $g_{MN} = \bar g_{MN}+h_{MN}$, with $\Lambda = 3L^{-2}$.
To quadratic order, the action~\eqref{fafterrfredee} then is
\be
S= \lambda^2\int\d^4x \,\bigg({1\over 4L^2} {\cal L}_{\rm FP,0}(h)- L^2 {\cal L}_{{\rm FP},2L^{-2}}(f)\bigg)\, ,
\ee
where the canonically normalized Fierz--Pauli Lagrangian for a massive spin-2 field is 
\begin{align}
\nonumber
\frac{{\cal L}_{{\rm FP},m^2}(X)}{\sqrt{-g}} =& -{1\over 2}\nabla_K X_{MN} \nabla^K X^{MN}+\nabla_K X_{MN} \nabla^N X^{MK}-\nabla_M X\nabla_N X^{MN}+\frac{1}{2} \nabla_M X\nabla^M X \\
&+3L^{-2}\left( X^{MN}X_{MN}-\frac{1}{2} X^2\right)-\frac{1}{2}m^2(X_{MN}X^{MN}-X^2)\, .\label{spin2lage}
\end{align}
We therefore see that the theory propagates a massless graviton, $h_{MN}$, and a PM graviton, $f_{MN}$, with $m^2=2L^{-2}$. Importantly, there is a relative sign between the two kinetic terms (regardless of the sign of $L^{2}$), reflecting the non-unitarity of the theory.

\vskip4pt
We can continue to expand~\eqref{fafterrfredee} to higher orders and find the interactions between $h_{MN}$ and $f_{MN}$. Since these interactions arise from expanding out a nonlinear gauge-invariant theory, they will manifestly be invariant under the relevant gauge transformations.  In Appendix~\ref{app:confgrav}, we perform this exercise in order to obtain the three-point interactions in conformal gravity.

	\section{Constraints on Interactions}
	\label{sec:Dynamics}
	
	In this section, we describe our  strategy for constraining consistent interactions of PM fields in (A)dS by studying the integrated Ward identities (``charge conservation identities") of the associated partially conserved currents.

\vskip 4pt
We start by constructing conserved charges from these conserved currents, and then explain how the action of these charges on other local operators lead to identities for correlation functions. We then explain how to parameterize the most general possible action of these charges on other operators (``current algebra"). Requiring that the current algebra be consistent with the charge conservation identities then places nonlinear constraints on the parameters of the ansatz, which cannot always be satisfied. We demonstrate the procedure in some simple examples, reproducing some classic features of consistent interactions in (A)dS. Finally, we explain how these consistency constraints are a curved space analog of Weinberg's soft constraints on the S-matrix~\cite{Weinberg:1964kqu}.
	
\subsection{Charge Conservation Identities}
	
As discussed in Section~\ref{sec:background}, 
 (partially) massless fields on (A)dS have corresponding (partially) conserved currents in the associated boundary CFT. These currents can be integrated to form topological surface operators, which are conserved charges. In this paper, we will analyze the constraints associated with the existence of the charge
\begin{equation}
		\label{equ:pmcharge}
		Q_{(s,t)}^{{\mu_1...\mu_t}}\left[\Sigma\right] \equiv \oint_\Sigma \text{d}\Sigma_{{\nu_{1}}}\, \partial_{\nu_{2}} \cdots \partial_{\nu_{s-t}} X_{(s,t)}^{{\mu_1 \cdots \mu_t}\, {\nu_{1}}\, {\nu_{2} \cdots \nu_{s-t}}}\,,
\end{equation}
where $\Sigma$ is a closed codimension-one surface.\footnote{\label{footnote:chargeops}This is not the only charge operator that can be constructed from the partially conserved current. The more general form of a charge operator is 
\beq
Q^\zeta_{(s,t)}[\Sigma]  = \oint_\Sigma \text{d}\Sigma_\mu J_\zeta^\mu\,,
\eeq
with $J^\mu_\zeta$ a conserved current built from $X_{(s,t)}^{{\mu_1 \cdots \mu_t}\, {\nu_{1}}\, {\nu_{2} \cdots \nu_{s-t}}}$ as 
\beq
\begin{aligned}
J^\mu_\zeta = &~\zeta_{\mu_1...\mu_t}\partial_{\nu_{2}} \cdots \partial_{\nu_{s-t}} X_{(s,t)}^{{\mu_1 \cdots \mu_t}\, {\mu}\, {\nu_{2} \cdots \nu_{s-t}}} - \partial_{\nu_{2}} \zeta_{\mu_1...\mu_t}\partial_{\nu_3}\cdots \partial_{\nu_{s-t}} X_{(s,t)}^{{\mu_1 \cdots \mu_t}\, {\mu}\, {\nu_{2} \cdots \nu_{s-t}}}\\[1pt]
&+\cdots +(-1)^{s-t-1}\partial_{\nu_{2}} \cdots \partial_{\nu_{s-t}}\zeta_{\mu_1...\mu_t} X_{(s,t)}^{{\mu_1 \cdots \mu_t}\, {\mu}\, {\nu_{2} \cdots \nu_{s-t}}}\,,
\end{aligned}
\eeq	 
where $\zeta$ is a traceless and symmetric solution of a generalization of the conformal Killing equation 
\begin{equation}
(\z\cdot \partial)^{s-t} \zeta =0\,, \quad\qquad \zeta(\x,\z) \equiv z_{\nu_1} \cdots z_{\nu_t} \zeta^{\nu_1...\nu_t}(\x)\,. 
\end{equation}
Apart from the constant solution, the fact that the Killing-like tensor $\zeta$ depends non-trivially on the position $\x$ implies the charge $Q^\zeta$  does not commute with the momentum generator $[Q^\zeta,P^\mu]\neq 0$. It is not clear if the charge conservation constraints associated with these generalized charges lead to further constraints. } 

\begin{figure}[t!]
	\centering
		\includegraphics[width = 0.9\textwidth]{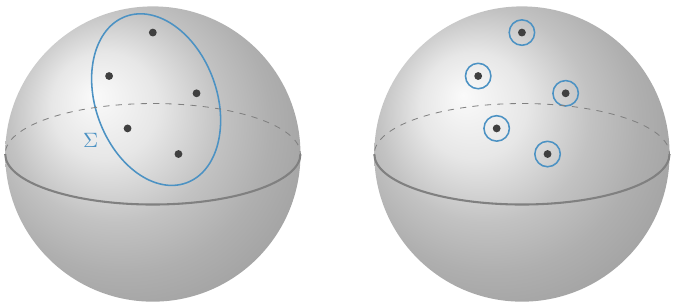}
	\caption{Insertion of a charge operator $Q[\Sigma]$ on a surface $\Sigma$ linking all local operator insertions. In the right-hand figure the surface is deformed to act locally on each operator $[Q,\O(x)]$. In this image the boundary CFT is depicted as being defined on $S^d$, the conformal compactification of $\mathds{R}^d$.	
 Deforming the surface in the left-hand figure to contract on the ``other side" of the sphere, the charge operator measures the charge of the vacuum, taken to be zero by assumption. }
	\label{fig: spheres}
	\end{figure}

\vskip 4pt
Inserting these charges in correlation functions with $\Sigma$ enclosing all local operator insertions gives \textit{integrated Ward identities}, also known as \textit{charge conservation identities}~\cite{Maldacena:2011jn}
\begin{equation}
		\label{eq:ccons}
		\langle Q[\Sigma]\O_1 \ldots \O_n\rangle \equiv \langle [Q, \O_1 \ldots \O_n] \rangle = \sum_{i=1}^n \langle \O_1 \ldots [Q,\O_i] \ldots \O_n \rangle = 0\,,
\end{equation}
	where  $[Q,\O_i]$ denotes a charge insertion with a surface that encloses only the operator $\O_i$ (see Fig.\,\ref{fig: spheres}). We can then deform the surface $\Sigma$ to infinity and contract it to a point, so that the charge is zero. 
This requires assuming that the vacuum is uncharged under the symmetry associated to the conserved charge---i.e.~the symmetry is not spontaneously broken---so that
	$Q |0\ra = 0$.
In the bulk, this corresponds to the assumption that the partially massless gauge fields are in the ``Coulomb phase". 
	
\vskip 4pt
The action of the charge on any operator should be writeable as a linear combination of local operators (assumed to be primaries and descendants), and can therefore be written schematically as 
	\begin{equation}
		\label{eq:sumrule}
		[Q,\O_i(\x)] \sim \sum_j a_{ij}\partial^{n_{ij}} \O_j(\x)\,.
	\end{equation}
	Since the charge operator (\ref{equ:pmcharge}) is constructed from a conserved current without explicit coordinate dependence, it commutes with the momentum generator $[P^\mu,Q] = 0$, which	in turn implies that the differential operators appearing on the right-hand-side of~\eqref{eq:sumrule} themselves cannot have any explicit coordinate dependence. If $\O_i(\x)$ is itself a conserved current, then such relations are called the \textit{current algebra} of the theory and they are further constrained by the requirement that they integrate to a consistent charge algebra.
	
	\vskip 4pt
	In this paper, we will analyze charge conservation identities of the form
	\begin{equation}\label{eq: general charge conservation constraint}
		\langle [Q,\O_1\O_2\O_3]\rangle = \langle [Q,\O_1]\O_2\O_3\rangle + \langle \O_1[Q,\O_2]\O_3\rangle + \langle \O_1 \O_2[Q,\O_3]\rangle = 0\,. 
	\end{equation}
	Combined with the current algebra (\ref{eq:sumrule}), these identities imply non-trivial constraints on combinations of CFT three-point functions:
	\beq
	\label{eq:sumrule3}
	\sum_j \Big[a_{1j} \partial^{n_{1j}} \langle \O_j\O_2\O_3 \rangle + a_{2j} \partial^{n_{2j}} \langle \O_1\O_j\O_3 \rangle  + a_{3j} \partial^{n_{3j}}\langle \O_1\O_2\O_j \rangle \Big] = 0\,.
	\eeq
	Note that although (\ref{eq:sumrule3}) is a sum rule on three-point functions, it originates from the assumption that there exists a consistent four-point function satisfying the required current conservation Ward identities. By definition, the expanded charge conservation identity (\ref{eq: general charge conservation constraint}) is equivalent to 
	\begin{equation}
        \label{eq:int4pt}
		\oint_\Sigma \text{d}\Sigma_{{\nu_{1}}} \partial_{\nu_{2}} \cdots \partial_{\nu_{s-t}} \langle X_{(s,t)}^{{\mu_1 \cdots \mu_t}\, {\nu_{1}}\, {\nu_{2} \cdots \nu_{s-t}}}(\x,\z)\O_1(\x_1) \O_2(\x_2) \O_3(\x_3)\rangle = 0\,.	
	\end{equation}
This constraint is therefore an integrated version of the local Ward identity satisfied by the four-point function.  Note that the integration cycle $\Sigma$ in \eqref{eq:int4pt} does not cross any local operator insertions. The integrated identity~\eqref{eq: general charge conservation constraint} is therefore insensitive to any contact terms present in the current conservation Ward identity (including those arising from 't Hooft anomalies). As such, the constraint is enforcing conformal invariance at separated points, as usual in position space.

\vskip 4pt
Boundary current conservation imposes only mild constraints on conformally invariant three-point functions, 
while much stronger constraints arise from the requirement that these interactions can be consistently combined into a four-point function.	
 Assuming an initial set of conformal data (i.e.~a spectrum of primary operators and OPE coefficients), the failure of the charge conservation constraint (\ref{eq:sumrule}) is an efficient way to diagnose the non-existence of such a four-point function and therefore an inconsistency in the assumed data. 
	
	\subsection{Current Algebra}
	\label{ssec:CurrentAlgebra}
	
	The explicit form of the charge conservation constraints requires an ansatz for the action of the charge $Q_{(s,t)}$ on the primary operators in the theory $\{X_{(s',t')}\}$, which gives the ``current algebra" of the model. It is convenient to uniformize our notation by using $X_{(s,t)}$ to denote a \textit{generic} operator of spin $s$ and depth $t$, whether-or-not the operator is (partially) conserved. For a non-conserved operator, we then take $\Delta = d-1+t$ as the definition of the ``depth" $t$, with $t<0$ for a massive field in the complementary series. Likewise, we will abuse terminology and refer to $[Q,X]$ as a ``current algebra" relation, even for the case where $X$ is not conserved. 
	
	\vskip 4pt
	In this subsection, we will restrict to the case of $t=0$, which is of particular relevance to the discussion in Section \ref{sec:PM-gravity}. The generalization to arbitrary depth is straightforward. 
  
	\vskip 4pt
	Using the index-free notation for symmetric and traceless tensors reviewed in Section~\ref{ssec:PM-Fields}, the differential operators that appear in the current algebra $\big[Q_{(s,0)},X_{(s',t')}\big]$ are composed of  $\square$, $\z\cdot\partial$ and $\partial\cdot D_z$. Acting on a local operator (not necessarily a primary), $\O^{(s)}_{\Delta}$, of dimension $\Delta$ and spin~$s$, these produce 
		\begin{equation}
	\square \O^{(s)}_{\Delta} \sim \O^{(s)}_{\Delta+2}\,, \quad
	\left(z\cdot \partial\right) \O^{(s)}_{\Delta} \sim \O^{(s+1)}_{\Delta+1}\,,\quad \left(\partial \cdot D_z\right) \O^{(s)}_{\Delta} \sim \O^{(s-1)}_{\Delta+1}\,.
	\label{equ:derivatives}
	\end{equation}
	An elementary constraint on the current algebra is that scaling dimensions and spins must match on both sides of the equation. (Note that the charge $Q_{(s,t)}$ has spin $t$ and dimension~$\Delta_Q = s-1$.) Using~\eqref{equ:derivatives}, it is straightforward to show that the most general form of the current algebra is
	\begin{equation}
		\label{eq:genCA}
		\big[Q_{(s,0)},X_{(s',t')}(\x,\z)\big] \supset \mathcal{P}_k\big(\square,(\z\cdot \partial) (\partial\cdot D_z)\big)
		\times\begin{cases} 
			\left(\partial \cdot D_z\right)^{s''-s'} X_{(s'',t'')}(\x,\z)  &\text{if} \;\;\; s''\geq s'\,, \\[10pt]
			(z \cdot \partial)^{s'-s''} X_{(s'',t'')}(\x,\z)  &\text{if} \;\;\; s'>s''\,,
		\end{cases}
	\end{equation}
	where $\mathcal{P}_k$ is a homogeneous polynomial of degree
	\begin{equation}
		k \equiv \frac{1}{2}\left(s-1-|s''-s'|+t'-t''\right) .
	\end{equation}
	The right-hand-side of (\ref{eq:genCA}) is a local operator only if $k\geq 0$, which can be rearranged into
	\begin{equation}
		\label{eq:CA1}
		t' -t''\geq |s''-s'| -s+1\,.
	\end{equation}
    In $d=3$, we might also consider \textit{parity-odd} primary operators and differential operators in the current algebra with Levi--Civita symbols. Extending the above to this case is, in principle, straightforward, but for simplicity we will assume that all operators are parity even.
    
    \vskip 4pt
	An important constraint on the operators that can appear in the current algebra  follows from the two-point charge conservation identity 
	\begin{equation}
		\label{eq:2ptCC}
		\big\langle \big[Q_{(s,t)},X_{(s',t')} X_{(s'',t'')}\big]\big\rangle = \big\langle \big[Q_{(s,t)},X_{(s',t')}\big] X_{(s'',t'')}\big\rangle + \big\langle X_{(s',t')} \big[Q_{(s,t)},X_{(s'',t'')}\big] \big\rangle = 0\,.
	\end{equation}
	Assuming the absence of non-trivial null states, this identity can only be satisfied by a cancellation between terms. Hence, $[Q_{(s,0)},X_{(s',t')}] \supset X_{(s'',t'')}$ implies  $[Q_{(s,0)},X_{(s'',t'')}] \supset X_{(s',t')}$; a property we will refer to as \textit{current algebra reciprocity}. By the same argument as above, we then have
	\begin{equation}
		\label{eq:CA2}
		t''-t' \geq |s''-s'|-s+1\,.
	\end{equation}
	Taken together, (\ref{eq:CA1}) and (\ref{eq:CA2}) imply that for any given current algebra relation, only a strictly finite subset of primary operators can appear, subject to the constraints
	\begin{equation}
		\label{eq:CAspectrumconstraints}
		\begin{aligned}
		&s''\in \mathds{Z}_{\geq 0}\,, \qquad t''\in \mathds{Z}\,,  \qquad s-1-|s''-s'|-|t''-t'| \in 2\mathds{Z}_{\geq 0}\,, \\[4pt]
		 & t'' \leq s''-1\,, \quad |s''-s'| \leq s-1\,.
	\end{aligned}
	\end{equation}
	Using this, we can explicitly parametrize the most general form of the  current algebra, with  a finite number of \textit{a priori} unknown coefficients appearing in the  polynomials $\mathcal{P}_k$ in (\ref{eq:genCA}).
	
	\vskip 4pt
	As an illustrative example---of relevance to Section~\ref{ssec:Spin2}---consider the current algebra describing the self-interaction of a partially massless spin-2 field, i.e.~$\big[Q_{(2,0)},X_{(2,0)}\big]$, so that~$s=2$, $t=0$, $s'=2$, $t'=0$. Solving the constraints~(\ref{eq:CAspectrumconstraints}), we find
	\begin{equation}
		(s'',t'') \in \{(1,0),(2,-1),(2,1),(3,0)\}\,.
	\end{equation}
	The general ansatz for the current algebra then is
	\beq
	\begin{aligned}
		\big[Q_{(2,0)},X_{(2,0)}(x,z)\big] &= a_1 (z \cdot \partial) X_{(1,0)}(x,z) + \left[a_2 \square + a_3 (z \cdot \partial)  (\partial\cdot D_z) \right] X_{(2,-1)}(x,z) \\
		&\hspace{7mm} + a_4 T_{(2,1)}(x,z)  + a_5 (\partial \cdot D_z) X_{(3,0)}(x,z)\,.
	\end{aligned}
	\eeq
To reintroduce indices into this index-free expression, 
one should act with the operator~\eqref{eq: Todorov operator} to strip off the appropriate number of auxiliary vectors.

	\subsection{Simple Examples}
	\label{sec:simpleex}
	Before considering charge conservation constraints on theories containing partially massless fields, it is instructive to demonstrate their power by looking at more familiar theories, involving strictly massless fields and the associated conserved currents. 
	
	\paragraph{Electrodynamics} We can start by considering the coupling of a conserved spin-1 operator $J_{(1,0)}$ to a set of scalar operators $\O_i$ of dimensions $\Delta_i$. Without loss of generality, we can assume that there exists an operator basis in which the action of the charge $Q_{(1,0)}$ is diagonal: 
	\begin{equation}
		\label{equ:electroexample}
		[Q_{(1, 0)}, \O_i(\x)] = q_i \O_i(\x)\,,
	\end{equation}
	where $q_i$ are some \textit{a priori} unconstrained constants.
	We will further assume that there exists a nonvanishing three-point interaction between the scalar operators, which has a unique conformally invariant form
	\begin{equation}
		\la \O_i(\x_1) \O_j(\x_2) \O_k(\x_3)\ra = \frac{\lambda_{ijk}}{P_{12}^{\Delta_i+\Delta_j-\Delta_k} P_{23}^{\Delta_j+\Delta_k-\Delta_i} P_{31}^{\Delta_k+\Delta_i-\Delta_j}}\,.
	\end{equation}
	Assuming this input, we impose the change conservation condition 
	\begin{equation}
		\la [Q_{(1, 0)}, \O_1(\x_1) \O_2(\x_2) \O_3(\x_3)]\ra = (q_1 + q_2 + q_3)\la \O_1(\x_1) \O_2(\x_2) \O_3(\x_3)\ra = 0\,.
	\end{equation}
	This can be satisfied in two ways: Either the ``charges" of the scalar operators sum to zero, $q_1+q_2+q_3=0$, or the cubic scalar coupling vanishes, $\lambda_{123}=0$. 
	
	\vskip 4pt
	The final step is to give a bulk interpretation to this constraint. The conserved spin-1 operator $J_{(1,0)}$ is dual to a massless bulk spin-1 field, and the scalar operators $\O_i$ are dual to elementary bulk scalar fields. The ansatz for the action of the charge (\ref{equ:electroexample}) corresponds to the assumption of a ``minimal" cubic coupling between the massless spin-1 field and the scalar fields proportional to the constants $q_i$. The constraint $q_1+q_2+q_3=0$ is then nothing but the familiar statement that the coupling constants of this minimal interaction are proportional to the electric charge of the matter fields, and that physical processes conserve the total charge.

	\paragraph{Gravity} We can perform a similar analysis for a conserved spin-2 operator $T_{(2,1)}^{\mu\nu}$ coupled to a set of scalars. The unique ansatz for the action of the charge is
	\begin{equation}
		[Q^\mu_{(2, 1)}, \O_i(\x)] = \kappa_i \partial^\mu \O_i(\x)\,,
	\end{equation}
	where $\kappa_i$ are constants. 
	The charge conservation constraint is then given by
\beq
\la[Q^\mu_{(2, 1)}, \O_1(\x_1) \O_2(\x_2) \O_3(\x_3)]\ra = (\kappa_1 \partial_1^\mu+\kappa_2 \partial_2^\mu+\kappa_3 \partial_3^\mu) \la \O_1(\x_1) \O_2(\x_2) \O_3(\x_3)\ra= 0\,,
\eeq
	which only vanishes if $\kappa_1 = \kappa_2 = \kappa_3\equiv \kappa$ (for $\lambda_{123}\neq 0$).  
	
	\vskip 4pt
The bulk interpretation of this result is the \textit{Einstein equivalence principle}. The conserved spin-2 operator $T_{(2,1)}$ is dual to a massless spin-2 field in the bulk, which is uniquely identified with the graviton and so couples to other bulk fields with uniform strength $\kappa \propto \sqrt{G}$.

	\paragraph{Massless higher spin} In~\cite{Maldacena:2011jn},  the charge conservation constraints were applied to operators dual to massless higher-spin fields in (A)dS. For concreteness, we consider a field of spin 3.
	The action of the charge is 
	\begin{equation}
		[Q_{(3,2)}(z), \O_i(\x)] = \tau_i\,(z\cdot \partial)^2 \O_i(\x)\,,
	\end{equation}
	where $Q_{(3,2)}(z)\equiv z_\mu z_\nu Q_{(3,2)}^{\mu\nu}$ and  $\tau_i$ are constants, so that
	the charge conservation constraint becomes
	\beq
	\begin{aligned}
		&\la [Q^{\mu \nu}_{(3,2)}, \O_1(\x_1) \O_2(\x_2) \O_3(\x_3) ]\ra \\
		&=  \left[\tau_1 \left(\partial_1^\mu \partial_1^\nu-\frac{1}{d}g^{\mu\nu}\square_1\right) +(1\leftrightarrow 2) +(1\leftrightarrow 3)\right] \la \O_1(\x_1) \O_2(\x_2) \O_3(\x_3)\ra= 0 \,. 
	\end{aligned}
	\eeq
It is straightforward to convince oneself (e.g.~by sampling random kinematic points) that the only solution to this constraint is to set $\tau_i=0$ for all $i$ (assuming that $\lambda_{123}\neq 0$). 

\vskip 4pt
Note that this constraint by itself does not rule out the existence of massless spin-3 particles in (A)dS, or even forbid ``minimal" interactions with scalars. We have simply learned that if such a minimal coupling to scalars exists, it would forbid a cubic scalar self-interaction. To expose the true constraints on higher-spin particles requires that we study charge conservation constraints associated with interactions that cannot be consistently set to zero (such as their universal couplings to gravity).

\subsection{Weinberg Soft Constraints}
\label{ssec:Weinberg}

The charge conservation conditions and associated constraints derived in the previous subsection have an obvious similarity with the Weinberg soft constraints~\cite{Weinberg:1964kqu,Grisaru:1977kk,Grisaru:1976vm,Porrati:2008rm,Elvang:2016qvq} obtained from flat-space scattering amplitudes by imposing gauge invariance in the soft limit. 
In the following, we show that this is not a coincidence.

\vskip 4pt
The connection to soft limits can be made manifest by beginning with the differential form of the Ward--Takahashi identity
\beq
\begin{aligned}
& z_{\mu_1} \cdots z_{\mu_t }\,\partial_{\nu_{1}} \cdots \partial_{\nu_{s-t}} \langle X_{(s,t)}^{\nu_1 \cdots\nu_{s-t} \mu_{1}\cdots \mu_t}(\x) \O_1(\x_1) \cdots \O_n(\x_n)\rangle \\
&\qquad\qquad~~~ = \sum_{i=1}^n \delta^{(d)}(\x-\x_i)\langle \O_1(\x_1) \cdots \big[Q_{(s,t)}(z),\O_i(\x_i)\big] \cdots \O_n(\x_n) \rangle\,.
\end{aligned}
\eeq
Integrating both sides over all of space, the right-hand-side vanishes as a consequence of the charge conservation identity (\ref{eq:ccons}). The integral of the left-hand-side, on the other hand, can be written as the limit of a Fourier transform at zero frequency: 
\beq
\begin{aligned}
     \lim_{k\rightarrow 0} &\int \text{d}^d x \; e^{i\k\cdot \x} \partial_{\nu_1} \ldots \partial_{\nu_{s-t}} \langle X_{(s,t)}^{\nu_1 \ldots\nu_{s-t} \mu_{1}\ldots \mu_t}(\x) \O_1(\x_1)\ldots \O_n(\x_n)\rangle \\
    &= (-i)^{s-t} \lim_{k\rightarrow 0} k_{\nu_1} \ldots k_{\nu_{s-t}} \langle X_{(s,t)}^{\nu_1 \ldots\nu_{s-t} \mu_{1}\ldots \mu_t}(k) \O_1(\x_1)\ldots \O_n(\x_n)\rangle\,.
\end{aligned}
\eeq
Fourier transforming the other operators, the momentum-space version of this constraint takes the form
\begin{equation}
    \lim_{k\rightarrow 0}\,k_{\nu_1}\cdots k_{\nu_{s-t}} z_{\mu_1} \cdots z_{\mu_t }\langle X_{(s,t)}^{\nu_1 \cdots\nu_{s-t} \mu_{1}\cdots \mu_t}(\k) \O_1(\k_1) \cdots \O_n(\k_n)\rangle = 0\,.
\end{equation}
We can make a more direct connection to Weinberg's soft theorem by using the observation of~\cite{Maldacena:2011nz,Raju:2012zr} that, for models in (A)dS with a sensible flat-space limit,  the momentum-space correlation function contains a singularity at $E \equiv |\k| +|\k_1|+...+|\k_n| \rightarrow 0$, with the coefficient of the most singular term being proportional to the flat-space scattering amplitude. In that case, the above constraint becomes  
\begin{equation}
    \lim_{k\rightarrow 0}\,\,k_{\nu_1}\cdots k_{\nu_{s-t}} z_{\mu_1} \cdots z_{\mu_t }\mathcal{A}_{n+1}^{\nu_1 \cdots\nu_{s-t} \mu_{1}\cdots \mu_t}\left(\Phi_{(s,t)}(\k),\Phi(\k_1),\,\cdots ,\Phi(\k_n)\right)  = 0\,,
    \label{equ:WSoft}
\end{equation}
where $\mathcal{A}_{n+1}$ denotes an $(n+1)$-point scattering amplitude, $\Phi_{(s,t)}$ is the spacetime field related to the partially conserved current $X_{(s,t)}$ and $\Phi$ is the spacetime field dual to the operator $O$.
Equation~\eqref{equ:WSoft} then simply states that the soft limit of a scattering amplitude with a longitudinal polarization tensor should vanish, which is precisely the Weinberg soft constraint. 

\vskip 4pt
It is instructive to see an explicit example. Consider the ``electrodynamics" example discussed in Section~\ref{sec:simpleex}. For simplicity, we will assume $d=3$ and take all scalars to be conformally coupled ($\Delta_i=2$). The explicit form of the momentum-space correlation function with a longitudinally polarized current is~\cite{Baumann:2020dch}   
\beq
\begin{aligned}
    k_\mu \langle J_{(1,0)}^\mu(\k) \O_1(\k_1) \O_2(\k_2) \O_3(\k_3)\rangle = \lambda_{123}& \bigg[\,q_1 \log\left(\frac{|\k+\k_1|+|\k_2|+ |\k_3|}{E_0}\right) \\
    &~+(1\leftrightarrow 2)+(1\leftrightarrow 3)\bigg]\,, 
    \end{aligned}
\eeq
where $E_0$ is an arbitrary energy scale. Taking the soft limit $k^\mu \rightarrow 0$, all  terms combine to give
\begin{equation}
    \lim_{k\rightarrow 0} k_\mu \langle J_{(1,0)}^\mu(\k) \O_1(\k_1) \O_2(\k_2) \O_3(\k_3)\rangle = \lambda_{123} (q_1+q_2+q_3) \log\left(\frac{E}{E_0}\right).
\end{equation}
In this case, the leading total energy singularity is $\log E$, whose coefficient agrees precisely with the Weinberg soft constraint 
\begin{equation}
    \lim_{k\rightarrow 0} k_\mu\mathcal{A}_4^\mu\left(\gamma(\k),\Phi_1(\k_1),\Phi_2(\k_2),\Phi_3(\k_3)\right)= (q_1+q_2+q_3)\mathcal{A}_3\left(\Phi_1(\k_1),\Phi_2(\k_2),\Phi_3(\k_3)\right) 
    = 0\,,
\end{equation}
where $\gamma$ is the bulk photon field.
The lesson is clear: If a sensible flat-space limit exists, then the charge conservation constraints on the (A)dS boundary correlators coincide exactly with the Weinberg soft constraints of the corresponding flat-space scattering amplitudes when expanded in the $E\rightarrow 0$ limit.

\vskip 4pt
For examples without a flat-space limit, such as theories involving partially massless fields, this final step must fail, possibly because the correlators lack an $E\rightarrow 0$ singularity.\footnote{In the flat-space limit, partially massless representations become reducible, and so the corresponding amplitudes presumably involve these reducible representations packaged into polarizations.} In these cases, the charge conservation constraints considered here are strictly more powerful and allow us to constrain intrinsically (A)dS models.

\vskip 4pt
There is an interesting difference between the derivation of the Weinberg soft and charge conservation constraints. In the former, we typically first derive a \textit{soft theorem} for physical (transverse) polarizations and then impose that this vanishes when the polarization tensor is taken to be longitudinal. In the latter, we instead immediately write down the constraint. This raises the natural question of whether the constraints we are considering in this paper are connected to an (A)dS version of a Weinberg soft theorem for PM fields. Recently, it was found~\cite{Mei:2025jko} that leading-order terms in the soft expansion of momentum-space correlators in (A)dS yield constraints similar to our charge conservation, while sub-leading orders give rise to soft theorems for the correlators.\footnote{See also \cite{Hinterbichler:2013dpa,Hinterbichler:2012nm,Creminelli:2012ed,Chowdhury:2024wwe,Albayrak:2024ddg,Chowdhury:2024snc}, for previous discussions of (A)dS soft theorems.} We leave  the interesting problem of understanding the general connection between soft theorems and the charge conservation constraints for PM fields to future work.

	\section{Partially Massless Electrodynamics}
	\label{sec:PM-electrodynamics}
	
As a warm-up to the more difficult problem of studying gravitationally coupled PM fields, we first consider the simpler case of a single Abelian partially conserved current $X_{(s,t)}$ (corresponding to a PM gauge field) coupled to a set of scalar primary operators $\O_i$ (corresponding to scalar matter fields) of dimension $\Delta_i$. The Abelian assumption means that the PM fields do not self-interact, which in terms of the current algebra means
\begin{equation}
	\label{equ:abelian}
    \left[Q_{(s,t)},X_{(s,t)}\right] = 0\,.
\end{equation}
For the special case $(s,t)=(1,0)$, this class of models includes ordinary scalar QED and we will therefore refer to them generally as \textit{partially massless electrodynamics}. (See~\cite{Deser:2006zx,Sleight:2021iix}, for previous discussions of Abelian PM models.) 

\subsection{Selection Rules for Matter Couplings}

Our fundamental assumption is that the PM charge (\ref{equ:pmcharge}) acts non-trivially on the scalar operators. Following the discussion in Section \ref{ssec:CurrentAlgebra}, the current algebra has the form 
\begin{equation}
    \left[Q_{(s,t)}(\z),\O_i(\x)\right] = \sum_j a_{ij} \,\square^{n_{ij}} (\z\cdot \partial)^t \O_j(\x)\,,
\end{equation}
where $n_{ij}\in \mathds{Z}_{\geq 0}$ in order for the right-hand-side to be a local operator. Since $[P^\mu,Q_{(s,t)}]=0$, the coefficients $a_{ij}$ do not depend on $x$ (and hence are dimensionless), and therefore matching scaling dimensions on both sides gives 
\begin{equation}
2n_{ij} =    \Delta_i - \Delta_j + s-1-t  \,.
    \label{equ:Dij}
\end{equation}
As a consequence of the reciprocity property discussed in Section \ref{ssec:CurrentAlgebra} ($a_{ij}\neq 0$ implies $a_{ji}\neq 0$), we must also have an identical relation with $i\leftrightarrow j$. Together these conditions imply the following constraints on the dimensions of the scalar operators 
\begin{equation}
\begin{aligned}
    \Delta_i-\Delta_j &\in 2\mathds{Z} + r\,, \\
     |\Delta_i-\Delta_j| &\leq r\,,
     \end{aligned}
     \label{equ:Delta-Constraints}
\end{equation}
where $r\equiv s-1-t$. These constraints on the coupling between PM fields and scalar matter fields were known from an analysis of bulk cubic vertices and their invariance under linearized gauge transformations \cite{Joung:2012hz}. Here, we see that the same constraints follow from elementary consistency properties of the boundary CFT. 

\vskip 4pt
From the above, we observe that there is a clear difference between the cases where $r$ is even or odd. If $r$ is even, then we can consider \textit{diagonal} couplings $\Delta_i=\Delta_j$, while if $r$ is odd then there are only \textit{non-diagonal} couplings $\Delta_i\neq \Delta_j$. The structure of the charge conservation constraints in the two cases are different, so we will consider them separately. 

\vskip 4pt
The solutions to these constraints can be illustrated in a few representative examples:
	\begin{itemize}
		\item $s=1$, $t=0$ ($r=0$). In this case, the constraints in \eqref{equ:Delta-Constraints} degenerate to the requirement that $\Delta_i=\Delta_j$ and we  therefore  recover the expected statement that a photon can only minimally couple to a pair of scalar fields if their masses are equal, $m_i^2=m_j^2$.
		\item $s=2$, $t=0$ ($r=1$). Since $|\Delta_i-\Delta_j|=1$, the scalar fields now \textit{cannot} have equal masses. Instead, the minimal coupling to a PM spin-2 field requires at least two scalar fields with a prescribed non-zero mass splitting. Furthermore, in order for the scaling dimensions to be integer spaced, the fields must belong to the complementary series (c.f.~Section \ref{ssec:PM-Fields}).
		
				\item $s=3$, $t=0$ ($r=2$). In this case, there are two allowed splittings of the dimensions $|\Delta_i-\Delta_j| = 0$ or $2$. The former is similar to the diagonal case $(s,t)=(1,0)$ discussed above, while the latter is similar to the non-diagonal case  $(s,t)=(2,0)$. 
	\end{itemize}

\subsection{Mass Diagonal Coupling}

We first consider the case of a partially massless field $X_{(s,t)}$, with $r\equiv s-t-1 \in 2\mathds{Z}_{\geq 0}$, and a mass diagonal coupling to two scalar fields.

\vskip 4pt 
In this case, there is a (unique)  three-point interaction, with $\Delta_i=\Delta_j\equiv \Delta$, given by
\begin{equation}
    \langle  X_{(s,t)}(\x_1,\z_1) \O_i(\x_2) \O_j(\x_3) \rangle = c_{ij} \frac{V_1^s}{ P_{12}^{(d-1+t-s)/2} P_{13}^{(d-1+t-s)/2} P_{23}^{(2\Delta-d+1-t-s)/2}}\,.
    \label{eq:3ptPMelectro}
\end{equation}
Bose symmetry requires this expression to be invariant under simultaneously exchanging $\x_2\leftrightarrow \x_3$ and $i\leftrightarrow j$ and therefore $c_{ji} = (-1)^s c_{ij}$. 
We assume that the two-point functions of the scalar operators are conventionally orthonormalized\footnote{In a generic non-unitary CFT, there is no requirement that the two-point functions of operators have a fixed sign. Here, again we need some physical input from the de Sitter bulk. We will assume that the leading-order large-$N$ two-point function is calculated in the usual way from a tree-level two-point Witten diagram of an elementary bulk scalar field. The sign of the boundary two-point function is then correlated with the sign of the bulk kinetic term, which for a unitary bulk theory we will assume is consistently chosen to be the same for all fields. } 
\begin{equation}
    \langle \O_i(\x_2)\O_j(\x_3)\rangle = \frac{\delta_{ij}}{P_{23}^\Delta}\,.
\end{equation}
We are then free to make a further orthogonal basis rotation to simplify the cubic couplings. 
For even spin, we can take $X_{(s,t)}$ to couple diagonally to a basis of real scalar operators, with $c_{ij} = g_i \delta_{ij}$, while, for odd spin, $X_{(s,t)}$ couples diagonally to a basis of \textit{complex} scalar operators, with $c_{ij}=i\lambda_i  \delta_{ij}$, where $\lambda_i\in \mathds{R}$ (and the indices $i,j$ run over a different range covering half as many fields). 

\vskip 4pt
The three-point functions determine the explicit form of the action of the charge $Q_{(s,t)}$ on the scalar operators (up to an overall numerical factor that we can absorb into the definition of the charge). This is because we can integrate~\eqref{eq:3ptPMelectro} to obtain the action of the charge operator on the two-point function.
Explicitly, this leads to
\begin{align}
    \left[Q_{(s,t)}(\z),\O_i(\x)\right] &= g_i \,\square^{(s-t-1)/2} (\z\cdot \partial)^t \O_i(\x)\,, \hspace{5mm} \text{if}\hspace{3mm} s \in  2\mathds{Z}_{\geq 0}\,, \\
    \left[Q_{(s,t)}(\z),\O_i(\x)\right] &= i\lambda_i\, \square^{(s-t-1)/2} (\z\cdot \partial)^t \O_i(\x)\,, \hspace{3.5mm} \text{if}\hspace{3mm} s  \in  2\mathds{Z}_{\geq 0}+1\,.
\end{align}
We will use these results in the charge conservation identity.
\begin{itemize}
\item For even spin,
we examine the charge conservation condition
\begin{equation}
    \langle \left[Q_{(s,t)}(\z), X_{(s,t)}(\x_1,\z_1)\O_i(\x_2) \O_i(\x_3) \right]\rangle = 0\,.
\end{equation}
Since we are assuming in (\ref{equ:abelian}) that the PM field does not self-interact, this identity has only two terms
\begin{equation}
    g_i \left(\square_2^{(s-t-1)/2} (\z\cdot \partial_2)^t+\square_3^{(s-t-1)/2} (\z\cdot \partial_3)^t\right)\langle X_{(s,t)}(\x_1,\z_1)  \O_i(\x_2) \O_i(\x_3) \rangle = 0\,.
\end{equation}
The only way for this equation to be true is to set $g_i=0$.\footnote{A simple way to see this is to examine the constraint in momentum space. For this expression to vanish would then require that the prefactor $|k_1|^{s-t-1}(\z\cdot \k_1)^t+|\k_2|^{s-t-1}(\z\cdot \k_2)^t$ vanishes for generic kinematics, which is clearly impossible, for any value of $s$ or $t$. For $t=s-1$, this momentum-space constraint is identical to the flat-space Weinberg soft theorem constraint that establishes the non-existence of long-range forces mediated by massless particles with spin $s>2$ \cite{Weinberg:1964kqu}. } 

\item Similarly, for odd spin, we get
\begin{equation}
    i\lambda_i \left(\square_2^{(s-t-1)/2} (\z\cdot \partial_2)^t-\square_3^{(s-t-1)/2} (\z\cdot \partial_3)^t\right)\langle X_{(s,t)}(\x_1,\z_1) \O_i(\x_2) \O^\dagger_i(\x_3) \rangle = 0\,.
\end{equation}
In this case, there is one non-trivial solution $(s,t)=(1,0)$ corresponding to ordinary scalar electrodynamics. For all other cases, the conclusion is the same as the even-spin case, namely that the only consistent solution is to set $\lambda_i=0$.
\end{itemize}
Taken together, these results establish a no-go theorem excluding mass-diagonally coupled Abelian partially massless fields. This extends a previous no-go theorem established using Mellin space methods for the case $r=2$ \cite{Sleight:2021iix}. 

\subsection{Mass Non-Diagonal Coupling}

Next, we turn to the less familiar case of a PM field $X_{(s,t)}$, with $r \in 2\mathds{Z}_{\geq 0}+1$, and mass non-diagonal couplings to two scalar fields. 

\vskip 4pt
We begin with $r=1$ ($t=s-2$), for which the mass splitting is  $|\Delta_i-\Delta_j| =1$. Schematically, the charge operator acts as 
\begin{equation}
    \left[Q_{(s,s-2)},\O_\Delta\right] \sim \O_{\Delta+1} + \O_{\Delta-1}\,,
\end{equation}
i.e.~the matter fields form non-trivial ``multiplets" composed of operators with unit-spaced dimensions. Since the charge $Q_{(s,s-2)}$ is bosonic, we can act with it on $\O_\Delta$ an arbitrary number of times and generate a multiplet composed of operators with arbitrarily large positive or negative dimensions. To proceed, we need some additional physical input. 

\vskip 4pt
Since the scalar operators have dimensions that are integer spaced, they must correspond to either the complementary series ($0<\Delta <d$) or exceptional type-I representations ($\Delta = d+\mathds{Z}_{\geq 0}$). We can therefore assume that for physically sensible unitary bulk fields $\Delta>0$. Since the dimensions of the operators are bounded from below there must exist some operator  in the multiplet of \textit{minimal} dimension,  $\O_\Delta$, which satisfies
\begin{equation}
    \left[Q_{(s,s-2)}(\z),\O_{\Delta}(\x)\right] = a\,(\z\cdot \partial)^{s-2} \O_{\Delta+1}(\x)\,.
\end{equation}
As above, we then examine the charge conservation constraint 
\begin{equation}
    \langle [Q_{(s,s-2)}(\z), X_{(s,s-2)}(\x_1,\z_1) \O_\Delta(\x_2) \O_{\Delta}(\x_3) ] \rangle = 0\,,
\end{equation}
or, explicitly, 
\begin{align}
    a\, (\z\cdot \partial_2)^{s-2}\langle X_{(s,s-2)}(\x_1,\z_1) \O_{\Delta+1}(\x_2) \O_{\Delta}(\x_3)  \rangle + \left(2\leftrightarrow 3\right)= 0\,.
\end{align}
Unlike in the mass diagonal case, this constraint doesn't have the form of a single differential operator acting on a single function. Nonetheless, for any given $s$, it is straightforward to check whether there are any values of $(\Delta,d)$ that make this expression  vanish non-trivially. We have explicitly checked all spins up to $s=10$ and have found no such non-trivial solutions. 

\vskip 4pt
For $r>1$, the situation is more complicated. Since there are multiple possible mass splittings, the action of the charge operator on the lowest-dimension scalar can be a non-trivial linear combination of operators of different dimensions. For example, for $r=3$ ($t=s-5$), the action of the charge may have the general form 
\begin{equation}
    \left[Q_{(s,s-5)}(\z),\O_\Delta(\x)\right] = a_1 \square (\z\cdot \partial)^{s-5}\O_{\Delta+1}(\x) + a_2 (\z\cdot \partial)^{s-5}\O_{\Delta+3}(\x)\,.
\end{equation}
We have not performed a systematic study of all possible multiplet structures, but for spins $s\leq 5$ have been unable to find any non-trivial solutions. We leave a thorough study of non-diagonal couplings to future work.
	
\section{Coupling to Einstein Gravity}
\label{sec:PM-gravity}
	
The charge conservation conditions (\ref{eq:ccons}) always admit a trivial solution corresponding to setting all of the three-point interactions to zero. By themselves, they can therefore never be used to completely rule out interacting partially massless fields, since models with interactions beginning at four points trivially evade the constraints. The situation is different, however, if we assume that our model contains Einstein gravity, since the equivalence principle then mandates the existence of a cubic coupling of the graviton to every field with uniform strength $\propto \sqrt{G}$.  In the boundary CFT, this implies the existence of a  conserved spin-2 operator $T_{(2,1)}$ (the stress tensor) with non-vanishing three-point functions $\langle T_{(2,1)} \O \O\rangle$, for all local operators $\O$. 

\vskip 4pt
In this section, we will first explain why coupling to gravity is so constraining, followed by a general algorithm for constraining theories with gravitationally coupled PM fields. We will explicitly carry out this procedure for the cases of spin 2 and spin 3, both of depth 0.

\subsection{Why Gravity is Constraining}
		
Our starting point will be to assume that the bulk theory contains \textit{at least} a graviton and a PM field, and that the boundary CFT therefore has  a stress tensor $T_{(2,1)}$ and a partially conserved current $X_{(s,t)}$ in the spectrum. We will call this the {\it minimal theory} of a PM field coupled to gravity. There are then at least two charge conservation constraints that must be satisfied:
	\begin{align}
		\label{eq:moller}
		\langle [Q_{(s,t)}\,,X_{(s,t)} X_{(s,t)} X_{(s,t)} ]\rangle &= 0\,,\\[6pt]
		\label{eq:compton}
		\langle [Q_{(s,t)}\,,X_{(s,t)} T_{(2,1)} T_{(2,1)}]\rangle &= 0\,.
		\end{align}	
Schematically, these constraints arise from the following combinations of four-point exchange diagrams:
\begin{align}
0 &\ = \quad \sum_{\O} \includegraphics[valign=c,scale=0.55]{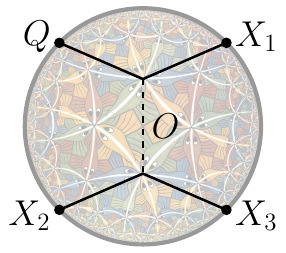} \quad + \quad 1 \leftrightarrow 2 \quad + \quad 1 \leftrightarrow 3\ , \\[6pt]
0 &\ = \quad \sum_{\O} \ \includegraphics[valign=c,scale=0.55]{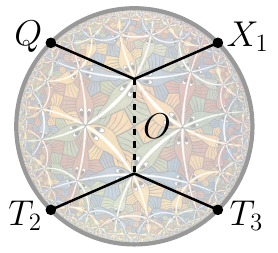} \quad + \quad \left(\ \ \sum_{{\O}^\prime} \includegraphics[valign=c,scale=0.55]{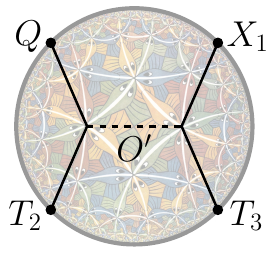}  \quad + \quad 2 \leftrightarrow 3\ \ \right)\ .
\end{align}
	The sums over $\O$ and $\O'$ are over all possible exchanged operators, 
and by assumption they must contain a non-zero contribution from $T_{(2,1)}$, because gravity couples to everything. 
    
    \vskip 4pt
    Before describing the solutions to these constraints, we will quickly review, from the boundary CFT perspective, why constructing consistent models of PM fields coupled to gravity is challenging. In a CFT, the dilatation generator $\hat D \propto \oint \text{d}\Sigma_\mu x_\nu T_{(2,1)}^{\mu\nu}$ counts the scaling dimensions of local operators. Using the definition (\ref{equ:pmcharge}), the PM charge operators $Q_{(s,t)}$ therefore satisfy the charge algebra relation
    \begin{equation}
    \label{eq:DQ}
    [\hat D,Q_{(s,t)}] = (s-1)Q_{(s,t)}\,.
    \end{equation}
The current algebra must integrate up to reproduce this relation. For $s>1$, we therefore must have $[Q_{(s,t)},T_{(2,1)}]\supset X_{(s,t)}$, and, due to the reciprocity property of the current algebra, also $[Q_{(s,t)},X_{(s,t)}]\supset T_{(2,1)}$. In the bulk, this corresponds to the statement that the graviton and PM gauge field form part of the same ``multiplet" and mix under the action of the PM gauge symmetry. On the boundary, the symmetries generated by the partially conserved currents must combine with the conformal generators to form a larger symmetry algebra.\footnote{This argument can be applied, essentially verbatim, to show that the presence of a spin-$3/2$ conserved current in a CFT implies that the conformal algebra is necessarily extended to a superconformal algebra, and therefore, in bulk AdS, the presence of a massless spin-$3/2$ field coupled to gravity is necessarily a model of supergravity.} 
 
 \vskip 4pt
 	As shown in \cite{Maldacena:2011jn}, for unitary CFTs in $d=3$ (extended to $d>3$ in \cite{Alba:2013yda, Alba:2015upa}), the presence of a single higher-spin current necessitates an \textit{infinite} extension of the conformal symmetry into a \textit{higher-spin algebra}~\cite{Eastwood:2002su,Vasiliev:2003ev}. The only models that realize this symmetry are essentially free (i.e.~the correlation functions coincide with those of either the free boson or free fermion vector models), establishing a CFT version of the Coleman--Mandula theorem \cite{Coleman:1967ad}. See~\cite{Fradkin:1986ka,Boulanger:2013zza}, for an algebraic perspective on this result. However, for the kind of non-unitary CFTs relevant for PM fields in de Sitter, the partially massless higher-spin algebras, and local field theories that realize them, are much less well understood.

	\subsection{Summary of the Algorithm}
	
The above considerations suggest a systematic approach
 to constrain PM fields coupled to gravity, which we will implement via the following algorithm: 
	\begin{enumerate}
	\item Assume an initial spectrum of operators consisting of at least the stress tensor and a partially conserved current $\{ T_{(2,1)}, X_{(s,t)},\, \cdots\}$. 
	\item Construct the most general form of the current algebra $[Q_{(s,t)},X_{(s,t)}]$ and $[Q_{(s,t)},T_{(2,1)}]$, with the assumed spectrum, using the procedure outlined in Section~\ref{ssec:CurrentAlgebra}. 
	\item Construct the most general conformally invariant three-point functions involving these operators, following the procedure outlined in Section~\ref{sec: Three-Point Functions}.
	\item Write down the charge conservation constraints (\ref{eq:moller}) and (\ref{eq:compton}). In practice, we sample a few hundred random rational kinematic points (subject to additional Gram determinant constraints in $d=3,4$). This gives a system of quadratic equations in the current algebra and OPE coefficients.\footnote{If the two-point functions are conventionally normalized, $\langle X_{(s,t)} X_{(s,t)}\rangle = {H_{12}^s/P_{12}^{d-1+t+s}}$, then the current algebra coefficients are not independent of the OPE coefficients---they are related by explicitly calculating  $\langle [Q_{(s,t)},X_{(s',t')}] X_{(s'',t'')}\rangle$ as a surface integral of a three-point function. We have found it simpler to instead let the two-point function normalizations (and hence the current algebra coefficients) float. This freedom can then be used to scale many of the current algebra coefficients to 1, reducing the number of nonlinear terms in the constraints.}
	\item Solve the equations subject to the additional constraint that 
	\begin{equation}
	\begin{aligned}
	\la T_{(2,1)} X_{(s, t)}X_{(s, t)} \ra &\neq 0\,, \\ 
	\la T_{(2,1)} T_{(2, 1)} T_{(2, 1)} \ra &\neq 0\,.
	\end{aligned}
	\end{equation}
 If there is {\it no} solution, then a model with the assumed spectrum is {\it rigorously
excluded}. 
	\item 
	Try to add new operators on the right-hand-side of the current algebra subject to the conditions (\ref{eq:CAspectrumconstraints}). If a non-trivial solution can be found by adding new operators, these will generate further constraints of the form (\ref{eq:moller}) and (\ref{eq:compton}) with new external states. Repeat the above steps. 
	\end{enumerate}
In the following sections, we will apply this algorithm 
to a few interesting examples. Details of the results can be found in a supplemental {\sc Mathematica} notebook at a GitHub repository for this paper~\href{https://github.com/CRTJones/PM-Consistency}{\faGithub}.

	\subsection{Constraints on Spin-2 Fields}
	\label{ssec:Spin2}
	
	In this section, we investigate the consistent coupling of a spin-2 PM field to Einstein gravity. We first demonstrate the inconsistency of the minimal theory, containing only gravity and the PM field. After this, we study which extensions of this theory allow for non-trivial solutions to the charge conservation constraints.
	
	\subsubsection{Minimal Theory}
	
	We begin with a minimal theory containing only the stress tensor $T_{(2,1)}$ and a partially conserved spin-2 operator $X_{(2,0)}$. With this restricted spectrum, following the analysis in Section~\ref{ssec:CurrentAlgebra}, the most general form of the action of the charge $Q_{(2, 0)}$ is 
	\begin{align}
		\label{eq:QX}
		[Q_{(2, 0)},X_{(2,0)}] &= a_1\hs T_{(2,1)}\,, \\[6pt]
		[Q_{(2, 0)},T_{(2,1)}] &= 
		b_1
		\left(\square - \frac{2}{d}(\z\cdot \partial) (\partial\cdot D_z) \right)X_{(2,0)}\,,  \label{eq:QT}
	\end{align}
	where the relative coefficient of the two differential operators in (\ref{eq:QT}) is determined by consistency with the conservation law for $T_{(2,1)}$. We then impose the constraints (\ref{eq:moller}) and (\ref{eq:compton}): 
	\begin{equation}
	\label{eq: charge conservation constraints spin-2 PM}
	\begin{aligned}
	\la [Q_{(2,0)}, X_{(2,0)}X_{(2,0)}X_{(2,0)}]\ra &= 0\,, \\ \la [Q_{(2,0)}, X_{(2,0)}T_{(2,1)}T_{(2,1)}]\ra &= 0\,.
	\end{aligned}
	\end{equation}
	Using the current algebra in (\ref{eq:QX}) and (\ref{eq:QT}), these charge conservation conditions take the following explicit form 
	\begin{align}
		0&\,=\, a_1   \la T_{(2,1)} X_{(2,0)} X_{(2,0)} \ra + (1\leftrightarrow 2) +   (1\leftrightarrow 3) \,, \label{equ:5-10}\\[6pt]
		0&\,=\, a_1 \la T_{(2,1)} T_{(2,1)} T_{(2,1)}\ra  \label{equ:5-11} \\
		&\qquad + \left\{ b_1\left( \square_2 - \frac{2}{d}(\z_2\cdot \partial_2) (\partial_2\cdot D_{z_2}) \right) \la X_{(2,0)}X_{(2,0)}T_{(2,1)}\ra + (2\leftrightarrow3)\right\} .	 \nonumber
	\end{align}
	As described in Section~\ref{sec: Three-Point Functions}, the three-point functions $\la T_{(2,1)}T_{(2,1)}T_{(2,1)}\ra$ and $\la T_{(2,1)}X_{(2,0)}X_{(2,0)}\ra$
	can be constructed in any dimension. Evaluating (\ref{equ:5-10}) and (\ref{equ:5-11})  for randomly generated rational kinematics then gives a system of simultaneous quadratic equations in terms of the parameters appearing in the current algebra and the three-point functions. We have checked that, in $d=3,4,5,6, 7$, these equations have \textit{no non-trivial solutions}. 
    In these dimensions, the minimal theory of a single PM spin-2 coupled to Einstein gravity is therefore ruled out. 
	
	\vskip 4pt
	Given the discussion in Section~\ref{sec: Examples of PM Theories}, we might have expected to find conformal gravity as a solution to the constraints (\ref{eq: charge conservation constraints spin-2 PM}) in $d=3$. 
	In Section~\ref{sec:SlightlyBroken}, we will explain why
 this theory is not found as a solution to the constraints.

	\subsubsection{Adding Extra Fields}
	
	We now proceed by enlarging the spectrum of operators appearing in the constraints (\ref{eq:moller}) and (\ref{eq:compton}) subject to the conditions described in Section  \ref{ssec:CurrentAlgebra}. The most general action of the charge $Q_{(2,0)}$ on the stress tensor $T_{(2,1)}$ and the partially conserved operator $X_{(2,0)}$ is 
	\begin{align}
		[Q_{(2,0)}, X_{(2, 0)}] &= a_1 T_{(2,1)} + a_2 (\partial \cdot D_z) X_{(3,0)} \\
		&\quad + a_3 \left(\square  - \frac{1}{d-1}(z\cdot \partial)(\partial \cdot D_z) \right)X_{(2,-1)} \,,\nonumber \\[6pt]
		[Q_{(2,0)}, T_{(2, 1)}] & = b_1 \left(\square - \frac{2}{d}(\z\cdot \partial)(\partial \cdot D_z) \right)X_{(2,0)}  + b_2 (\partial \cdot D_z)  X_{(3,1)}\,.
	\end{align}
Following the discussion in Section  \ref{ssec:CurrentAlgebra}, we might have expected a contribution of the form $(z\cdot \partial)X_{(1,0)}$ in the first equation. However, the conditions (\ref{eq:CAspectrumconstraints}) are necessary, but not sufficient, for an operator to appear in the current algebra. We also require the existence of a corresponding three-point function $\la X_{(1,0)} X_{(2,0)} X_{(2,0)}\ra$. In this case, it is straightforward to show that no such three-point function exists consistent with the required conformal invariance, conservation conditions and Bose symmetry, and so the operator $X_{(1,0)}$ cannot appear in the current algebra.

\vskip 4pt	
The only fields that can be added to the bulk theory (that will non-trivially contribute to the charge conservation constraints) are then a spin-3 PM field of depth 0 or 1, and a (non-conserved) massive spin-2 field with scaling dimension~$\Delta = d-2$. These correspond to the boundary operators $X_{(3,0)}$, $X_{(3,1)}$ and $X_{(2,-1)}$, respectively. 
	With this extended current algebra, the charge conservation constraints \eqref{eq: charge conservation constraints spin-2 PM}  become
	\begin{align}
		0&\,=\, a_1   \la T_{(2,1)} X_{(2,0)} X_{(2,0)} \ra + a_3\left( \square_1 - \frac{1}{d-1}(\z_1\cdot \partial_1) (\partial_1\cdot D_{z_1})\right)\la X_{(2,-1)}X_{(2,0)}X_{(2,0)}\ra \\
		&\qquad + a_2 (\partial_1 \cdot D_{z_1})\la X_{(3,0)}X_{(2,0)}X_{(2,0)} \ra+ (1\leftrightarrow 2) +  (1\leftrightarrow 3) \,, \nonumber \\[6pt]
		0&\,=\,a_1   \la T_{(2,1)} T_{(2,1)} T_{(2,1)} \ra + a_3\left( \square_1 - \frac{1}{d-1}(\z_1\cdot \partial_1) (\partial_1\cdot D_{z_1})\right)\la X_{(2,-1)}T_{(2,1)}T_{(2,1)}\ra \\
		&\qquad + \Bigg\{b_1 \left(\square_2 -\frac{2}{d} (\z_2\cdot \partial_2) (\partial_2\cdot D_{z_2}) \right) \la X_{(2,0)}X_{(2,0)}T_{(2,1)}\ra \nonumber \\ &\hspace{15mm}+ b_2 (\z_2\cdot \partial_2) (\partial_2\cdot D_{z_2})\la X_{(2,0)}X_{(3,1)}T_{(2,1)}\ra + (2\leftrightarrow3) \Bigg\}  \nonumber \,, 
	\end{align}
	where, in the first equation, we have used the fact that there is no consistent three-point function of the form $\la X_{(3,0)} T_{(2,1)}T_{(2,1)}\ra$.
	
		\begin{figure}[t!]
	\centering
	\includegraphics[width= 0.98\textwidth]{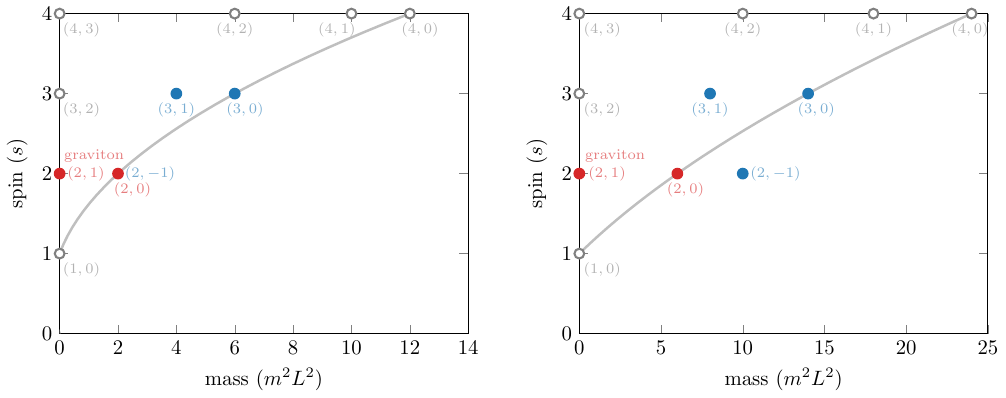}
	\caption{Illustration of the candidate spectra for $d=3$ ({\it left}) and $d = 7$ ({\it right}). In $d=3$, the masses of the fields dual to the operators $X_{(2,0)}$ and $X_{(2,-1)}$ coincide---the latter is marginally non-unitary, leaving a unique candidate spectrum with the additional operators $\{X_{(3,0)},X_{(3,1)}\}$. In $d>3$, candidate consistent spectra exist where the additional operators are either $X_{(2,-1)}$ or $\{X_{(3,0)},X_{(3,1)}\}$.}
	\label{fig:spin2}
	\end{figure}
	
	\vskip 4pt
	Contrary to the minimal theory, this extended theory does have non-vanishing solutions to the charge conservation constraints (see~\href{https://github.com/CRTJones/PM-Consistency}{\faGithub}). The details are dimension dependent and depicted in Figure~\ref{fig:spin2}. The results can be summarized as follows:
	
	\begin{itemize}
		\item $\boldsymbol{d=3}$: In three dimensions, non-trivial solutions exist only if the spectrum is enlarged with at least two out of the three operators  $\{X_{(2, -1)}, X_{(3,0)},X_{(3,1)}\}$:
 \begin{align}
 {\rm S1}:& \quad \big\{{\color{Red} T_{(2,1)}}, {\color{Red} X_{(2,0)}}, {\color{Blue} X_{(2,-1)}}, {\color{Blue} X_{(3,0)}}\big\}\ ,  \label{equ:S1}\\
  {\rm S2}:& \quad \big\{{\color{Red} T_{(2,1)}}, {\color{Red} X_{(2,0)}}, {\color{Blue} X_{(2,-1)}}, {\color{Blue} X_{(3,1)}}\big\}\ ,  \label{equ:S2}\\
  {\rm S3}:& \quad \big\{{\color{Red} T_{(2,1)}}, {\color{Red} X_{(2,0)}}, {\color{Blue} X_{(3,0)}}, {\color{Blue} X_{(3,1)}}\big\}\ .
 \label{equ:S3}
 \end{align}
However, in $d=3$, $X_{(2, -1)}$ has scaling dimension $\Delta = 1$, which is marginally below the unitarity bound of the spinning complementary series of dS$_{4}$ (c.f.~Section \ref{ssec:dS-Rep}). We interpret this to mean that the solutions with $X_{(2, -1)}$ in the spectrum may exist as gauge-invariant field theories in (A)dS, but not as physically acceptable unitary quantum field theories. 
 
 This leaves the solution S3 in  \eqref{equ:S3} as a single candidate for a physically acceptable extension to the minimal theory. 
  An interesting aspect of this solution is that it uniquely predicts the allowed three-point function of the stress tensor, $\langle T_{(2,1)} T_{(2,1)} T_{(2,1)}\rangle$, to be precisely that of Einstein gravity (i.e.~without a contribution from the higher-derivative Weyl-cubed operator). As we will find in Section~\ref{ssec:Spin3}, however, the solution in (\ref{equ:S3})  is ruled out when we look at the constraints coming from the current algebra of $X_{(3,0)}$.
  We have therefore obtained a sharp \textit{no-go} result for PM spin-2 in $d=3$.

		\item $\boldsymbol{d >3}$: In higher dimensions, 
		we find two distinct solutions to the charge conservation constraints, with at least the following operators:	
		\begin{align}
	{\rm S1}:& \quad	\big\{{\color{Red} T_{(2,1)}}, {\color{Red} X_{(2,0)}}, {\color{Blue} X_{(2,-1)}}\big\}\ ,\\[4pt]
		{\rm S2}:& \quad	\big\{{\color{Red} T_{(2,1)}}, {\color{Red} X_{(3,0)}},  {\color{Blue} X_{(3,0)}}, {\color{Blue} X_{(3,1)}}\big\}\ .
		\end{align}
		Note that, in $d>3$, the operator $X_{(2,-1)}$ corresponds to a unitary representation in the complementary series, and therefore the solution S1 may correspond to a physical theory in de Sitter, where the minimal spectrum of bulk fields is extended to include a spin-2 field with mass  $m^2 L^2 = 2(d-2)$. 
	\end{itemize}

	\vskip 4pt	
Since in all cases a non-trivial solution requires introducing additional operators, we now have additional charge conservation constraints that should be analyzed. In particular, the introduction of any additional (partially) conserved operators in the spectrum implies another set of charge conservation constraints associated with the new charge operators. As shown in \cite{Maldacena:2011jn}, restricting to singly conserved operators (massless higher-spin fields) this process will not terminate.  In particular, the new constraints require the addition of ever higher-spin operators with the eventual consistent spectrum consisting of an infinite tower. 
Whether our candidate solutions also require an infinite tower of operators remains an interesting open problem.

	\subsection{Constraints on Spin-3 Fields}
	\label{ssec:Spin3}
	In this section, we investigate the consistent coupling of a spin-3 PM field to Einstein gravity. 
	For simplicity, we restrict ourselves to the case of a depth-0 field.
	We will first demonstrate the inconsistency of the minimal theory, containing only gravity and the PM field. After this, we study which extensions of this theory allow for non-trivial solutions to the charge conservation constraints.

	\subsubsection{Minimal Theory}
	
	As above, we begin with a minimal theory consisting of only the stress tensor $T_{(2,1)}$ and a partially conserved spin-3, depth-0 operator $X_{(3,0)}$. With this restricted spectrum, the current algebra is 
	\begin{align}
			[Q_{(3,0)}, X_{(3,0)}] & = a_1 \hs (\z\cdot \partial )T_{(2,1)}\,,\label{equ:QX3} \\[6pt]
			[Q_{(3,0)}, T_{(2,1)}] & = b_1 \left(\square (\partial \cdot D_z)  - \frac{2}{d} (\z\cdot \partial)(\partial \cdot D_z)^2 \right)X_{(3,0)}\,,
			\label{equ:QT3}
	\end{align}
	where the relative coefficient between the two differential operators in (\ref{equ:QT3}) is determined by consistency with the conservation law for $T_{(2,1)}$. We are again interested in the charge conservation constraints (\ref{eq:moller}) and (\ref{eq:compton}):
	\begin{equation}\label{eq: charge conservation constraints spin-3 PM}
		\begin{aligned}
		\la [Q_{(3,0)}, X_{(3,0)}X_{(3,0)}X_{(3,0)}]\ra &= 0\,, \\
		\la [Q_{(3,0)}, X_{(3,0)}T_{(2,1)}T_{(2,1)}]\ra &= 0\,.
	\end{aligned}
	\end{equation}
	Using  (\ref{equ:QX3}) and (\ref{equ:QT3}), we get
	\begin{align}\label{eq: charge conservation eqs for spin-3}
		0&\,=\,a_1 (\z_1 \cdot \partial_1) \la T_{(2,1)}X_{(3,0)}X_{(3,0)}\ra + (1\leftrightarrow2) + (1\leftrightarrow3) \,,\\[6pt]
		0&\,=\,a_1 (\z_1 \cdot \partial_1) \la T_{(2,1)}T_{(2,1)}T_{(2,1)}\ra  \\
		&\qquad + \left\{ b_1\left(\square_2 (\partial_2 \cdot D_{z_2})  - \frac{2}{d}  (\z_2\cdot \partial_2)(\partial_2 \cdot D_{z_2})^2\right) \la X_{(3,0)}X_{(3,0)}T_{(2,1)}\ra + (2\leftrightarrow 3)\right\}  .  \nonumber
	\end{align}
	Again, following the procedure presented in Section~\ref{sec: Three-Point Functions}, ans\"atze for the three-point functions $\la T_{(2,1)}T_{(2,1)}T_{(2,1)}\ra$ and $\la T_{(2,1)}X_{(3,0)}X_{(3,0)}\ra$ can be constructed in any dimension. However, after substituting these into the above constraints, there are \textit{no non-trivial solutions} in $d=3,4,5,6,7$. This means that, just like the minimal theory for the spin-2 PM field, the minimal theory for the spin-3 PM field is ruled out. The next step is to investigate which fields can be added to the spectrum to induce a non-trivial solution to the charge conservation constraints.
	
	\subsubsection{Adding Extra Fields}
	\label{subsec:spin3extra}
	
	To find non-trivial solutions to the charge conservation constraints \eqref{eq: charge conservation constraints spin-3 PM}, we add extra operators to the current algebra consistent with the constraints (\ref{eq:CAspectrumconstraints}).  We find that the most general action of the charge $Q_{(3,0)}$ on the partially conserved operator $X_{(3,0)}$ and the stress tensor $T_{(2,1)}$ is 
	\begin{align}\label{equ:5-26}
		[Q_{(3,0)}, X_{(3,0)}]  = &\,\, a_1 \hs (\z\cdot \partial )T_{(2,1)}  \\[6pt]
		& +  \bigg[a_2 \left(\square^2 - \frac{2}{3d(d-1)} (\z\cdot \partial)^2 (\partial \cdot D_z)^2\right) \nonumber \\
		&\quad + a_3 \left(\square (\z\cdot \partial) (\partial \cdot D_z) - \frac{1}{d-1}(\z\cdot \partial)^2 (\partial \cdot D_z)^2 \right) \bigg] X_{(3,-2)}  \nonumber \\
		& + a_4 (\partial \cdot D_z) X_{(4,1)} + a_5 \left( \square (\z\cdot \partial) -\frac{2}{3d} (\z\cdot \partial)^2 (\partial \cdot D_z) \right) X_{(4,-1)} \nonumber  \\
		& +  a_6 \left(\square (z\cdot\partial)-\frac{1}{d-1}(z\cdot\partial)^2 (\partial\cdot D_z)\right)X_{(2,-1)} \nonumber \\
		&+ a_7 (\partial \cdot D_z)^2 X_{(5,0)}\,, \nonumber \\[8pt]
		[Q_{(3,0)}, T_{(2,1)}]  = &\,\,  b_1 \left(\square (\partial \cdot D_z)  - \frac{2}{d} (\z\cdot \partial)(\partial \cdot D_z)^2 \right)X_{(3,0)}   \label{equ:5-27} \\
		& +  \bigg[ b_2 \left( \square^2   -\frac{2}{d} \square (\z\cdot \partial)(\partial \cdot D_z)\right) \nonumber \\
		&\quad +  b_3 \left(\square^2 + \frac{2}{d(d-1)}(\z\cdot \partial)^2(\partial \cdot D_z)^2\right) \bigg] \tilde{X}_{(2,-1)}  \nonumber  \\[2pt]
		& + b_4 (\partial \cdot D_z)^2 \tilde{X}_{(4,1)}\, .  \nonumber
	\end{align}
	Again, while the constraints (\ref{eq:CAspectrumconstraints}) allow for a contributions of the form $(\z\cdot \partial )^2 X_{(1,0)} $ and $X_{(3,2)}$ in the first equation, there are no conformal three-point functions $\la X_{(1,0)}X_{(3,0)}X_{(3,0)}\ra$ and $\la X_{(3,2)}X_{(3,0)}X_{(3,0)}\ra$ consistent with the relevant conservation laws. The constraints (\ref{eq:CAspectrumconstraints}) also allow for an operator $X_{(3,2)}$ in~\eqref{equ:5-27}, but since this would have to be of the form $(\partial \cdot D_z) X_{(3,2)}$ it vanishes identically. Note that operators with $(s,t) = (2,-1)$ or $(4,1)$  can consistently appear on the right-hand-side of both current algebra relations. However, there is no requirement that these are the same operator, and so we have indicated the second instance by $\tilde{X}_{(2,-1)}$ and $\tilde{X}_{(4,1)}$.
    Finally, as discussed above, $X_{(2,-1)}$ and $X_{(4,-1)}$ have scaling dimensions $\Delta =  1$ in $d= 3$, which is marginally outside the range of the complementary series, and so they do not correspond to physical fields in the bulk. Similarly, $ X_{(3,-2)}$ is unphysical for $d \leq 4$.
	
	\vskip 4pt
	With this extended current algebra, the charge conservation constraints \eqref{eq: charge conservation constraints spin-3 PM}  become  
		\begin{align}
		0&\,=\,a_1 (\z_1 \cdot \partial_1) \la T_{(2,1)}X_{(3,0)}X_{(3,0)}\ra \\
		&\qquad + \bigg[a_2 \left(\square_1^2 - \frac{2}{3d(d-1)} (\z_1\cdot \partial_1)^2 (\partial_1 \cdot D_{z_1})^2\right) \nonumber \\
		&\qquad\quad+ a_3 \left(\square_1 (\z_1\cdot \partial_1) (\partial_1 \cdot D_1) - \frac{1}{d-1}(\z_1\cdot \partial_1)^2 (\partial_1 \cdot D_{z_1})^2 \right) \bigg] \la X_{(3,-2)} X_{(3,0)} X_{(3,0)} \ra \nonumber \\
		&\qquad + a_4 (\partial_1 \cdot D_{z_1})  \la X_{(4,1)}X_{(3,0)}X_{(3,0)}\ra \nonumber \\
		&\qquad + a_5 \left( \square_1 (\z_1\cdot \partial_1) -\frac{2}{3d} (\z_1\cdot \partial_1)^2 (\partial_1 \cdot D_{z_1}) \right) \la X_{(4,-1)} X_{(3,0)} X_{(3,0)}\ra \nonumber  \\
        &\qquad + a_6 \left(\square_1 (z_1\cdot \partial_1) -\frac{1}{d-1}(z_1\cdot \partial_1)^2 (\partial_1\cdot D_{z_1})\right)\langle X_{(2,-1)} X_{(3,0)} X_{(3,0)}\rangle \nonumber\\
		&\qquad +  a_7 (\partial_1 \cdot D_{z_1})^2  \la X_{(5,0)}X_{(3,0)}X_{(3,0)}\ra +   (1\leftrightarrow2) + (1\leftrightarrow3) \,,\nonumber
	\end{align}
	\begin{align}
		0&\,=\, a_1 (\z_1 \cdot \partial_1) \la T_{(2,1)}T_{(2,1)}T_{(2,1)}\ra  \\
		&\qquad + \bigg[a_2 \left(\square_1^2 - \frac{2}{3d(d-1)} (z_1\cdot \partial_1)^2 (\partial_1 \cdot D_{z_1})^2\right)  \nonumber \\
		&\qquad\quad+ a_3 \left(\square_1 (\z_1\cdot \partial_1) (\partial_1 \cdot D_{z_1}) - \frac{1}{d-1}(\z_1\cdot \partial_1)^2 (\partial_1 \cdot D_{z_1})^2 \right) \bigg] \la X_{(3,-2)} T_{(2,1)}T_{(2,1)} \ra \nonumber \\
		&\qquad + a_4 (\partial_1 \cdot D_{z_1})  \la X_{(4,1)}T_{(2,1)}T_{(2,1)}\ra \nonumber \\
		&\qquad + a_5 \left( \square_1 (\z_1\cdot \partial_1) -\frac{2}{3d} (\z_1\cdot \partial_1)^2 (\partial_1 \cdot D_{z_1}) \right) \la X_{(4,-1)} T_{(2,1)}T_{(2,1)}\ra \nonumber  \\
            &\qquad + a_6 \left(\square_1 (z_1\cdot \partial_1) -\frac{1}{d-1}(z_1\cdot \partial_1)^2 (\partial_1\cdot D_{z_1})\right)\langle X_{(2,-1)} T_{(2,1)} T_{(2,1)}\rangle \nonumber\\
		&\qquad +  a_7 (\partial_1 \cdot D_{z_1})^2  \la X_{(5,0)}T_{(2,1)}T_{(2,1)}\ra \nonumber\\
		&\qquad + \Bigg\{b_1 \left(\square_2 (\partial_2 \cdot D_{z_2})  - \frac{2}{d} (\z_2\cdot \partial_2)(\partial_2 \cdot D_{z_2})^2 \right)\la X_{(3,0)}X_{(3,0)}T_{(2,1)}\ra  \nonumber\\
		&\qquad\quad +  \bigg[ b_2 \left( \square^2_2   -\frac{2}{d} \square_2 (\z_2\cdot \partial_2)(\partial_2 \cdot D_{z_2})\right) \nonumber\\
		&\qquad\quad  +  b_3 \left(\square_2^2 + \frac{2}{d(d-1)}(\z_2\cdot \partial_2)^2(\partial_2 \cdot D_{z_2})^2\right) \bigg] \la X_{(3,0)} \tilde{X}_{(2,-1)} T_{(2,1)}\ra   \nonumber\\
		&\qquad\quad + b_4 (\partial_2 \cdot D_{z_2})^2 \la X_{(3,0)}\tilde{X}_{(4,1)} T_{(2,1)} \ra + (2 \leftrightarrow3)\Bigg\} \,.  \nonumber 
	\end{align}
        Note that although formally the three-point functions $\la X_{(3,-2)}T_{(2,1)}T_{(2,1)} \ra$ and $\la X_{(5,0)}T_{(2,1)}T_{(2,1)} \ra$ appear in the above identity, there are no non-vanishing expressions with the expected conservation and Bose symmetry properties.
        
        There are again multiple ways of solving the charge conservation constraints \eqref{eq: charge conservation constraints spin-3 PM} in a non-trivial  way (see~\href{https://github.com/CRTJones/PM-Consistency}{\faGithub}). These candidate theories depend on the dimensionality and are depicted in Figure~\ref{fig:spin3}. The results can be summarized as follows:
	
	\begin{itemize}
		\item $\boldsymbol{d=3}$: In three dimensions, we find two distinct solutions to the charge conservation constraints, with the following spectrum of operators:	
		\begin{align}
	{\rm S1}:& \quad	\big\{{\color{Red} T_{(2,1)}}, {\color{Red} X_{(3,0)}}, {\color{Blue} X_{(4,-1)}}, {\color{Blue} X_{(4,1)}}\big\}\ ,\\[4pt]
		{\rm S2}:& \quad	\big\{{\color{Red} T_{(2,1)}}, {\color{Red} X_{(3,0)}}, {\color{Blue} X_{(2,-1)}},{\color{Blue} \tilde{X}_{(2,-1)}}, {\color{Blue} X_{(3,-2)}},  {\color{Blue} X_{(4,1)}}\big\}\ .
		\end{align}
		However, as noted above, both of these solutions contain unphysical non-unitary operators. If we restrict the spectrum of additional operators to only those with a physical, unitary interpretation, $X_{(4,1)}$ and $X_{(5,0)}$, then we find no solutions.  This implies a \textit{no-go result for PM spin-3 depth-0 in $d=3$}. Since the operator $X_{(3,0)}$ also appears in the candidate solution  \eqref{equ:S3}, it also rules out that solution, giving a \textit{no-go result for PM spin-2 in $d=3$}.
		
		\begin{figure}[t!]
	\centering
	\includegraphics[width= 0.98\textwidth]{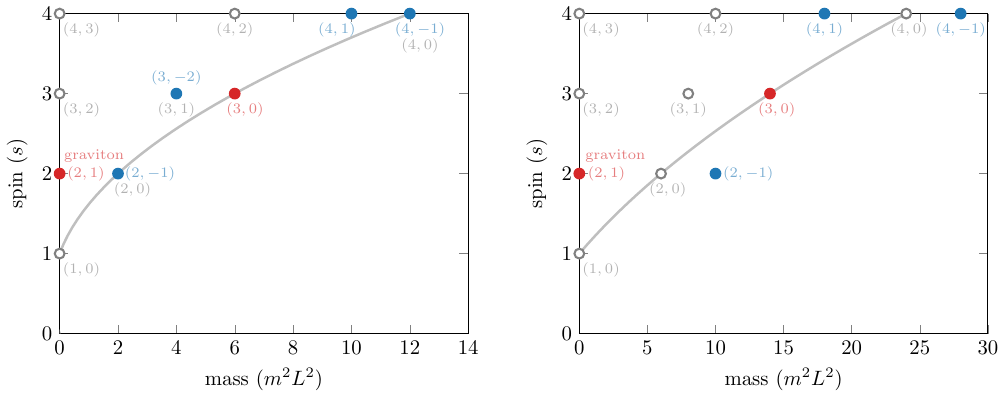}
	\caption{Illustration of the candidate spectra for $d=3$ ({\it left}) and $d = 7$ ({\it right}). In $d=3$, the spectra contain the additional operators $\big\{X_{(4,-1)}, X_{(4,1)}\big\}$ or $\big\{X_{(2,-1)},\tilde{X}_{(2,-1)}, X_{(3,-2)}, X_{(4,1)}\big\}$. However, since the operators $X_{(2,-1)}$, $\tilde{X}_{(2,-1)}$ and $ X_{(4,-1)}$ are marginally non-unitary, there is no physically acceptable extension of the minimal theory. In $d>3$, the extended spectra contain any two out of the operators $\big\{X_{(2,-1)}, X_{(4, -1)}, X_{(4,1)}\big\}$, where $X_{(2,-1)}$ and $ X_{(4, -1)}$ now correspond to unitary representations.}
	\label{fig:spin3}
	\end{figure}
	
		\item $\boldsymbol{d>3}$: In higher dimensions, the constraints are less restrictive and candidate physical solutions do exist. There are three distinct extensions to the spectrum---any two out of the three operators $X_{(2,-1)}, X_{(4, -1)}$ and, $X_{(4,1)}$ give a non-trivial solution:
	\begin{align}
	{\rm S1}:& \quad	\big\{{\color{Red} T_{(2,1)}}, {\color{Red} X_{(3,0)}}, {\color{Blue} X_{(4,-1)}}, {\color{Blue} X_{(4,1)}}\big\}\ ,\\[4pt]
		{\rm S2}:& \quad	\big\{{\color{Red} T_{(2,1)}}, {\color{Red} X_{(3,0)}}, {\color{Blue} X_{(2,-1)}},   {\color{Blue} X_{(4,1)}}\big\}\ ,\\[4pt]
		{\rm S3}:& \quad	\big\{{\color{Red} T_{(2,1)}}, {\color{Red} X_{(3,0)}}, {\color{Blue} X_{(2,-1)}},   {\color{Blue} X_{(4,-1)}}\big\}\ .
		\end{align}	
		 Note that in $d \neq 3$, $X_{(2,-1)}$ and $ X_{(4, -1)}$ correspond to unitary representations in the complementary series. Interestingly, in $d=6$, the solution S3  does not exist. 

	\end{itemize}
	As a consistency check, we have verified that these constraints are solved by the $\square^2$ theory presented in Section~\ref{ssec:Box2}. In generic dimensions ($d=5$ and $d\geq 7$), this theory contains a subset of the allowed primary operators appearing in the current algebra (\ref{equ:5-26}) and (\ref{equ:5-27}):
	\begin{equation}
		\label{eq:box2spectrum}
	\{X_{(2,-1)},T_{(2,1)},X_{(3,0)},X_{(4,1)}\}\,.
	\end{equation}
	Restricting to this spectrum, and substituting the explicit current algebra coefficients of this theory---derived in Appendix~\ref{sec: Current Algebra of Box-Squared CFT}---into the charge conservation constraints, uniquely fixes the OPE coefficients appearing in the constraints to their expected values. 
		
	\vskip 4pt
As reviewed in Section~\ref{ssec:special}, the $\square^2$ theory possesses peculiarities (perhaps pathologies) in $d = 3, 4, 6$. Consistent with these expectations, for $d=3$, restricting to the spectrum (\ref{eq:box2spectrum}), we find the charge conservation constraints have no solution. However, for $d = 4, 6$, the charge conservation constraints \textit{do} allow for a non-trivial solution even for the restricted spectrum. For these cases, we expect that an inconsistency will be found if more constraints are included.\footnote{The peculiar nature of $d=3,4,6$ is evident in the explicit $\square^2$ current algebra relations (\ref{eq: box-squared CA1}) as singularities in the coefficients. In the present analysis, however, we are considering only constraints with contributions from $[Q_{(3,0)},X_{(3,0)}]$ and $[Q_{(3,0)},T_{(2,1)}]$, for which only a singularity as $d\rightarrow 3$ is evident. By extending this analysis to include further constraints that are sensitive to singularities as $d\rightarrow 4,6$, we expect that these cases will be similarly excluded. } 

\vskip4pt
It is worth mentioning that the approach we have taken in this section places constraints only on partially massless interactions with a non-Abelian algebra of symmetries. For example, for any set of PM fields it is always possible to write interactions using powers of the linearized field strengths~\cite{Deser:2006zx,Skvortsov:2009zu,Gover:2014vxa,Cherney:2015jxp,Hinterbichler:2016fgl}. These will be invariant under linearized gauge transformations, which do not generate a non-trivial current algebra.  These interactions---like their flat-space massless higher-spin counterparts---contain too many derivatives to generate long-range forces.
There also exist interactions which cannot be written in terms of field strengths, and for which the gauge transformations are not equivalent up to field re-definition to their free counterparts, but for which the algebra is still undeformed, so that there is still no non-trivial current algebra (e.g.~the multi-field interacting PM theory of~\cite{Boulanger:2019zic} is of this type). This can be seen explicitly for the case of a single spin-2 field. In that case, there is no non-trivial current algebra that can be written, so that the PM symmetry is always Abelian (as was seen from the bulk in~\cite{Garcia-Saenz:2014cwa}). Thus, the algebraic method we have employed cannot see, for example, the obstruction to quartic completion of the two-derivative PM cubic vertex in $D=4$ encountered in~\cite{deRham:2013wv}. However, in light of arguments in flat space~\cite{Porrati:2012rd}, it is natural to expect that any current algebra involving the stress tensor will similarly require a deformation of the partially massless symmetry algebra, and so interactions of this type that don't deform the algebra must live in a decoupled sector.

\section{Higgsed Symmetry}
\label{sec:SlightlyBroken}

The analysis of the previous sections 
assumes the existence of {\it exactly} conserved currents, which lead to a non-trivial charge algebra. In Section~\ref{sec:background}, we described why these currents are expected as a consequence of the gauge invariance of interactions involving bulk (partially) massless fields.
In four spacetime dimensions, we then saw that there was no theory involving both a single partially conserved spin-2 current and a conserved stress tensor. This raises a sharp puzzle: Why did we fail to (re)discover conformal gravity? After all, as reviewed in Section~\ref{ssec:confgrav}, conformal gravity has precisely the field content and gauge invariances that we would expect to lead to such a pattern of conserved currents.

\vskip4pt
The resolution of this apparent paradox is both subtle and interesting. In order for a conserved current to exist on a boundary, it is necessary both that the bulk theory {\it and} the boundary conditions for the field be gauge invariant. 
It is the latter requirement that fails for conformal gravity. Correspondingly, boundary conservation of the partially conserved current is broken in this case.
This situation is not without precedent. In the AdS context, there are well-known examples~\cite{Karch:2000ct,Porrati:2001db,Burrington:2003gv,Karch:2023wui,Porrati:2024zvi} where one can make a choice of boundary conditions that break the 
 gauge invariance. In these cases, this leads to a (weakly) non-conserved current on the boundary, which is interpreted as a {\it Higgsing} of the bulk gauge field.
The novel feature in the de Sitter context is that we do not get to choose boundary conditions at the de Sitter boundary, but instead pick them in the far past by starting with 
Bunch--Davies initial conditions.
In order to derive observables, we then integrate over {\it all} possible future boundary conditions. As such, the de Sitter generalization of the Higgs phenomenon described in~\cite{Karch:2000ct,Porrati:2001db,Burrington:2003gv,Karch:2023wui,Porrati:2024zvi} has some novelties compared to the AdS case. In this section, we provide a brief discussion, but leave a more systematic analysis for future work.

\vskip4pt
Interestingly, we find that, for conformal gravity,  there is {\it no} choice of boundary conditions that preserve the bulk gauge invariance. Hence, even in AdS space, it does not lead to a boundary theory with both a partially conserved and exactly conserved current. Instead, conservation of the partially conserved current is necessarily weakly broken. As such, conformal gravity is {\it not} covered by the analysis of Section~\ref{sec:PM-gravity}. It is then natural to ask whether we can generalize our analysis to detect the existence of such bulk theories that exist only in a Higgs phase. Indeed we can. By using the fact that conservation of the boundary current is only softly broken, it is possible to continue to constrain the form of correlation functions, as we explore in this section.

		\subsection{Double Trace Deformations}
	\label{sec:doubletrace}	
	
	We begin by considering a PM field $\Phi_{M_1\ldots M_s}$ of spin $s$ and depth $t$ coupled to a pair of real scalar fields $\phi_1$ and $\phi_2$, with masses $m_1^2$ and $m_2^2$, respectively. Modulo integration-by-parts and equation of motion redundancies, there is a unique cubic interaction
	\begin{equation}
	S_{(s,t)} = \lambda\int \text{d}^{d+1}y \sqrt{-g} \; \Phi^{M_1\ldots M_s} \phi_1 \nabla_{M_1} \ldots \nabla_{M_s}\phi_2\,.
	\label{equ:Sst}
	\end{equation}
	On-shell, the linearized PM gauge transformation (\ref{equ:gauge}) of the action takes the form
	\begin{equation}
		\label{eq:dSst}
		\delta_\xi S_{(s,t)} = \lambda\mathcal{P}_{(s,t)}(m_1^2,m_2^2) \int \text{d}^{d+1}y \sqrt{-g} \; \xi^{M_1\ldots M_t} \phi_1 \nabla_{M_1} \ldots \nabla_{M_t}\phi_2\,,
	\end{equation}
	where $\mathcal{P}_{(s,t)}$ is a polynomial in $m_1^2$ and $m_2^2$ of degree $2s$. Invariance under \textit{small} gauge transformations\footnote{We assume that $\xi$ vanishes sufficiently quickly at the boundary of spacetime, so that the integral in (\ref{eq:dSst}) is finite.} then requires 
	\begin{equation}
		\label{eq:pst}
		\mathcal{P}_{(s,t)}(m_1^2,m_2^2) = 0\,.
	\end{equation}
	Equivalently, we can write this as a condition on the boundary dimensions $\Delta_i$, where as always $m_i^2 = \Delta_i (d-\Delta_i)$ (if we are setting $L\equiv 1$). For a fixed value of $\Delta_2$, the condition (\ref{eq:pst}) gives a polynomial constraint on $\Delta_1$ with $2s$ (not necessarily distinct) solutions. Some of these solutions can be found by imposing conservation of the corresponding boundary correlator. From the discussion in Section~\ref{sec:PM-electrodynamics}, we expect that there are $s$ distinct values of $\Delta_1$ that preserve the conservation of $\langle  O_1 O_2 X_{(s,t)}\rangle$. The remaining solutions can be deduced by observing that the constraint (\ref{eq:pst}) is trivially invariant under $\Delta_2 \rightarrow d-\Delta_2$, replacing one of the scalar operator dimensions by its \textit{shadow}. This therefore gives an additional $s$ solutions and determines the form of $\mathcal{P}_{(s,t)}(m_1^2,m_2^2)$ up to an overall coefficient.
	
	\vskip 4pt
As an explicit example, consider the case $s=2$, $t=0$. The polynomial $\mathcal{P}_{(2,0)}$ then is
\beq
	\mathcal{P}_{(2,0)} = \left[(\Delta_1-\Delta_2)^2-1\right] \left[(\Delta_1+\Delta_2-d)^2-1\right] .
	\label{eq:P20poly}
\eeq
Notice that this vanishes if either of the two factors in brackets vanishes. However, from the discussion in Section \ref{sec:PM-electrodynamics}, we know that the boundary correlator $\langle O_1 O_2  X_{(2,0)} \rangle$ is only conserved if $\Delta_1-\Delta_2=\pm 1$. In certain dimensions, there can then be gauge-invariant bulk vertices for which the second factor in~\eqref{eq:P20poly} vanishes, that do not actually generate a conserved correlator.
As an example, in $d=3$, there is a non-trivial mass-diagonal solution $m_1^2 = m_2^2 =2$, or, equivalently, $\Delta_1=\Delta_2 = 2$, which is the dimension of a conformally coupled scalar.  Evidently, this solution causes the term in the second bracket in~\eqref{eq:P20poly} to vanish.
However,  since this solution does not satisfy  $\Delta_1-\Delta_2=\pm 1$, the corresponding boundary correlator (surprisingly) is not conserved.
  		
	\vskip 4pt
	In general, such ``shadow" solutions correspond to bulk interactions that are locally gauge invariant, but fail to be invariant under \textit{large} gauge transformations. It is instructive to see in detail how this failure of boundary conservation arises. The explicit form of the boundary correlator in the above example is 
	\begin{equation}
		\label{eq:XOO}
	 \langle \O_1 \O_2X_{(2,0)} \rangle = \lambda \int \text{d}^{d+1}y \sqrt{-g} \, K_{\Delta_1}(x_1;y)  \nabla_{M_1} \nabla_{M_2} K_{\Delta_2}(x_2;y) K^{M_1 M_2}_{d-1}(z_3,x_3;y) \,,
	\end{equation}
	where $K^{M_1\ldots M_s}_\Delta$ is the bulk-to-boundary propagator for a symmetric and traceless tensor field of spin~$s$ and dimension $\Delta$ \cite{Costa:2014kfa}. At the PM values of the dimension, this satisfies
	\begin{equation}
		(\partial_x \cdot D_z)^{s-t} K^{M_1\ldots M_s}_{d-1+t} (z,x;y) \propto \nabla^{(M_{t+1}}\ldots \nabla^{M_s} K^{M_1\ldots M_t)}_{d-1+s}(x,y) - \text{traces}\,.
	\end{equation}
	This implies that the boundary divergence of a spin-$s$ depth-$t$ bulk-to-boundary propagator is ``pure gauge" with the gauge parameter given by a bulk-to-boundary propagator of spin $t$ and dimension $d-1+s$. 
	
	\vskip 4pt
	Taking the double-divergence of (\ref{eq:XOO}) gives
	\beq
	\begin{aligned}
	(\partial_3\cdot D_{z_3})^2 \langle \O_1 \O_2 X_{(2,0)} \rangle \, \propto \ &\left[(\Delta_1-\Delta_2)^2-1\right]  \left[(\Delta_1+\Delta_2-d)^2-1\right] \\
	&\times \int \text{d}^{d+1}y \sqrt{-g} \; K_{\Delta_1}(x_1;y)  K_{\Delta_2}(x_2;y) K_{d+1}(x_3,y) \,. 
	\end{aligned}
	\eeq	 
	This is of the same form as (\ref{eq:dSst}) with the bulk fields replaced with corresponding bulk-to-boundary propagators. The explicit value of this integral is well-known and given in (\ref{eq:bulk3pt1}).  Importantly, we observe that
	\begin{align}
	\int \text{d}^{d+1}y \sqrt{-g} \; K_{\Delta_1}(x_1;y)  K_{\Delta_2}(x_2;y) K_{d+1}(x_3,y) \propto \Gamma\left(\frac{1}{2}(\Delta_1+\Delta_2-d-1)\right).
	\end{align}
	The values of the conformal dimensions for which this integral diverges define so-called \textit{extremal correlators} \cite{DHoker:1999jke,Castro:2024cmf}. We see that precisely for the values of $\Delta_1$ and $\Delta_2$ in which the shadow prefactor vanishes, the bulk integral diverges, leaving a non-zero remainder. Moreover, this remainder has an interesting analytic structure, since the usual singularity in the OPE limit $|x_{12}|\rightarrow 0$ is absent. For the case $\Delta_1+\Delta_2=d+1$, we find 
	\begin{equation}
	(\partial_3\cdot D_{z_3})^2 \langle \O_1 \O_2 X_{(2,0)} \rangle  \propto \langle O_1(x_1) O_1(x_3)\rangle \langle O_2(x_2) O_2(x_3)\rangle = \frac{1}{P_{23}^{\Delta_2} P_{31}^{\Delta_1}}\,.
	\end{equation}
The interpretation is that boundary conservation is broken by a \textit{double-trace} operator\footnote{In the special case $O_1=O_2$, and therefore $\Delta_1=\Delta_2 = \frac{1}{2}(d+1)$, the double-trace operator on the right-hand-side is defined as the normal-ordered product $:O(x)O(y):\; \equiv \; O(x)O(y) - \langle O(x)O(y)\rangle $.}
\begin{equation}
	(\partial\cdot D_{z})^2 X_{(2,0)}(z,x) \propto O_1(x) O_2(x)\,.
\end{equation}
Here, the notion of ``double-trace" uses the fact that the boundary correlators, constructed holographically from a weakly coupled bulk, exhibit large-$N$ factorization
\begin{equation}
\langle O_1(x_1) O_2(x_2) O_1(x) O_2(y)\rangle = \langle O_1(x_1) O_1(x) \rangle \langle O_2(x_2) O_2(y)\rangle + \text{connected contributions}\,.
\end{equation}
The structure of the above example is generic. Shadow solutions to the local gauge invariance condition (\ref{eq:pst}) correspond to boundary correlators where current conservation is broken by a double-trace operator. This was interpreted in \cite{Porrati:2001db,Porrati:2003sa} as a dynamical Higgsing of the bulk gauge invariance. In AdS, if $-\frac{d^2}{4}<m_2^2< -\frac{d^2}{4}+1$, it is possible to construct a different boundary theory from the same bulk Lagrangian, with the expected boundary conservation, by imposing \textit{alternate quantization} boundary conditions on $\phi_2$ \citep{Klebanov:1999tb}, which effectively   exchanges a shadow solution for a regular solution. In dS, however, the only sensible boundary condition is on the choice of vacuum state. In this context, ``alternate quantization" simply reflects the freedom to express the wavefunction in a different basis of field eigenstates \cite{Anninos:2012ft} and therefore does not modify the physical system under consideration.    
	
	\subsection{Conformal Gravity}
	\label{sec:confgrav}
	
	Let us now employ these lessons to find conformal gravity from the algebraic perspective.
As reviewed in Section \ref{ssec:confgrav}, when expanded around an (A)dS$_4$ background, the spectrum of physical states in conformal gravity consists of a massless spin-2 field and a PM spin-2 field, dual to boundary operators $T$ and $X$, respectively.\footnote{In this subsection, we will drop the $(s,t)$ subscripts for reasons that will become apparent.} Naively, we should expect a solution to the charge conservation constraints for the minimal spin-2 spectrum in $d=3$, but, as discussed in Section~\ref{ssec:Spin2}, no such solution exists. 
	
	\vskip 4pt
	In Appendix \ref{app:confgrav}, the three-point function $\langle X X X\rangle$ is calculated from the bulk action (\ref{Lfff}), with the explicit result given as (\ref{eq:XXXbulk}). Taking the double-divergence, we find
        \begin{equation}
		\label{eq:ddXOO}
		\partial_\mu \partial_\nu \langle X^{\mu\nu} XX\rangle = \frac{9}{32\pi^4 M_{\text{Pl}}} \frac{\left(H_{23}+2 V_2 V_3\right)^2}{P_{12}^2 P_{23}^2 P_{31}^2}\,.
	\end{equation}
	Writing this out in real space, we find that the right-hand-side is regular in the limit $|x_{23}|\rightarrow 0$. As in the scalar example above, even though the bulk interaction is invariant under local gauge transformations, the boundary conditions implied by the choice of Bunch--Davies vacuum state have Higgsed the PM gauge symmetry.  {There is a mismatch between the number of bulk PM gauge-invariant cubic structures in $D=4$, and the number of exactly doubly conserved three-point structures in $d=3$ CFT (this mismatch does not occur in higher dimensions). We interpret the semi-analytic structure of (\ref{eq:ddXOO}) as the breaking of boundary conservation by a double-trace operator: 
\beq
	\partial_\mu\partial_\nu   X^{\mu\nu} =\ \frac{1}{2M_{\rm Pl}} : X^{\mu\nu}X_{\mu\nu}:\,,\label{doubletrX}
\eeq
which is derived in Appendix~\ref{app:confgrav}.

\vskip 4pt
	An important consequence of this relation is that $X$ acquires an {\it anomalous dimension}, $\Delta_X = 2 + \gamma$.	 As explained in more detail in Section~\ref{subsec:anomalous}, beginning with the unique conformally invariant two-point function 
\beq	\label{eq:ddX}
	\LA XX\RA = - \frac{3}{\pi^2}\frac{H_{12}^2}{(-2P_{12})^{4+\gamma}}\,,
\eeq
	taking the double-divergence of both operators using (\ref{doubletrX}) 
	and expanding both sides to leading order in $1/N$, gives 
 \begin{equation}
		\gamma =  \frac{3}{4\pi^2M_{\rm Pl}^2}\, .
        \label{eq:gamma}
	\end{equation}
	Equivalently, this corresponds to a correction of the mass of the bulk field, $\delta m^2 = -\gamma$. By counting powers of the coupling, we see that this is a dynamical effect corresponding to a one-loop calculation in the bulk. We conclude that while classically conformal gravity is a model of a PM spin-2 field coupled to Einstein gravity, quantum mechanically the PM gauge symmetry is broken. It would be interesting to reproduce this in an explicit bulk loop calculation.\footnote{It would also be interesting to explore the analogous phenomenon in the context of conformal supergravity, whose PM interactions were recently studied in~\cite{Boulanger:2024hrb,Boulanger:2024qrn}.}
	
	\vskip 4pt
	The physical interpretation of this dynamical mass generation is somewhat curious. The classically PM bulk field is in a short exceptional series representation of the conformal group with four degrees of freedom. To gain an anomalous dimension, and therefore form a long complementary series representation, there must be a non-trivial recombination of conformal multiplets; the PM spin-2 must eat a spin-0 Goldstone mode. In this example, the Goldstone is a bound state of the bulk fields dual to the single-trace constituents of the double-trace operator $:X^{\mu\nu}X_{\mu\nu}:$\,. However, these bulk fields are precisely the transverse modes of the field that is being Higgsed, which leads to a novel kind of \textit{self-Higgsing}: A field that gains a mass correction by eating itself.\footnote{For massless higher-spin fields (fully conserved higher-spin boundary currents) such a situation is straightforwardly impossible. The corresponding double-trace breaking of conservation would have to be of the schematic form $\partial \cdot J^{(s)} \sim J^{(s)} \partial^k J^{(s)}$, for some $k\geq 0$. However, since $[J^{(s)}] = d-2+s$, matching scaling dimensions and requiring $s\geq 1$ gives no solutions for $d> 2$.} It is worth noting that this is an intrinsically curved-space phenomenon. There is no flat-space counterpart, since bound states cannot form at weak coupling. In AdS space, the gravitational potential makes it possible to form effective bound states even at zero coupling. In de Sitter, the fact that massive particles redshift into the cosmic rest frame also allows them to become effectively bound in perturbation theory.

	\section{Conclusions and Outlook}
	\label{sec:conclusions}

While particles in Minkowski space are either massive or massless, the
representation theory of de Sitter space allows for an extra category of exotic partially massless 
particles. If these particles existed during inflation, they would lead to distinct imprints in inflationary correlators~\cite{Baumann:2017jvh,Goon:2018fyu,Baumann:2019oyu}, providing a direct probe of the de Sitter era of the early universe.
It is therefore important to understand what interactions involving light particles are consistent in cosmological spacetimes.

\vskip 4pt
In this paper, we have studied the consistency of theories involving partially massless fields in (A)dS, especially when coupled to gravity.  To derive constraints on these theories, we used the fact that PM fields are naturally associated with partially conserved currents on the boundary of the spacetime.
 These currents can be used to construct symmetry operators, whose presence greatly constrains correlation functions of charged operators.  If an assumed spectrum of currents is inconsistent with the associated ``charge conservation" constraints, the corresponding bulk theory cannot exist. 
These constraints are especially stringent when the theory is coupled to gravity, which demands the existence of the stress tensor in the spectrum of conserved currents. 

\vskip 4pt
The no-go results derived in this paper are only as strong as their assumptions. To recapitulate, we have assumed:
\begin{enumerate}
    \item The bulk theory consists of weakly interacting fields propagating on an \textit{exact} dS$_{d+1}$ background with Bunch--Davies initial conditions (\ref{eq:BDinitial}). 
    \item The boundary correlation functions (wavefunction coefficients) satisfy the same kinematic properties (Ward identities) as correlation functions in a Euclidean CFT. We do {\it not} assume any other CFT properties, such as a convergent OPE expansion. 
    \item Elementary bulk fields are dual to conformal primary operators corresponding to the UIRs of SO$(d+1,1)$ enumerated in Section \ref{ssec:dS-Rep}. This restriction of admissible operator dimensions is the \textit{only} use of unitarity. 
    \item The action of the charge (\ref{equ:pmcharge}) on a local operator $[Q,O(x)]$ is another local operator. All local operators are assumed to be linear combinations of \textit{primaries} and \textit{descendants}. 
    \item In Section \ref{sec:PM-gravity}, we assume that the bulk theory contains Einstein gravity. In the boundary CFT this means we are assuming the existence of a (unique) conserved spin-2 stress tensor $T_{\mu\nu}$ with corresponding charge operators that generate the $d$-dimensional conformal group. 
    \item In dS$_4$, we have restricted to \textit{parity-even} operators (tensor fields) and assumed parity preserving interactions.\footnote{Even if we assume parity as a symmetry of the system, the exchange of parity-odd operators (pseudo-tensor fields) may give a non-trivial contribution to the charge conservation constraints (\ref{eq:moller}) and (\ref{eq:compton}). We thank Sasha Zhiboedov for pointing this out. }
\end{enumerate}
With these assumptions, we then obtained the following results:
In four-dimensional de Sitter space, we found a strict no-go for PM fields of spin 2 or 3 coupled to gravity. 
In particular, we showed that the charge conservation constraints require the addition of extra massive fields, but that these fields are  non-unitary. In higher dimensions, on the other hand, the constraints can be solved without any violation of unitarity. These solutions require the existence of additional PM fields, which lead to additional charge conservation constraints. 
Whether this implies an infinite tower of higher-spin PM fields, like in the case of ordinary higher-spin currents~\cite{Maldacena:2011jn}, is still an open problem.

\vskip 8pt
We end with a list of future directions:
\begin{itemize}
\item Our strongest results were for four-dimensional de Sitter space, the case of physical interest. In that case, gravitationally coupled PM fields were ruled out because they demanded the existence of extra fields with masses just below the unitarity bound. However, since this unitarity bound was derived in perfect de Sitter space, it is still conceivable that the theory is consistent in an inflationary background where this bound could be slightly shifted. In that case, it would lead to an interesting prediction: If the signal of a PM field were to be observed, it must come together with the imprint of another field with a fixed mass. It would be interesting to explicitly construct an inflationary model with this precise spectrum.

\item Relatedly, it would be phenomenologically interesting to study the weak breaking of the symmetries associated to PM fields. The inflationary signatures of massive particles become larger as the fields get lighter. Since
the depth-$0$ PM point is the lightest that a field can be in de Sitter space while remaining unitary, it is interesting to study how fields behave as we dial their masses to this value. In all cases that we have studied, the presence of PM fields requires additional higher-spin particles to be present in the spectrum. This suggests that the gap between a light particle near the depth-$0$ mass and the next lightest state cannot be parametrically large. It would be very useful to turn this into a sharp bound.

\item In Section \ref{sec:SlightlyBroken}, we have made a preliminary investigation of the possibility that the PM symmetry may be weakly Higgsed in the Bunch--Davies vacuum. This corresponds to the breaking of current conservation by a double-trace operator. In~\cite{Maldacena:2011jn,Maldacena:2012sf} (see also \cite{Skvortsov:2015pea,Giombi:2016zwa,Sezgin:2017jgm,Sharapov:2018kjz,Ponomarev:2019ltz,Skvortsov:2020wtf,Gerasimenko:2021sxj}), it was shown that the charge conservation identities (\ref{equ:CCI}) can be extended to \textit{pseudo-charge} (non-)conservation identities. By parametrizing all possible double-trace operators constructible from an assumed spectrum, it should be possible to use these identities to generalize the approach of Section \ref{sec:PM-gravity} to carve out the space of consistent weakly Higgsed models.  In the bulk, it would be interesting to reproduce the predicted anomalous dimension of the PM spin-2 field appearing in conformal gravity from an explicit loop calculation along the lines of~\cite{Porrati:2003sa,Duff:2004wh,Karch:2023wui}. 

\item Mathematically, we have solved a set of polynomial equations of the schematic form
\beq
\sum_{i,j}a_i c_j F_{ij}(x_1,x_2,x_3) = 0\,,
\eeq
where $F_{ij}(x_1,x_2,x_3)$ are known functions,  $a_i$ are coefficients in the current algebra, and $c_j$ are coefficients in the conformal three-point functions.  In practice, this has been cumbersome, since the number of equations and parameters quickly becomes very large. To derive no-go results, however, we only have to be able to prove that a given set of equations has no non-trivial solutions, without trying to explicitly solve them. We hope to revisit this problem with more mathematical sophistication in the future. For larger values of spin, even constructing the necessary (partially) conserved three-point functions is computationally non-trivial. It would be preferable to have a constructive approach to generating these expressions or even a closed-form solution for $\langle X_{(s,t)} X_{(s',t')} X_{(s'',t'')}\rangle$,  generalizing similar results for fully conserved correlators \cite{Giombi:2011rz}. Recent developments in the application of twistor space methods to 3d CFTs~\cite{Baumann:2024ttn} may be useful here. Such closed-form expressions may be essential to generalizing the powerful light-cone limit analysis of \cite{Maldacena:2011jn}.

\item In higher-dimensional de Sitter space, we have found ``candidate solutions" describing a PM field of spin 2 or 3 coupled to gravity. These solutions involve additional PM fields, leading to new charge conservation identities, whose solutions may require additional PM fields, and so on. It is unclear whether this process ever terminates or whether we get an infinite tower of fields (as for ordinary higher-spin currents~\cite{Maldacena:2011jn}). It is tempting to conjecture that, by adding successively more constraints, the consistent solutions will converge to the spectrum and interactions identified with the non-unitary generalizations of the free large-$N$ $U(N)$ vector models reviewed in Section \ref{ssec:Box2} (and their fermionic generalizations). This would be a more-or-less direct analog of the Maldacena--Zhiboedov uniqueness result \cite{Maldacena:2011jn}, or, equivalently, a generalization of the Coleman--Mandula theorem to non-unitary CFTs. 
\end{itemize}

 \vspace{0.2cm}
 \paragraph{Acknowledgments} We thank Veronica Calvo Cortes, Harry Goodhew, Johannes Henn, Diego Hofman, Claire de Korte, Gordon Lee, Juan Maldacena, Dmiitri Pavlov, Kostas Skenderis, Bernd Sturmfels and Alexander Zhiboedov for helpful discussions.
 
 \vskip 4pt
The research of DB, CRTJ, JM and NM  is funded by the European Union (ERC,  \raisebox{-2pt}{\includegraphics[height=0.9\baselineskip]{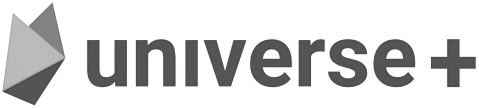}}, 101118787). DB is further supported by a Yushan Professorship at National Taiwan University (NTU) funded by the Ministry of Education (MOE) NTU-112V2004-1. He also holds the Chee-Chun Leung Chair of Cosmology at NTU.
 KH acknowledges support from DOE award DE-SC0009946.
 AJ is supported by DOE award DE-SC0025323 and by the Kavli Institute for Cosmological Physics
at the University of Chicago. HL is supported in part by DOE award DE-SC0013528.

	\newpage
	
	\appendix
	\section{Aspects of \texorpdfstring{$\boldsymbol{\Box^2}$}\ \hskip 2pt CFT }
	\label{sec: Current Algebra of Box-Squared CFT}
	
	In this appendix, we present the spectrum of primary operators and derive the current algebra relations for the $\square^2$ generalization of the large-$N$ free bosonic $U(N)$ vector model relevant for the discussion in Section~\ref{subsec:spin3extra}. 
	We also comment on some special features of these models in dimensions $d=3,4,6$.
		
	\subsection{Spectrum of Operators}
	
	Up to an arbitrary normalization, the lowest-dimension primary operators have the explicit form: 
	\begin{align}
		X_{(0,-3)} &= \sum_{a=1}^N \phi_a^\dagger \phi_a \,,\\[4pt]
		X_{(0,-1)} &= \sum_{a=1}^N \left[\phi_a^\dagger \square\phi_a + \frac{2}{d-4}\partial_z\phi_a^\dagger \partial_z\phi_a \right] + \text{h.c.}\,,  \\[4pt]
		X_{(1,-2)} &= i\sum_{a=1}^N \phi_a^\dagger \partial_z\phi_a + \text{h.c.}\,,\\[4pt]
		X_{(1,0)} &=i\sum_{a=1}^N \left[\phi_a^\dagger \square\partial_z\phi_a +\frac{d}{d-4} \partial_z\phi_a^\dagger \square\partial_z\phi_a+\frac{4}{d-4} \partial_\mu\phi_a^\dagger \partial^\mu\partial_z\phi_a \right] + \text{h.c.}\,,\\[4pt]
		X_{(2,-1)} &= \sum_{a=1}^N \left[\phi_a^\dagger \partial_z^2\phi_a - \frac{d-2}{d-4}\partial_z\phi_a^\dagger \partial_z\phi_a \right]+ \text{h.c.}\,,\\[4pt]
		T_{(2,1)} &= \sum_{a=1}^N \left[\phi^\dagger_a \square \partial_z^2 \phi_a -\frac{2 (d+2)}{d-4}\partial_z \phi_a^\dagger \square \partial_z \phi_a +\frac{d(d+2)}{(d-2)(d-4)} \partial_z^2 \phi_a^\dagger \square \phi_a  \right. \\
		&\left. \hspace{15mm}+\frac{4}{d-4} \partial_z^2\partial_\mu\phi^\dagger_a \partial^\mu\phi_a  -\frac{4 d}{(d-2)(d-4)} \partial_\mu \partial_z\phi_a^\dagger \partial^\mu \partial_z \phi_a \right] + \text{h.c.}\,,  \nonumber \\[6pt]
		X_{(3,0)} &= i\sum_{a=1}^N \left[\phi_a^\dagger \partial_z^3\phi_a -\frac{3d}{d-4}\partial_z\phi_a^\dagger \partial_z^2\phi_a\right] + \text{h.c.} \,,\\[4pt]
		X_{(3,2)} &= i\sum_{a=1}^N \left[\phi_a^\dagger \square \partial_z^3 \phi_a  -\frac{3 (d+4)}{d-4} \partial_z^2 \phi_a^\dagger \square \partial_z \phi_a +\frac{3 (d+2) (d+4)}{(d-4) (d-2)}\square\partial_z^2 \phi^\dagger_a  \partial_z \phi_a \right. \\
		&\hspace{15mm}\left.-\frac{(d+2) (d+4)}{(d-4) (d-2)}\partial_z^3 \phi^\dagger_a \square \phi_a +\frac{4}{d-4}\partial_\mu\phi^\dagger_a \partial^\mu \partial_z^3 \phi_a\right. \nonumber \\
		&\hspace{15mm}\left.-\frac{12 (d+2)}{(d-4) (d-2)}\partial_\mu \partial_z^2\phi^\dagger_a \partial^\mu \partial_z\phi_a\right] + \text{h.c.}\,, \nonumber\\[4pt]
		X_{(4,1)} &= \sum_{a=1}^N \left[\phi_a^\dagger \partial_z^4\phi_a - \frac{4(d+2)}{d-4}\partial_z\phi_a^\dagger \partial_z^3\phi_a +\frac{3d(d+2)}{(d-2)(d-4)}\partial_z^2\phi_a^\dagger \partial_z^2\phi_a \right]+ \text{h.c.}\,,\\[6pt]
		X_{(4,3)} &= \sum_{a=1}^N \left[\phi_a^\dagger \square \partial_z^4 \phi_a -\frac{4 (d+6)}{d-4}\partial_z \phi^\dagger_a \square \partial_z \phi_a-\frac{4 (d+2) (d+4) (d+6)}{(d-4) (d-2) d}\square \partial_z^2 \phi_a^\dagger \partial_z^2 \phi_a\right. \\
		&\hspace{15mm}\left.+\frac{(d+4) (d+6)}{(d-4) (d-2)}\partial_z^4\phi_a^\dagger \square \phi_a +\frac{6 (d+4) (d+6)}{(d-4) (d-2)}\partial_z\phi_a^\dagger\square\partial_z\phi_a +\frac{4}{d-4}\partial_\mu \phi_a^\dagger \partial^\mu\partial_z^4\phi_a\right. \nonumber\\
		&\hspace{15mm}\left.-\frac{16 (d+4)}{(d-4) (d-2)}\partial_\mu \partial_z \phi_a^\dagger \partial^\mu \partial_z^3 \phi_a+\frac{12 (d+2) (d+4)}{(d-4) (d-2)
			d} \partial_\mu \partial_z^2 \phi^\dagger_a \partial^\mu \partial_z^2 \phi_a\right] + \text{h.c.}\,,\nonumber
	\end{align}
	where we have used the shorthand $\partial_z \equiv z\cdot\partial$.

	\subsection{Current Algebra}
	
	We are interested in the action of the translationally invariant charge operator corresponding to the lowest-dimension partially conserved current~$X_{(3,0)}$. 
	To compute the current algebra, we use the action of this charge on the fundamental field
	\begin{equation}
		\label{eq:box2transformations}
		\begin{aligned}
		\big[Q_{(3,0)},\phi_a\big] &= i\hs \delta\phi_a =  i\hs \square \phi_a\,, \\
		\big[Q_{(3,0)},\phi^\dagger_a\big] &= -i\hs\delta\phi^\dagger_a = -i\hs \square \phi^\dagger_a\,.
		\end{aligned}
	\end{equation}
	It is straightforward to verify that this is a symmetry of the action (\ref{eq:box2action}). 
	The current algebra is found by applying this transformation to the above explicit results for the single-trace operators. The resulting expression is a large sum of bilinears that can be rewritten back in terms of the known primaries and their descendants in a unique way. The result is
	\begin{align}\label{eq: box-squared CA1}
		\left[Q_{(3,0)},X_{(0,-3)}\right] &= \frac{2}{d-2}(\partial \cdot D_z) X_{(1,-2)}\,,\\[4pt]
		\left[Q_{(3,0)},X_{(0,-1)}\right] &= \frac{4}{(d-2)(d-4)}\square (\partial\cdot D_z)X_{(1,-2)}\,,\\[4pt]
		\left[Q_{(3,0)},X_{(1,-2)}\right] &= -\frac{2}{(d-3)(d-6)}\square \partial_z X_{(0,-3)} + \frac{d-4}{d(d-3)}(\partial\cdot D_z)X_{(2,-1)}\\[4pt]
		&\hspace{5mm} + \frac{2(d-4)}{d(d-6)}\partial_z X_{(0,-1)}\,, \nonumber\\[6pt]
		\left[Q_{(3,0)},X_{(1,0)}\right] &= \frac{2}{d(d-2)(d-3)}\left((d-2)\square-2\partial_z (\partial\cdot D_z)\right)(\partial\cdot D_z)X_{(2,-1)}\,,\\[4pt]
		\left[Q_{(3,0)},X_{(2,-1)}\right] &= -\frac{4}{(d-1)(d-4)^2}\partial_z\left((d-3)\square-\partial_z (\partial\cdot D_z)\right) X_{(1,-2)} \\[4pt]
		&\hspace{5mm} +\frac{2(d-3)}{3(d-1)(d+2)}(\partial\cdot D_z)X_{(3,0)} +\frac{4(d-3)}{(d-4)(d+2)}\partial_z X_{(1,0)}\,, \nonumber \\[6pt]
		\left[Q_{(3,0)},T_{(2,1)}\right] &= \frac{4}{3(d-2)d(d+2)}\left(d\hs \square -2\partial_z(\partial\cdot D_z)\right)(\partial\cdot D_z)X_{(3,0)}\,,\\[4pt]
		\left[Q_{(3,0)},X_{(3,0)}\right] &= \frac{6}{d+4}\partial_z T_{(2,1)} + \frac{d-2}{2(d+1)(d+4)} (\partial \cdot D_z)X_{(4,1)} \\[4pt]
		&\hspace{5mm}- \frac{6}{(d-3)(d-2)(d+1)}\partial_z((d-1)\square -\partial_z(\partial\cdot D_z))X_{(2,-1)}\,, \nonumber \\[6pt]
		\left[Q_{(3,0)},X_{(3,2)}\right] &= \frac{1}{(d-1)(d+2)(d+4)}\left((d+2)\square-2\partial_z (\partial\cdot D_z)\right)(\partial\cdot D_z) X_{(4,1)}\,.
	\end{align}
	The action of the charge on the rest of the infinite tower of operators can be calculated analogously, but will not be of interest to us.

	\subsection{Special Cases}
	\label{ssec:special}

	The  current algebra in the previous subsection was derived while taking the dimension $d$ to be generic. However, at the special values $d=3,4,6$, the theory develops peculiarities that discontinuously change the spectrum of primary operators. 	
	\begin{itemize}
		\item In \textbf{$d = 3$}, there is an accidental degeneracy of operators 
			\begin{equation}
		X_{(2,-1)} \overset{d\rightarrow 3}{\propto} \partial_z^2 X_{(0,-3)}\,.
	\end{equation}
	To maintain completeness of the spectrum, a new so-called \textit{extension operator} appears that is neither primary nor descendant.\footnote{In a unitary CFT, the statement that all operators are primaries or descendants is a theorem \cite{Simmons-Duffin:2016gjk}. In a non-unitary CFT, however, this is no longer guaranteed and the $\square^2$ model provides perhaps the simplest explicit counter-example in $d>2$.} It is unclear if the bulk interpretation of such an unusual conformal multiplet is physically acceptable, so our analysis excludes this possibility by assumption. 
		 
	\item In \textbf{$d = 4$}, the fundamental field $\phi_a$ has scaling dimension $\Delta_\phi = 0$ and therefore 
	\begin{equation}
	\langle \phi_a^\dagger(x) \phi_b(0)\rangle \overset{d\rightarrow 4}{=} 1\,.
	\end{equation}
	Every operator in the spectrum, other than $X_{(0, -3)}$, including the stress tensor $T_{(2,1)}$ has a vanishing two-point function and therefore corresponds to a null state. This theory is therefore not a CFT in the usual sense and cannot be dual to a bulk theory containing Einstein gravity.
	
	\item In \textbf{$d = 6$}, there is again an accidental degeneracy of operators 
				\begin{equation}
		X_{(0, -1)} \overset{d\rightarrow 6}{\propto} \square X_{(0, -3)}\,.
	\end{equation}
	As in the case $d=3$, this theory contains an extension operator and so is excluded from our analysis by assumption.
	
	\end{itemize}		
For $d=5$ and $d\geq 7$, the spectrum is continuous and (at least plausibly) has a physically sensible bulk dual interpretation. A more detailed discussion can be found in \cite{Brust:2016gjy}.

	\newpage
	\section{Computations in Conformal Gravity}
	\label{app:confgrav}

In this appendix, we compute the three-point function of partially massless spin-2 fields in conformal gravity (reviewed in Section~\ref{ssec:confgrav}). 
We begin by canonically normalizing the fields and extracting the relevant cubic vertices from the action. 
The resulting three-point function is then computed using weight-shifting operators in Euclidean AdS, which is equivalent to the de Sitter calculation up to an overall phase.

\subsection{Cubic Interactions}

First, we introduce the canonically normalized spin-two fields via 
\begin{align}
h_{MN} = \frac{2 L}{|\lambda|} \,\hat h_{MN}\,,\quad f_{MN} = \frac{1}{|\lambda| L}\, \hat f_{MN}\,.
\end{align}
We expand the action of conformal gravity to cubic order in the field. 
After performing the field redefinitions as described in the main text, the Lagrangian up to cubic order separates into contributions from a massless graviton and a partially massless field:
\beq  {\cal L}=  {\cal L}_{\text{FP},2L^{-2}}(\hat h)- {\cal L}_{\rm FP,0}(\hat f)+{\cal L}_{\hat h \hat h \hat h}+{\cal L}_{\hat h \hat f \hat f}+{\cal L}_{\hat f\hat f\hat f}\,.
 \eeq
The explicit form of the cubic Lagrangians, put as on-shell as possible, is given by 
\begin{align} 
\frac{{\cal L}_{hhh}}{\sqrt{-g}} &= -{1\over M_{\rm Pl}}\Big(  2 h^{MN} \nabla_N h_{KL}\nabla^L h_M^{\ K}+h_{MN} h^{KL}\nabla_K\nabla_L h^{MN}+2L^{-2} h^M_{\ N}h^N_{\ K}h^K_{\ M}  \Big)\, ,\\[4pt] 
 \frac{{\cal L}_{hff}}{\sqrt{-g}} &=  {1\over M_{\rm Pl}}\Big(  2h_{MN} f^{KL}\nabla_K\nabla_L f^{MN} +f_{MN} h^{KL}\nabla_K\nabla_L f^{MN}  +2 f^{MK}\nabla_M f^{N L} \nabla_L h_{KN}  \\
  & \ \ \ + 4 h^{MK}\nabla_K f_{NL}\nabla^L f_M^{\ N}+6L^{-2}  h^M_{\ N}f^N_{\ K}f^K_{\ M} \Big)\,, \nonumber\\[4pt]
    \frac{{\cal L}_{ f  f  f}}{\sqrt{-g}} &=   -{2\over M_{\rm Pl}}\Big(  2 f^{MN} \nabla_N f_{KL}\nabla^L f_M^{\ K}+f_{MN} f^{KL}\nabla_K\nabla_L f^{MN}+2L^{-2} f^M_{\ N}f^N_{\ K}f^K_{\ M}  \Big)\, ,\label{Lfff}
\end{align}
where we have identified the Planck mass, $M_{\rm Pl}= |\lambda| /L$,
by matching the $hhh$ cubic structure to that obtained from the Einstein--Hilbert action. 
We see that the cubic structures are those of ordinary Einstein gravity for $hhh$, minimal coupling of the PM field in $hff$, and the special two-derivative cubic PM vertex $fff$ for the PM field (which is only PM invariant in $D=4$~\cite{deRham:2013wv}). There is no $fhh$ coupling.

\vskip 4pt
Note that the Planck scale is separated from the background curvature scale by the overall coupling $\lambda^2$, so $\lambda^2\gg 1$ is weak coupling. 
We can also see manifestly from \eqref{fafterrfredee} that we can truncate by setting $f_{MN}=0$ and killing the PM mode, leaving ordinary Einstein gravity. 

\subsection{Weight-Shifting Operators}

We now compute the three-point function in Euclidean AdS, with line element $\text{d}s^2 = z^{-2}(\text{d}z^2+\text{d} x^2)$ and 
AdS radius $L\equiv 1$, which is equivalent to the dS result up to an overall phase. 
The calculation is organized using the weight-shifting formalism in embedding space~\cite{Karateev:2017jgd}.

\vskip 4pt
The basic ingredient is the scalar bulk-to-boundary propagator given by
\begin{align}
    \Pi_\Delta =  \frac{{\cal C}_\Delta}{(-2P\cdot Y)^\Delta}\,,\quad {\cal C}_\Delta \equiv \frac{\Gamma(\Delta)}{2\pi^{\frac{d}{2}}\Gamma(\Delta-\frac{d}{2}+1)}\,,\label{scalarPi}
\end{align}
where $Y$ is a bulk coordinate in AdS$_{d+1}$ parameterized as~\cite{Costa:2014kfa}
\begin{align}
	Y^A = (Y^+,Y^-,Y^\mu) = \frac{1}{z}(1+z^2+|x|^2,x^\mu)\,,
\end{align}
with $Y^2=-1$. 
To construct spinning propagators, we use the following weight-shifting operators~\cite{Li:2022tby, Lee:2023qqx}:
\begin{align}
	\cE^A &= N_\cE \Big((\Delta+\ell) \delta_B^A + P^A \partial_{P^B}\Big) Z^B\label{cE2}\,,\\[-3pt]
	\cP^A &= N_\cP \Big( c_1 \delta^A_B + P^A \partial_{P^B} \Big)
\Big( c_2 \delta^B_C + Z^B \partial_{Z^C} \Big)
\Big( c_3 \delta^C_D - \partial^C_{Z} Z_{D}  \Big) \partial_{P_D} \,,\label{cP2}
\end{align}
 which raise the spin and dimension by one unit, respectively.
The  normalization factors are
\begin{align}
	N_\cE \equiv \frac{\Delta-\ell}{\Delta(\Delta-1)}\,,\quad N_\cP \equiv \frac{2i}{(\ell+1)(1-\Delta)(d-\Delta-2)(d-2\Delta-2)}\,,
\end{align}
and the coefficients in (\ref{cP2}) are~\cite{Costa:2018mcg} 
\be
c_1 \equiv 2 - d + 2 \Delta\,, \quad
c_2 \equiv 2 - d + \Delta - \ell\,, \quad
c_3 \equiv \Delta + \ell \,.
\ee
These operators satisfy a number of useful identities, such as
\be
\begin{aligned}
    (Y\cdot \cE) \Pi_{\Delta}&=0\,,\\[-2pt] (Y\cdot \cP) \Pi_{\Delta-1} &= i(d-\Delta)\Pi_{\Delta}\,,\\
    (Y\cdot\cE) \cP^A\Pi_{\Delta-1} &= i\cE^A \Pi_{\Delta}\,.\label{Pi2id2}
\end{aligned}
\ee
The spin-2 bulk-to-boundary propagator can then be obtained from the scalar one by
\begin{align}
\Pi_{\Delta}^{AB} & =  \cE^{A}\cE^{B}\Pi_\Delta \nonumber\\
& = {\cal C}_{\Delta,2}\,\frac{4(P^A\, Y\cdot Z-Z^A\, Y\cdot P)(P^B\, Y\cdot Z-Z^B\, Y\cdot P)}{(-2Y\cdot P)^{\Delta+2}} \,,
\end{align}
where the normalization of the bulk-to-boundary propagator in embedding space is~\cite{Costa:2014kfa}
\begin{align}
	{\cal C}_{\Delta,J} = \frac{(J+\Delta-1)\Gamma(\Delta)}{2\pi^{\frac{d}{2}}(\Delta-1)\Gamma(\Delta+1-\frac{d}{2})} \,.
\end{align}
Covariant derivatives of spinning propagators can also be written as
\begin{align}
	\nabla_C \Pi_{\Delta}^{AB} &= i\cE^A\cE^B \cP_C \Pi_{\Delta-1} - \Big[(d-\Delta) Y_C \cE^A\cE^B + Y^{(A}\cE^{B)}\cE_C\Big] \Pi_\Delta\,,\label{Pi2id}
\end{align}
which allows us to express all derivatives in terms of scalar propagators and boundary differential operators.

\subsection{Three-Point Function}

We now compute the boundary three-point function $\LA XXX\RA$ for a PM spin-2 field. The relevant cubic Lagrangian \eqref{Lfff} leads to the following integral for the three-point function:
\beq
\begin{aligned}
	\LA XXX\RA &= -\frac{2}{M_{\rm Pl}}\int\limits_{\text{AdS}_{d+1}} \hspace{-0.25cm} \Big(2\Pi^{CD}\nabla_D \Pi_{AB}\nabla^B {\Pi_C}^A  - \Pi^{AB}\nabla_A\Pi_{CD}\nabla_B \Pi^{CD} \\[-6pt]
	&\qquad\qquad\qquad\qquad\quad
	- 2{\Pi^A}_B{\Pi^B}_C{\Pi^C}_A\Big)\,,
\end{aligned}
\eeq
where the integral is performed over the bulk point $Y$ and $\Pi_{AB}$ denotes the PM spin-2 propagator with $\Delta=2$. In conformal gravity, this canonically normalized PM field has the wrong (negative) sign for the kinetic term, so that the propagator acquires an overall minus sign, $\Pi^{AB}=-\cE^A\cE^B \Pi_{d-1}$.
Note that we have integrated  the second term in \eqref{Lfff} by parts and the last term has a minus sign since we are doing the calculation in AdS.

\vskip 4pt
Now, using the identity \eqref{Pi2id}, we can turn all bulk covariant derivatives into boundary differential operators, up to some leftover $Y$ dependence that is fully cancelled after further using the identities \eqref{Pi2id2}. 
As a result, we can write the above spinning three-point function as a purely boundary differential operator acting on a scalar three-point function as
\begin{align}
	\langle XXX\rangle = -\frac{4}{M_{\rm Pl}}{\cal D}_{XXX} \langle \phi\phi\phi\rangle\,,
\end{align}
where $\langle \phi \phi \phi \rangle$ is a tree-level scalar three-point function\footnote{It is understood that each differential operator acts on a scalar three-point function with the correct scaling dimensions, so that we land on the correct dimension $\Delta=2$ for the PM three-point function. We omit these labels here for brevity.} and $\mathcal{D}_{XXX}$ is a differential operator built from weight-shifting operators that act only on boundary points. 
Explicitly, the latter takes the form
\beq
\begin{aligned}
	{\cal D}_{XXX} &= (\cE_1\cdot\cE_2)^2\,(\cE_3\cdot \cP_1)^2 + 2\cE_1\cdot\cE_2\,\cE_1\cdot\cE_3\,\cE_3\cdot \cP_1\,\cE_2\cdot \cP_3 + \text{cyc.} \\[3pt]
	&=\ :\!(\cE_1\cdot\cE_2\,\cE_3\cdot \cP_1 + \text{cyc.})^2\!:\,,
\end{aligned}
\label{equ:DXXX}
\eeq
where $:(\cdots):$ denotes ``normal ordering,'' with all $\mathcal{E}$ operators placed to the left of all $\mathcal{P}$ operators (this ordering is necessary due to the non-commutative nature of the differential operators involved). Interestingly, the operator in (\ref{equ:DXXX}) has a differential double copy structure, similar to the one found for the graviton three-point function~\cite{Lee:2022fgr,Li:2022tby,Lee:2023qqx}.\footnote{This is somewhat expected since the cubic Lagrangians for the two fields $f$ and $h$ are the same. More precisely, comparing this with the  differential operator for the graviton (stress tensor) three-point function, $\cD_{TTT}$, the difference is given by
\begin{align}
	{\cal D}_{XXX} - {\cal D}_{TTT} \propto\cE_1\cdot\cE_2\,\cE_1\cdot\cE_3\,\cE_2\cdot\cE_3\,.
\end{align}
This discrepancy is due to the slightly different translation between bulk derivatives and boundary differential operators depending on their scaling dimensions, even though their cubic Lagrangians take the same form.
}

\vskip 4pt
The scalar three-point function has the well-known form
\begin{align}
	\label{eq:bulk3pt1}
	\LA \phi_{\Delta_1}\phi_{\Delta_2}\phi_{\Delta_3}\RA &= \frac{c_{\Delta_1\Delta_2\Delta_3}}{(-2P_{12})^{\frac{\Delta_1+\Delta_2-\Delta_3}{2}}(-2P_{23})^{\frac{\Delta_2+\Delta_3-\Delta_1}{2}}(-2P_{31})^{\frac{\Delta_3+\Delta_1-\Delta_2}{2}}}\,,
\end{align}
where the normalization is
\begin{align}
	\label{eq:bulk3pt2}
	c_{\Delta_1\Delta_2\Delta_3}&\equiv {\cal C}_{\Delta_1} {\cal C}_{\Delta_2} {\cal C}_{\Delta_3} \frac{\pi^{\frac{d}{2}}}{2}\frac{\Gamma(\frac{\Delta_1+\Delta_2+\Delta_3-d}{2})\Gamma(\frac{\Delta_1+\Delta_2-\Delta_3}{2})\Gamma(\frac{\Delta_2+\Delta_3-\Delta_1}{2})\Gamma(\frac{\Delta_3+\Delta_1-\Delta_2}{2})}{\Gamma(\Delta_1)\Gamma(\Delta_2)\Gamma(\Delta_3)}\,,
\end{align}
where ${\cal C}_{\Delta}\equiv {\cal C}_{\Delta,0}$. Finally, acting with the differential operators, we obtain
\begin{align}
	\label{eq:XXXbulk}
	\LA XXX\RA = \frac{3}{128\pi^4 M_{\rm Pl}}\frac{12Q_1-2Q_2+Q_3+8Q_4}{P_{12}^2P_{13}^2P_{23}^2}\,,\quad Q_n = \begin{pmatrix}
		V_1^2V_2^2V_3^2\\
		H_{13}H_{23}V_1V_2+\text{cyc.}\\
		H_{12}^2V_3^2+\text{cyc.}\\
		H_{12}V_1V_2V_3^2+\text{cyc.}
	\end{pmatrix},
\end{align}
where $Q_n$ are the independent symmetric tensor structures.

\subsection{Anomalous Dimension}
\label{subsec:anomalous}

We now turn to the question of whether the PM field acquires an anomalous dimension.  
This effect originates from the presence of a double-trace deformation which explicitly breaks partial conservation. 
Concretely, the double-divergence of the PM operator is sourced by the composite operator
\begin{align}
	\partial_\mu \partial_\nu   X^{\mu\nu} =\ \frac{g}{\sqrt N} : X_{\mu\nu}X^{\mu\nu}:\,,\label{doubletr}
\end{align}
where $g$ is a numerical constant.  
As a consequence, the two-point function of the PM field develops an anomalous dimension $\gamma$, taking the form\footnote{Note that this result follows from the embedding-space normalization of a canonically normalized bulk field.  
This choice differs from that of \cite{Goon:2018fyu}, where the normalization was fixed in momentum space.  
Taking the inverse Fourier transform of their result, the (non-canonically normalized) two-point function can be written as
\begin{align}
	\langle XX\rangle 
	&= -i^{\,d+1}\,\tilde c_X\,\frac{H_{12}^2}{(-2P_{12})^{d+1}} \, , 
	\qquad 
	\tilde c_X \;=\; \left(\frac{M_{\rm Pl}}{H}\right)^{d-1}\,
	\frac{\Gamma(d+1)}{4\pi^{\frac{d}{2}}(d-1)\,\Gamma(\tfrac{d}{2}-1)} \, ,
\end{align}
where the factor of $i$ reflects the analytic continuation from AdS to dS. 
The normalization constant $\tilde c_X$ thus differs from $c_X$ by $d$-dependent numerical factors.  
For instance, in $d=3$, one finds
\beq
	\tilde c_X \;=\; \frac{3}{4\pi^2}\left(\frac{M_{\rm Pl}}{H}\right)^{d-1}.
\eeq
To keep everything self-consistent, we have instead adopted the embedding-space normalization throughout. 
 }
\beq
	\LA XX\RA = - c_X \frac{H_{12}^2}{(-2P_{12})^{4+\gamma}}\,,\quad\text{with}\quad c_X = {\cal C}_{2,2} = \frac{3}{\pi^2}\,,
	\label{equ:TWO}
\eeq
where the negative overall sign reflects the wrong-sign kinetic term of the PM field.  
Our goal is to determine $\gamma$ in terms of $g/\sqrt N$ and in turn in terms of $M_{\rm Pl}$. 
We accomplish this by comparing the double-divergence of the three-point function to the product of two-point functions.

\subsubsection{Three-Point Function}

We begin by fixing the ratio $g/\sqrt N$ using the three-point function computed in the previous subsection.  
There are two equivalent ways to compute its double-divergence.  
First, by directly acting with the derivatives on the explicit bulk expression \eqref{eq:XXXbulk}, one obtains
\begin{align}
	\partial_1^\mu \partial_1^\nu \LA  X_{\mu\nu}(x_1) X(x_2,z_2)X(x_3,z_3) \RA = \frac{9}{32\pi^4M_{\rm Pl}}\frac{(H_{12}+2V_1V_2)^2}{P_{12}^2P_{13}^2P_{23}^2}\,,
\end{align}
where we used
\beq
	X^{\mu\nu}(x) = \frac{2}{3}D_z^\mu D_z^\nu X(x,z)\,.
\eeq
Alternatively, we may use the operator relation \eqref{doubletr}.  
Inserting the composite operator and performing the Wick contractions gives
\begin{align}
	\partial_1^\mu\partial_1^\nu \LA  X_{\mu\nu}(x_1) X(x_2,z_2)X(x_3,z_3) \RA 
	&= \frac{g}{\sqrt N}\LA :\!X^{\mu\nu}(x_1)X_{\mu\nu}(x_1)\!:X(x_2)X(x_3)\RA\nonumber\\
	&= \frac{2g}{\sqrt N}\LA X^{\mu\nu}(x_1)X(x_2)\RA\LA X_{\mu\nu}(x_1)X(x_3)\RA\nonumber\\
	&=\frac{c_X^2 g}{32\sqrt N} \frac{(H_{12}+2V_1V_2)^2}{P_{12}^2P_{13}^2P_{23}^2}\,.
\end{align}
Comparing the two expressions above, we deduce that
\begin{align}
	\frac{g}{\sqrt N} = \frac{1}{M_{\rm Pl}}\,,
\end{align}
which is consistent with the expected scaling of $N$ with the Planck scale.

\subsubsection{Two-Point Function}

Next, we determine the anomalous dimension $\gamma$ by analyzing the two-point function.  
Taking the double-divergence of both operators in \eqref{equ:TWO} yields
\begin{align}
	\partial_1^{\mu}\partial_1^{\nu} \partial_2^{\rho}\partial_2^{\sigma}\LA X_{\mu\nu}(x_1)X_{\rho\sigma}(x_2)\RA =  \frac{5\hs c_X}{2}\frac{\gamma}{P_{12}^4} + {\cal O}(\gamma^2)\,.
\end{align}
On the other hand, using the double-trace relation \eqref{doubletr}, we find
\begin{align}
	\partial_1^{\mu}\partial_1^{\nu} \partial_2^{\rho}\partial_2^{\sigma}\LA X_{\mu\nu}(x_1)X_{\rho\sigma}(x_2)\RA &= \frac{g^2}{N}\LA :\!X^{\mu\nu}(x_1)X_{\mu\nu}(x_1)\!:\, :\!X^{\rho\sigma}(x_2)X_{\rho\sigma}(x_2)\!:\RA\nonumber\\
	&=\frac{2g^2}{N}\LA X^{\mu\nu}(x_1)X^{\rho\sigma}(x_2)\RA\LA X_{\mu\nu}(x_1)X_{\rho\sigma}(x_2)\RA\nonumber\\
	&=\frac{5\hs c_X^2}{8} \frac{g^2}{N}\frac{1}{P_{12}^4}+ {\cal O}(\gamma^2)\,.
\end{align}
By comparing the two expressions, we obtain the anomalous dimension
\begin{align}
	\gamma = \frac{c_X}{4} \frac{g^2}{N} = \frac{3}{4\pi^2M_{\rm Pl}^2}\,.\label{eq:gamma2}
\end{align}
The PM field thus acquires a {\it positive} anomalous dimension suppressed by the Planck scale, confirming the breaking of partial conservation.\footnote{While the PM two-point function flips sign when going from AdS to dS, the overall minus sign in the exponent of $\Psi$ (as opposed to the plus sign multiplying the two-point function in $Z_{\rm AdS}$) compensates for this, so the conclusion that $\gamma>0$ remains unaffected.} In de Sitter space, $\gamma>0$ implies a 
\textit{negative} mass shift of the PM field.
This is the result quoted in \eqref{eq:gamma} in the main text.

	\newpage
	\phantomsection
	\addcontentsline{toc}{section}{References}
	\bibliographystyle{utphys}
	{\linespread{1.075}
	\bibliography{PM-Refs}
	}
	
\end{document}